\documentstyle[preprint,aps,eqsecnum]{revtex}
\input epsf
\tightenlines
\begin{document}
\draft

\preprint{\vbox{\baselineskip=12pt{\hbox{CALT-68-2021}
\hbox{quant-ph/9602016}}}}

\title{Efficient networks for quantum factoring}
\author{David Beckman\footnote{\tt beckman@theory.caltech.edu}, Amalavoyal N.
Chari\footnote{\tt chari@cco.caltech.edu}, Srikrishna Devabhaktuni\footnote{\tt
dsrikris@cco.caltech.edu}, and John Preskill\footnote{\tt
preskill@theory.caltech.edu}}
\address{California Institute of Technology,
Pasadena, CA 91125}

\maketitle

\begin{abstract}

We consider how to optimize memory use and computation time in operating a
quantum computer.  In particular, we estimate the number of memory qubits and
the number of operations required to perform factorization, using the algorithm
suggested by Shor.  A $K$-bit number can be factored in time of order $K^3$
using a machine capable of storing $5K+1$ qubits. Evaluation of the modular
exponential function (the bottleneck of Shor's algorithm) could be achieved
with about $72 K^3$ elementary quantum gates; implementation using a linear ion
trap would require about $396K^3$ laser pulses.  A proof-of-principle
demonstration of quantum factoring (factorization of 15) could be performed
with only 6 trapped ions and 38 laser pulses. Though the ion trap may never be
a useful computer, it will be a powerful device for exploring experimentally
the properties of entangled quantum states.
\end{abstract}


\narrowtext
\parskip=5pt 
\section{Introduction and Summary}

Recently, Shor \cite{shor} has exhibited a probabilistic algorithm that enables
a quantum computer to find a nontrivial factor of a large composite number $N$
in a time bounded from above by a polynomial in $\log(N)$.  As it is widely
believed that no polynomial-time factorization algorithm exists for a classical
Turing machine, Shor's result indicates that a quantum computer can efficiently
perform interesting computations that are intractable on a classical computer,
as had been anticipated by Feynman \cite{feynman}, Deutsch \cite{deutsch}, and
others \cite{others}.

Furthermore, Cirac and Zoller \cite{cirac} have suggested an ingenious scheme
for performing quantum computation using a potentially realizable device.  The
machine they envisage is an array of cold ions confined in a linear trap, and
interacting with laser beams. Such linear ion traps have in fact been built
\cite{wineland}, and these devices are remarkably well protected from the
debilitating effects of decoherence. Thus, the Cirac-Zoller proposal has
encouraged speculation that a proof-of-principle demonstration of quantum
factoring might be performed in the reasonably near future.

Spurred by these developments, we have studied the computational resources that
are needed to carry out the factorization algorithm using the linear ion trap
computer or a comparable device.  Of particular interest is the inevitable
tension between two competing requirements.  Because of practical limitations
on the number of ions that can be stored in the trap, there is a strong
incentive to minimize the number of qubits in the device by managing memory
resources frugally.  On the other hand, the device has a characteristic
decoherence time scale, and the computation will surely crash if it takes much
longer that the decoherence time. For this reason, and because optimizing speed
is desirable anyway, there is a strong incentive to minimize the total number
of elementary operations that must be completed during the computation.  A
potential rub is that frugal memory management may result in longer computation
time.

One of our main conclusions, however, is that substantial squeezing of the
needed memory space can be achieved without sacrificing much in speed.  A
quantum computer capable of storing $5K+1$ qubits can run Shor's algorithm to
factor a $K$-bit number $N$ in a time of order $K^3$.  Faster implementations
of the algorithm are possible for asymptotically large $N$, but these require
more qubits, and are relatively inefficient for values of $N$ that are likely
to be of practical interest.  For these values of $N$, a device with unlimited
memory using our algorithms would be able to run only a little better than
twice as fast as fast as a device that stores $5K+1$ qubits. Further squeezing
of the memory space is also possible, but would increase the computation time
to a higher power of $K$.

Shor's algorithm (which we will review in detail in the next section) includes
the evaluation of the modular exponential function; that is, a unitary
transformation $U$ that acts on elements of the computational basis as
\begin{equation}
\label{exp}
U: |a \rangle_i |0\rangle_o \longmapsto |a \rangle_i |x^a (mod ~N)\rangle_o \ .
\end{equation}
Here $N$ is the $K$-bit number to be factored, $a$ is an $L$-bit number (where
usually $L\approx 2K$), and $x$ is a randomly selected positive integer less
than $N$ that is relatively prime to $N$; $|\cdot\rangle_i$ and
$|\cdot\rangle_o$ denote the states of the ``input'' and ``output'' registers
of the machine, respectively.  Shor's algorithm aims to find the period of this
function, the {\it order} of $x$ mod $N$. From the order of $x$, a factor of
$N$ can be extracted with reasonable likelihood, using standard results of
number theory.

To perform factorization, one first prepares the input register in a coherent
superposition of all possible $L$-bit computational basis states:
\begin{equation}
{1\over 2^{L/2}}\sum_{a=0}^{2^{L}-1}|a\rangle_i \ .
\end{equation}
Preparation of this state is relatively simple, involving just $L$ one-qubit
rotations (or, for the Cirac-Zoller device, just $L$ laser pulses applied to
the ions in the trap). Then the modular exponential function is evaluated by
applying the transformation $U$ above.  Finally, a discrete Fourier
transformation is applied to the input register, and the input register is
subsequently measured.  From the measured value, the order of $x$ mod $N$ can
be inferred with reasonable likelihood.

Shor's crucial insight was that the discrete Fourier transform can be evaluated
in polynomial time on a quantum computer.  Indeed, its evaluation is remarkably
efficient.  With an improvement suggested by Coppersmith \cite{coppersmith} and
Deutsch \cite{deutschFFT}, evaluation of the $L$ bit Fourier transform is
accomplished by composing $L$ one-qubit operations and ${1\over 2}L(L-1)$
two-qubit operations.  (For the Cirac-Zoller device, implementation of the
discrete Fourier transform requires $L(2L-1)$ distinct laser pulses.)

The bottleneck of Shor's algorithm is the rather more mundane task of
evaluating the modular exponential function, {\it i.e.}, the implementation of
the transformation $U$ in Eq.\ (\ref{exp}).  This task demands far more
computational resources than the rest of the algorithm, so we will focus on
evaluation of this function in this paper.  There is a well-known (classical)
algorithm for evaluating the modular exponential that involves O$(K^3)$
elementary operations, and we will make use of this algorithm here.

The main problem that commands our attention is the management of the
``scratchpad'' space that is needed to perform the computation; that is, the
extra qubits aside from the input and output registers that are used in
intermediate steps of the computation.  It is essential to erase the scratchpad
before performing the discrete Fourier transform on the input register.  Before
the scratchpad is erased, the state of the machine will be of the form
\begin{equation}
\label{garbage}
{1\over 2^{L/2}}\sum_a |a \rangle_i |x^a (mod ~N)\rangle_o |g(a)\rangle_s \ ,
\end{equation}
where $|g(a)\rangle_s$ denotes the ``garbage'' stored in the scratchpad.  If we
were now to perform the discrete Fourier transform on $|\cdot\rangle_i$, we
would be probing the periodicity properties of the function $x^a~ ({\rm mod}
{}~N)\otimes g(a)$, which may be quite different than the periodicity
properties of $x^a ~(mod ~N)$ that we are interested in.  Thus, the garbage in
the scratchpad must be erased, but the erasure is a somewhat delicate process.
To avoid destroying the coherence of the computation, erasure must be performed
as a reversible unitary operation.

In principle, reversible erasure of the unwanted garbage presents no
difficulty.  Indeed, in his pioneering paper on reversible computation, Bennett
\cite{bennett} formulated a general strategy for cleaning out the scratchpad:
one can run the calculation to completion, producing the state Eq.\
(\ref{garbage}), copy the result from the output register to another ancillary
register, and then run the computation backwards to erase both the output
register and the scratchpad.  However, while this strategy undoubtedly works,
it may be far from optimal, for it may require the scratchpad to be much larger
than is actually necessary.  We can economize on scratchpad space by running
subprocesses backwards at intermediate stages of the computation, thus freeing
some registers to be reused in a subsequent process. (Indeed, Bennett himself
\cite{bennett_trade} described a general procedure of this sort that greatly
reduces the space memory requirements.) However, for this reduction in required
scratchpad space, we may pay a price in increased computation time.

One of our objectives in this paper is to explore this tradeoff between memory
requirements and computation time.  This tradeoff is a central general issue in
quantum computation (or classical reversible computation) that we have
investigated by studying the implementation of the modular exponential
function, the bottleneck of Shor's factorization algorithm.  We have
constructed a variety of detailed quantum networks that evaluate the modular
exponential, and we have analyzed the complexity of our networks. A convenient
(though somewhat arbitrary) measure of the complexity of a quantum algorithm is
the number of laser pulses that would be required to implement the algorithm on
a device like that envisioned by Cirac and Zoller.  We show that if $N$ and $x$
are $K$-bit classical numbers and $a$ is an $L$-bit quantum number, then, on a
machine with $2K+1$ qubits of scratch space, the computation of $x^a~({\rm
mod}~N)$ can be achieved with $198 L\left[K^2 +O(K)\right]$ laser pulses.  If
the scratch space of the machine is increased by a single qubit, the number of
pulses can be reduced by about 6\% (for $K$ large), and if $K$ qubits are
added, the improvement in speed is about 29\%.  We also exhibit a network that
requires only $K+1$ scratch qubits, but where the required number of pulses is
of order $LK^4$.

The smallest composite number to which Shor's algorithm may be meaningfully
applied is $N$=15.  (The algorithm fails for $N$ even and for $N=p^\alpha$, $p$
prime.) Our general purpose algorithm (which works for any value of $N$), in
the case $N=15$ (or $K=4$, $L=8$), would require 21 qubits and about 15,000
laser pulses.  In fact, a much faster special purpose algorithm that exploits
special properties of the number 15 can also be constructed---for what it's
worth, the special purpose algorithm could ``factor 15'' with 6 qubits and only
38 pulses.

The fastest modern digital computers have difficulty factoring numbers larger
than about 130 digits (432 bits). According to our estimates, to apply Shor's
algorithm to a number of this size on the ion trap computer (or a machine of
similar design) would require about 2160 ions and $3\times 10^{10}$ laser
pulses. The ion trap is an intrinsically slow device, for the clock speed is
limited by the frequency of the fundamental vibrational mode of the trapped
ions.  Even under very favorable conditions, it seems unlikely that more than
$10^4$ operations could be implemented per second. For a computation of
practical interest, the run time of the computation is likely to outstrip by
far the decoherence time of the machine. It seems clear that a practical
quantum computer will require a much faster clock speed than can be realized in
the Cirac-Zoller design.  For this reason, a design based on cavity quantum
electrodynamics (in which processing involves excitation of photons rather than
phonons) \cite{kimble,pellizzari} may prove more promising in the long run.

Whatever the nature of the hardware, it seems likely that a practical quantum
computer will need to invoke some type of error correction protocol to combat
the debilitating effects of decoherence \cite{noise}.  Recent progress in the
theory of error-correcting quantum codes \cite{error_correct} has bolstered the
hope that real quantum computers will eventually be able to perform interesting
computational tasks .

Although we expect that the linear ion trap is not likely to ever become a
practical computer, we wish to emphasize that it is a marvelous device for the
experimental studies of the peculiar properties of entangled quantum states.
Cirac and Zoller \cite{cirac} have already pointed out that maximally entangled
states of $n$ ions \cite{greenberger} can be prepared very efficiently.  Since
it is relatively easy to make measurements in the Bell operator basis for any
pair of entangled ions in the trap \cite{ekert_deutsch}, it should be possible
to, say, demonstrate the possibility of quantum teleportation
\cite{teleportation} (at least from one end of the trap to the other).

In Sec.\ II of this paper, we give a brief overview of the theory of quantum
computation and describe Shor's algorithm for factoring.  Cirac and Zoller's
proposed implementation of a quantum computer using a linear ion trap is
explained in Sec.\ III.  Sec.\ IV gives a summary of the main ideas that guide
the design of our modular exponentiation algorithms; the details of the
algorithms are spelled out in Sec.\ V, and the complexity of the algorithms is
quantified in Sec.\ VI. The special case $N=15$ is discussed in Sec.\ VII.  In
Sec.\ VIII, we propose a simple experimental test of the quantum Fourier
transform.  Finally, in Appendix A,  we describe a scheme for further improving
the efficiency of our networks.

Quantum networks that evaluate the modular exponential function have also been
designed and analyzed by Despain \cite{despain}, by Shor \cite{shor_long} and
by Vedral, Barenco, and Ekert \cite{vedral}.  Our main results are in
qualitative agreement with the conclusions of these authors, but the networks
we describe are substantially more efficient.

\section{Quantum Computation and Shor's Factorization Algorithm}

\subsection{Computation and physics}

The theory of computation would be bootless if the computations that it
describes could not actually be carried out using physically realizable
devices.  Hence it is really the task of physics to characterize what is
computable, and to classify the efficiency of computations.  The physical world
is quantum mechanical.  Therefore, the foundations of the theory of computation
must be quantum mechanical as well.  The classical theory of computation ({\it
e.g}, the theory of the universal Turing machine) should be viewed as an
important special case of a more general theory.

A ``quantum computer'' is a computing device that invokes intrinsically
quantum-mechanical phenomena, such as interference and
entanglement.\footnote{For a lucid review of quantum computation and Shor's
algorithm, see \cite{ekert_jozsa}.}  In fact, a Turing machine can simulate a
quantum computer to any desired accuracy (and vice versa); hence, the classical
theory and the more fundamental quantum theory of computation agree on what is
computable \cite{deutsch}.  But they may disagree on the classification of {\it
complexity}; what is easy to compute on a quantum computer may be hard on a
classical computer.

\subsection{Bits and qubits}

In classical theory, the fundamental unit of information is the {\it bit}---it
can take either of two values, say 0 and 1.  All classical information can be
encoded in bits, and any classical computation can be reduced to fundamental
operations that flip bits (changing 0 to 1 or 1 to 0) conditioned on the values
of other bits.

In the quantum theory of information, the bit is replaced by a more general
construct---the {\it qubit}.  We regard $\bigl |{0}\bigr\rangle$ and $ \bigl
|{1}\bigr\rangle$ as the orthonormal basis states for a two-dimensional complex
vector space.  The state of a qubit (if ``pure'') can be any normalized vector,
denoted
\begin{equation}
\label{qubit_state}
c_0 \bigl |{0}\bigr\rangle + c_1 \bigl |{1}\bigr\rangle \ ,
\end{equation}
where $c_0$ and $c_1$ are complex numbers satisfying $|c_0|^2+|c_1|^2=1$.
A classical bit can be viewed as the special case in which the state of the
qubit is always either $c_0=1,~c_1=0$ or $c_0=0,~c_1=1$.

The possible pure states of a qubit can be parametrized by two real numbers.
(The overall phase of the state is physically irrelevant.)  Nevertheless, only
one bit of classical information can be stored in a qubit and reliably
recovered.  If the value of the qubit in the state Eq.\ (\ref{qubit_state}) is
measured, the result is 0 with probability $|c_0|^2$ and 1 with probability
$|c_1|^2$; in the case $|c_0|^2=|c_1|^2={1\over 2}$, the outcome of the
measurement is a random number, and we recover no information at all.

A string of $n$ classical bits can take any one of $2^n$ possible values.  For
$n$ qubits, these $2^n$ classical strings are regarded as the basis states for
a complex vector space of dimension $2^n$, and a pure state of $n$ qubits is a
normalized vector in this space.

\subsection{Processing}
In a quantum computation, $n$ qubits are initially prepared in an
algorithmically simple input state, such as
\begin{equation}
\bigl |{\rm input}\bigr\rangle=\bigl |{0}\bigr\rangle\bigl
|{0}\bigr\rangle\bigl |{0}\bigr\rangle\dots \bigl |{0}\bigr\rangle \ .
\end{equation}
Then a unitary transformation $U$ is applied to the input state, yielding an
output state
\begin{equation}
\bigl |{\rm output}\bigr\rangle = U\bigl |{\rm input}\bigr\rangle
\end{equation}
Finally, a set of commuting observables ${\cal O}_1, {\cal O}_2, {\cal O}_3,
\dots$ is measured in the output state.  The measured values of these
observables constitute the outcome of the computation.  Since the output state
is not necessarily an eigenstate of the measured observables, the quantum
computation is not deterministic---rather, the same computation, performed many
times, will generate a probability distribution of possible outcomes.

(Note that the observables that are measured in the final step are assumed to
be simple in some sense; otherwise the transformation $U$ would be superfluous.
 Without loss of generality, we may specify that the values of all qubits (or a
subset of the qubits) are measured at the end of the computation; that is, the
$j$th qubit $\bigl |{\cdot}\bigr\rangle_j$ is projected onto the
``computational basis'' $\lbrace \bigl |{0}\bigr\rangle_j~, ~ \bigl
|{1}\bigr\rangle_j \rbrace $.)

To characterize the complexity of a computation, we must formulate some rules
that specify how the transformation $U$ is constructed.  One way to do this is
to demand that $U$ is expressed as a product of elementary unitary
transformations, or ``quantum gates,'' that act on a bounded number of qubits
(independent of $n$).  In fact, it is not hard to see \cite{universal} that
``almost any'' two-qubit unitary transformation, together with qubit swapping
operations, is universal for quantum computation.  That is, given a generic
$4\times 4$ unitary matrix $\tilde U$, let $\tilde U^{(i,j)}$ denote $\tilde U$
acting on the $i$th and $j$th qubits according to
\begin{equation}
\tilde U^{(i,j)}: \>\>\bigl |{\epsilon_i}\bigr\rangle_i\bigl
|{\epsilon_j}\bigr\rangle_j\longmapsto \tilde
U_{\epsilon_i\epsilon_j,\epsilon'_i\epsilon'_j}
|{\epsilon'_i}\bigr\rangle_i\bigl |{\epsilon'_j}\bigr\rangle_j
\end{equation}
Then any $2^n\times 2^n$ unitary transformation $U$ can be approximated to
arbitrary precision by a finite string of $\tilde U^{(i,j)}$'s,
\begin{equation}
U\simeq \tilde U^{(i_T,j_T)}\dots\tilde U^{(i_2,j_2)}\tilde U^{(i_1,j_1)}
\end{equation}
The length $T$ of this string (the ``time'') is a measure of the complexity of
the quantum computation.

Determining the precise string of $\tilde U^{(i,j)}$'s that is needed to
perform a particular computational task may itself be computationally
demanding.  Therefore, to have a reasonable notion of complexity, we should
require that a conventional computer (a Turing machine) generates the
instructions for constructing the unitary transformation $U$.  The complexity
of the computation is actually the sum of the complexity of the classical
computation and the complexity of the quantum computation.  Then we may say
that a problem is {\it tractable} on a quantum computer if the computation that
solves the problem can be performed in a time that is bounded from above by a
polynomial in $n$, the number of qubits contained in the quantum register.
This notion of tractability has the nice property that it is largely
independent of the details of the design of the machine---that is, the choice
of the fundamental quantum gates.  The quantum gates of one device can be
simulated to polynomial accuracy in polynomial time by the quantum gates
of another device.

It is also clear that a classical computer can simulate a quantum computer to
any desired accuracy---all that is required to construct the state $\bigl |{\rm
output}\bigr\rangle$ is repeated matrix multiplication, and we can simulate the
final measurement of the observables by expanding $\bigl |{\rm
output}\bigr\rangle$ in a basis of eigenstates of the observables.  However,
the classical simulation may involve matrices of exponentially large size ($U$
is a $2^n\times 2^n$ matrix), and so may take an exponentially long time.  It
was this simple observation that led Feynman \cite{feynman} to suggest that
quantum computers may be able to solve certain problems far more efficiently
than classical computers.
\subsection{Massive Parallelism}

Deutsch \cite{deutsch} put this suggestion in a more tangible form by
emphasizing that a quantum computer can exploit ``massive quantum
parallelism.''  Suppose we are interested in studying the properties of a
function $f$ defined on the domain of  nonnegative integers
$0,1,2,\dots,2^L-1$.  Imagine that a unitary transformation $U_f$ can be
constructed that efficiently computes $f$:
\begin{eqnarray}
\label{f_compute}
U_f:\>\> &&\bigl | \left(i_{L-1}i_{L-2}\dots i_1
i_0\right)\bigr\rangle_{in}\bigl |\left(00\dots00\right)\bigr\rangle_{out}
\nonumber\\
\longmapsto &&\bigl | \left(i_{L-1}i_{L-2}\dots i_1
i_0\right)\bigr\rangle_{in}\bigl |f(i_{L-1}i_{L-2}\dots i_1
i_0)\bigr\rangle_{out} \ .
\end{eqnarray}
Here $\left(i_{L-1}i_{L-2}\dots i_1 i_0\right)$ is an integer expressed in
binary notation, and $\bigl | \left(i_{L-1}i_{L-2}\dots i_1
i_0\right)\bigr\rangle$ denotes the corresponding basis state of $L$ qubits.
Since the function $f$ might not be invertible, $U_f$ has been constructed to
leave the state in the $\bigl | \cdot\bigr\rangle_{in}$ register undisturbed,
to ensure that it is indeed a reversible operation.

Eq.\ (\ref{f_compute}) defines the action of $U_f$ on each of $2^L$ basis
states, and hence, by linear superposition, on all states of a
$2^L$-dimensional Hilbert space.  In particular, starting with the state $\bigl
| \left(00\dots00\right)\bigr\rangle_{in}$, and applying single-qubit unitary
transformations to each of the $L$ qubits, we can easily prepare the state
\begin{eqnarray}
\label{rotate_bits}
\left({1\over\sqrt{2}}\bigl | 0\bigr\rangle + {1\over\sqrt{2}}\bigl |
1\bigr\rangle\right)^L \>&&=\>{1\over 2^{L/2}}\sum_{i_{L-1}=0}^1\dots
\sum_{i_{1}=0} ^1\sum_{i_{0}=0}^1 \bigl | \left(i_{L-1}i_{L-2}\dots i_1
i_0\right)\bigr\rangle_{in}\nonumber\\
&&\equiv \>{1\over 2^{L/2}}\sum_{x=0}^{2^L-1}\bigl |x\bigr\rangle_{in}\ ,
\end{eqnarray}
an equally weighted coherent superposition of all of the $2^L$ distinct basis
states.  With this input, the action of $U_f$ prepares the state
\begin{equation}
\label{massive}
\bigl |\psi_f\bigr\rangle\> \equiv \>{1\over 2^{L/2}}\sum_{x=0}^{2^L-1}\bigl
|x\bigr\rangle_{in}\bigl |f(x)\bigr\rangle_{out}\ .
\end{equation}
The highly entangled quantum state Eq.\ (\ref{massive}) exhibits what Deutsch
called ``massive parallelism.''  Although we have run the computation (applied
the unitary transformation $U_f$) only once, in a sense this state encodes the
value of the function $f$ for each possible value of the input variable $x$.
Were we to measure the value of all the qubits of the input register, obtaining
the result $x=a$, a subsequent measurement of the output register would reveal
the value of $f(a)$.  Unfortunately, the measurement will destroy the entangled
state, so the procedure cannot be repeated.  We succeed, then, in unambiguously
evaluating $f$ for only a single value of its argument.

\subsection{Periodicity}
\label{sec:ft}

Deutsch \cite{deutsch} emphasized, however, that certain global properties of
the function $f$ {\it can} be extracted from the state Eq.\ (\ref{massive}) by
making appropriate measurements.  Suppose, for example, that $f$ is a {\it
periodic} function (defined on the nonnegative integers), whose period $r$ is
much less than $2^L$ (where $r$ does not necessarily divide $2^L$), and that we
are interested in finding the period.  In general, determining $r$ is a
computationally difficult task (for a classical computer) if $r$ is large.
Shor's central observation is that a quantum computer, by exploiting quantum
interference, can determine the period of a function efficiently.

Given the state Eq.\ (\ref{massive}), this computation of the period can be
performed by manipulating (and ultimately measuring) only the state of the
input register---the output register need not be disturbed.  For the purpose of
describing the outcome of such measurements, we may trace over the unobserved
state of the output register, obtaining the mixed density matrix
\begin{equation}
\label{periodic_density}
\rho_{in,f} \> \equiv \> {\rm tr}_{out}\left(\bigl | \psi_f\bigr \rangle
\bigl\langle \psi_f \bigr | \right) \> = \> {1\over r}\sum_{k=0}^{r-1} \bigl
|\psi_k \bigr \rangle\bigl\langle \psi_k \bigr | \ ,
\end{equation}
where
\begin{equation}
\label{periodic_state}
 \bigl |\psi_k \bigr \rangle \>=\> {1\over\sqrt{{\cal N}}} \sum_{j=0}^{{\cal
N}-1} \bigl |x=k+rj \bigr \rangle\>_{in}
\end{equation}
is the coherent superposition of all the input states that are mapped to a
given output. (Here ${\cal N}-1$ is the greatest integer less than
$(2^L-k)/r$.)

Now, Shor showed that the unitary transformation
\begin{equation}
\label{ft}
FT:\>\> \bigl | x\bigr \rangle \longmapsto {1\over 2^{L/2}} \sum_{y=0}^{2^L-1}
e^{2\pi i xy/2^L} \bigl | y\bigr \rangle
\end{equation}
(the Fourier transform) can be composed from a number of elementary quantum
gates that is bounded from above by a polynomial in $L$.  The Fourier transform
can be used to probe the periodicity properties of the state Eq.\
(\ref{periodic_density}).  If we apply $FT$ to the input register and then
measure its value $y$, the outcome of the measurement is governed by the
probability distribution
\begin{equation}
\label{ft_prob}
P(y)\> = \> {{\cal N}\over 2^L}\left| {1\over{\cal N}} \sum_{j=0}^{{\cal N}-1}
e^{2\pi iyrj/2^L}\right|^2 \ .
\end{equation}
This probability distribution is strongly peaked about values of $y$ of the
form
\begin{equation}
\label{peak_cond}
{y\over 2^L}\> = \>{{\rm integer}\over r} \pm O(2^{-L}) \ ,
\end{equation}
where the integer is a random number less than $r$.  (For other values of $y$,
the phases in the sum over $j$ interfere destructively.)  Now suppose that the
period $r$ is known to be less than $2^{L/2}$. The minimal spacing between two
distinct rational numbers, both with denominator less than $2^{L/2}$ is
$O(2^{-L})$.  Therefore, if we measure $y$, the rational number with
denominator less than $2^{L/2}$ that is closest to  $y/2^L$ is reasonably
likely to be a rational number with denominator $r$, where the numerator is a
random number less than $r$.  Finally, it is known that if positive integers
$r$ and $s<r$ are randomly selected, then $r$ and $s$ will be relatively prime
with a probability of order $1/\log\log r$.  Hence, even after the rational
number is reduced to lowest terms, it is not unlikely that the denominator will
be $r$.

We conclude then (if $r$ is known to be less than $2^{L/2}$), that each time we
prepare the state Eq.\ (\ref{massive}), apply the $FT$ to the input register,
and then measure the input register, we have a probability of order $1/\log\log
r> 1/\log L$ of successfully inferring from the measurement the period $r$ of
the function $f$.  Hence, if we carry out this procedure a number of times that
is large compared to $\log L$, we will find the period of $f$ with probability
close to unity.

All that remains to be explained is how the construction of the unitary
transformation $FT$ is actually carried out.  A simpler construction than the
one originally presented by Shor \cite {shor} was later suggested by
Coppersmith \cite{coppersmith} and Deutsch \cite{deutschFFT}.  (It is, in fact,
the standard fast Fourier transform, adapted for a quantum computer.)  In their
construction, two types of elementary quantum gates are used.  The first type
is a single-qubit rotation
\begin{equation}
\label{qubit_rot}
U^{(j)}: \>\> \pmatrix{\bigl|0\bigr\rangle_j\cr \bigl|1\bigr\rangle_j}
\longmapsto {1\over\sqrt{2}}\pmatrix{1&1\cr 1&-1}
\pmatrix{\bigl|0\bigr\rangle_j\cr \bigl|1\bigr\rangle_j}
\ ,
\end{equation}
the same transformation that was used to construct the state Eq.\
(\ref{rotate_bits}).  The second type is a two-qubit conditional phase
operation
\begin{equation}
\label{cond_phase}
V^{(j,k)}(\theta): \>\> \bigl|\epsilon\bigr\rangle_j \bigl|\eta\bigr\rangle_k
\longmapsto e^{i\epsilon\eta \theta} \bigl|\epsilon\bigr\rangle_j
\bigl|\eta\bigr\rangle_k \ .
\end{equation}
That is, $V^{(j,k)}(\theta)$ multiplies the state by the phase $e^{i\theta}$ if
both the $j$th and $k$th qubits have the value 1, and acts trivially otherwise.

It is not difficult to verify that the transformation
\begin{eqnarray}
\label{hat_ft}
\hat{FT}&& \equiv \left\{U^{(0)}V^{(0,1)}(\pi/2)V^{(0,2)}(\pi/4)\cdots
V^{(0,L-1)}(\pi/2^{L-1})\right\}\cdots\nonumber\\ &&
\cdots\left\{U^{(L-3)}V^{(L-3,L-2)}(\pi/2)V^{(L-3,L-1)}
(\pi/4)\right\}\cdot\left\{U^{(L-2)}V^{(L-2,L-1)}(\pi/2)\right\}
\cdot\left\{U^{(L-1)}\right\}
\end{eqnarray}
acts as specified in Eq.\ (\ref{ft}), except that the order of the qubits in
$y$ is reversed.\footnote{For a lucid explanation, see \cite{shor_long}.}
(Here the transformation furthest to the {\it right} acts first.) We may act on
the input register with $\hat {FT}$ rather than $FT$, and then reverse the bits
of $y$ after the measurement.  Thus, the implementation of the Fourier
transform is achieved by composing altogether $L$ one-qubit gates and
$L(L-1)/2$ two-qubit gates.

Of course, in an actual device, the phases of the $V^{(j,k)}(\theta)$ gates
will not be rendered with perfect accuracy.  Fortunately, the peaking of the
probability distribution in Eq.\ (\ref{ft_prob}) is quite robust.  As long as
the errors in the phases occurring in the sum over $j$ are small compared to
$2\pi$, constructive interference will occur when the condition Eq.\
(\ref{peak_cond}) is satisfied.  In particular, the gates in Eq.\
(\ref{hat_ft}) with small values of $\theta=\pi/2^{|j-k|}$ can be omitted,
without much affecting the probability of finding the correct period of the
function $f$.  Thus (as Coppersmith \cite{coppersmith} observed), the time
required to execute the $FT$ operation to fixed accuracy increases only
linearly with $L$.

\subsection{Factoring}
\label{sec:factoring}
The above observations show that a quantum computer can find the prime factors
of a number efficiently, for it is well known that factoring can be reduced to
the problem of finding the period of a function.  Suppose we wish to find a
nontrivial prime factor of the positive integer $N$.  We choose a random number
$x<N$.  We can efficiently check, using Euclid's algorithm, whether $x$ and $N$
have a common factor.  If so, we have found a factor of $N$, as desired.  If
not, let us compute the period of the modular exponential function
\begin{equation}
f_{N,x}(a)\> \equiv \> x^a ~({\rm mod}~ N) \ .
\end{equation}
The period is the smallest positive $r$ such that $x^r\equiv 1 ~({\rm mod}~
N)$,  called the {\it order} of $x$ mod $N$.  It exists whenever $N$ and $x<N$
have no common factor.

Now suppose that $r$ is even, and that $x^{r/2}\not\equiv - 1 ~({\rm mod}~ N)$.
 Then, since $N$ divides the product $\left(x^{r/2}+1\right)
\left(x^{r/2}-1\right)=x^r-1$, but does not divide either one of the factors
$\left(x^{r/2}\pm 1\right)$, $N$ must have a common factor with each of
$\left(x^{r/2}\pm 1\right)$.  This common factor, a nontrivial factor of $N$,
can then be efficiently computed.

It only remains to consider how likely it is, given a random $x$ relatively
prime to $N$, that the conditions $r$  even and $x^{r/2}\not\equiv - 1 ~{(\rm
mod)}~N$ are satisfied.  In fact, it can be shown \cite{shor_long,ekert_jozsa}
that, for $N$ odd, the probability that these conditions are met is at least
$1/2$, except in the case where $N$ is a prime power ($N=p^\alpha$, $p $
prime).  (The trouble with $N=p^\alpha$ is that in this case $\pm 1$ are the
{\it only} ``square roots'' of 1 in multiplication mod $N$, so that, even if
$r$ is even, $x^{r/2}\equiv -1~ ({\rm mod}~N)$ will always be satisfied.)
Anyway, if $N$ is of this exceptional type (or if $N$ is even), it can be
efficiently factored by conventional (classical) methods.

Thus, Shor formulated a probabilistic algorithm for factoring $N$ that will
succeed with probability close to 1 in a time that is bounded from above by a
polynomial in $\log N$.  To factor $N$ we choose $L$ so that, say, $N^2 \le
2^L<2N^2$.  Then, since we know that $r<N<2^{L/2}$ we can use the method
described above to efficiently compute the period $r$ of the function
$f_{N,x}$.  We generate the entangled state Eq.\ (\ref{massive}), apply the
Fourier transform, and measure the input register, thus generating a candidate
value of $r$.  Then, a classical computer is used to find $gcd(x^{r/2}-1,N)$.
If there is a nontrivial common divisor, we have succeeded in finding a factor
of $N$.  If not, we repeat the procedure until we succeed.

Of course, it is implicit in the above description that the evaluation of the
function $f_{N,x}$ can be performed efficiently on the quantum computer.  The
computational complexity of $f_{N,x}$ is, in fact, the main topic of this
paper.

\subsection{Outlook}

It is widely believed that no classical algorithm can factor a large number in
polynomially bounded time (though this has never been rigorously demonstrated).
 The existence of Shor's algorithm, then, indicates that the classification of
complexity for quantum computation differs from the corresponding classical
classification.  Aside from being an interesting example of an intrinsically
hard problem, factoring is also of some practical interest---the security of
the widely used RSA public key cryptography scheme \cite{rsa} relies on the
presumed difficulty of factoring large numbers.

It is not yet known whether a quantum computer can efficiently solve
``NP-complete'' problems, which are believed to be intrinsically more difficult
than the factoring problem.  (The ``traveling salesman problem'' is a notorious
example of an NP-complete problem.)  It would be of great fundamental interest
(and perhaps of practical interest) to settle this question.  Conceivably, a
positive answer could be found by explicitly exhibiting a suitable algorithm.
In any event, better characterizing the class of problems that can be solved in
``quantum polynomial time'' is an important unsolved problem.

The quantum factoring algorithm works by coherently summing an exponentially
large number of amplitudes that interfere constructively, building up the
strong peaks in the probability distribution Eq.\ (\ref{ft_prob}).
Unfortunately, this ``exponential coherence'' is extremely vulnerable to the
effects of noise \cite{noise}.  When the computer interacts with its
environment, the quantum state of the computer becomes entangled with the state
of the environment;  hence the pure quantum state of the computer decays to an
incoherent mixed state, a phenomenon known as decoherence. Just as an
illustration, imagine that, after the coherent superposition state Eq.\
(\ref{periodic_state}) is prepared, each qubit has a probability $p<<1$ of
decohering completely before the $FT$ is applied and the device is measured; in
other words, $pL$ of the $L$ qubits decohere, and the state of the computer
becomes entangled with $2^{pL}$ mutually orthogonal states of the environment.
Thus, the number of terms in the coherent sum in Eq.\ (\ref{ft_prob}) is
reduced by the factor $2^{-pL}$, and the  peaks in the probability distribution
are weakened by the factor $2^{-2pL}$.  For any nonzero $p$, then, the
probability of successfully finding a factor decreases exponentially as $L$
grows large.

Interaction with the environment, and hence decoherence, always occur at some
level.  It seems then, that the potential of a quantum computer to solve hard
problems efficiently can be realized only if suitable schemes are found that
control the debilitating effects of decoherence.  In some remarkable recent
developments \cite{error_correct}, clever error correction schemes have been
proposed for encoding and {\it storing} quantum information that sharply reduce
its susceptibility to noise.  Some remaining challenges are: to incorporate
error correction into the operation of a quantum network (so that it can
operate with high reliability in spite of the effects of decoherence), and to
find efficient error-correction schemes that can be implemented in realistic
working devices.

\section{The Linear Ion Trap}

\subsection{A realizable device}

The hardware for a quantum computer must meet a variety of demanding criteria.
A suitable method for storing qubits should be chosen such that: (1) the state
of an individual qubit can be controlled and manipulated,  (2) carefully
controlled strong interactions between distinct qubits can be induced (so that
nonlinear logic gates can be constructed), and (3) the state of a qubit can be
read out efficiently.  Furthermore, to ensure effective operation:  (1)  the
storage time for the qubits must be long enough so that many logical operations
can be performed, (2) the machine should be free of imperfections that could
introduce errors in the logic gates, and (3) the machine should be well
isolated from its environment, so that the characteristic decoherence time is
sufficiently long.

Cirac and Zoller \cite{cirac} proposed an incarnation of a quantum computer
that meets these criteria remarkably well and that may be within the grasp of
existing technology.  In their proposal, ions are collected in a linear
harmonic trap.  The internal state of each ion encodes one qubit:  the ground
state $\bigl | g\bigr\rangle$ is interpreted as $\bigl | 0\bigr\rangle$, and a
long-lived metastable excited state $\bigl | e\bigr\rangle$ is interpreted as
$\bigl | 1\bigr\rangle$.  The quantum state of the computer in this basis can
be efficiently read out by the ``quantum jump method'' \cite{jump}. A laser is
tuned to a transition from the state $\bigl | g\bigr\rangle$ to a short-lived
excited state that decays back to $\bigl | g\bigr\rangle$; when the laser
illuminates the ions, each qubit with value $\bigl | 0\bigr\rangle$ fluoresces
strongly, while the qubits with value $\bigl | 1\bigr\rangle$ remain dark.

Coulomb repulsion keeps the ions sufficiently well separated that they can be
{\it individually} addressed by pulsed lasers \cite{wineland}.  If a laser is
tuned to the frequency $\omega$, where $\hbar\omega$ is the energy splitting
between $\bigl | g\bigr\rangle$ and $\bigl | e\bigr\rangle$, and is focused on
the the $i$th ion, then Rabi oscillations are induced between $\bigl |
0\bigr\rangle_i$ and $\bigl | 1\bigr\rangle_i$.  By timing the laser pulse
properly, and choosing the phase of the laser appropriately, we can prepare the
$i$th ion in an arbitrary superposition of $\bigl | 0\bigr\rangle_i$ and $\bigl
| 1\bigr\rangle_i$.  (Of course, since the states $\bigl | g\bigr\rangle$ and
$\bigl | e\bigr\rangle$ are nondegenerate, the relative phase in this linear
combination rotates with time as $e^{-i\omega t}$ even when the laser is turned
off.  It is most convenient to express the quantum state of the qubits in the
interaction picture, so that this time-dependent phase is rotated away.)

Crucial to the functioning of the quantum computer are the quantum gates that
induce entanglement between distinct qubits.  The qubits must interact if
nontrivial quantum gates are to be constructed. In the ion trap computer, the
interactions are effected by the Coulomb repulsion between the ions.
Because of the mutual Coulomb repulsion, there is a spectrum of coupled normal
modes for the ion motion.  When an ion absorbs or emits a laser photon, the
center of mass of the ion recoils.  But if the laser is properly tuned, then
when a single ion absorbs or emits, a normal mode involving many ions will
recoil coherently (as in the M\"ossbauer effect).

The vibrational mode of lowest frequency (frequency $\nu$) is the
center-of-mass (CM) mode, in which the ions oscillate in lockstep in the
harmonic well of the trap. The ions can be laser cooled to a temperature much
less than $\nu$, so that each vibrational normal mode is very likely to occupy
its quantum-mechanical ground state.  Now imagine that a laser tuned to the
frequency $\omega-\nu$ shines on the $i$th ion.  For a properly timed pulse (a
$\pi$ pulse, or a $k\pi$ pulse for $k$ odd), the state $\bigl |
e\bigr\rangle_i$ will rotate to $\bigl | g\bigr\rangle_i$, while the CM
oscillator makes a transition from its ground state $\bigl | 0\bigr\rangle_{\rm
CM}$ to its first excited state $\bigl | 1\bigr\rangle_{\rm CM}$ (a CM
``phonon'' is produced).  However, the state $\bigl | g\bigr\rangle_i\bigl |
0\bigr\rangle_{\rm CM} $ is not on resonance for any transition, and so is
unaffected by the pulse.  Thus, with a single laser pulse, we may induce the
unitary transformation
\begin{equation}
\label{Wphonon}
W^{(i)}_{\rm phon}:\>\> \left\{\begin{array}{lll}&\bigl | g\bigr\rangle_i|
0\bigr\rangle_{\rm CM}\longmapsto &| g\bigr\rangle_i| 0\bigr\rangle_{\rm
CM}\nonumber\\
&\bigl | e\bigr\rangle_i| 0\bigr\rangle_{\rm CM}\longmapsto -i&|
g\bigr\rangle_i| 1\bigr\rangle_{\rm CM}\end{array}\right\}\nonumber\\
\end{equation}
This operation removes a bit of information that is initially stored in the
internal state of the $i$th ion, and deposits the bit in the CM phonon mode.
Applying $W^{(i)}_{\rm phon}$ again would reverse the operation (up to a
phase), removing the phonon and reinstating the bit stored in ion $i$.
However, all of the ions couple to the CM phonon, so that once the information
has been transferred to the CM mode, this information will influence the
response of ion $j$ if a laser pulse is subsequently directed at that ion.  By
this scheme, nontrivial logic gates can be constructed, as we will describe in
more detail below.

An experimental demonstration of an operation similar to $W^{(i)}_{\rm phon}$
was recently carried out by Monroe {\it et al.} \cite{monroe}. In this
experiment, a single $^9 Be^+$ ion occupied the trap.  In earlier work, a
linear trap was constructed that held 33 ions, but these were not cooled down
to the vibrational ground state.  The effort to increase the number of qubits
in a working device is ongoing.

Perhaps the biggest drawback of the ion trap is that it is an intrinsically
slow device.  Its speed is ultimately limited by the energy-time uncertainty
relation; since the uncertainty in the energy of the laser photons should be
small compared to the characteristic vibrational splitting $\nu$, the pulse
must last a time large compared to $\nu^{-1}$.  In the Monroe {\it et al.}
experiment, $\nu$ was as large as 50 MHz, but it is likely to be orders of
magnitude smaller in a device that contains many ions.

In an alternate version of the above scheme (proposed by the Pellizzari {\it et
al.} \cite{pellizzari}) many atoms are stored in an optical cavity, and the
atoms interact via the cavity photon mode (rather than the CM vibrational
mode).  In principle, quantum gates in a scheme based on cavity QED could be
intrinsically much faster than gates implemented in an ion trap.  An
experimental demonstration of a rudimentary quantum gate involving photons
interacting with an atom in a cavity was recently reported by Turchette {\it et
al.} \cite{kimble}.


\subsection{Conditional phase gate}
An interesting two-qubit gate can be constructed by applying three laser pulses
\cite{cirac}.
After a phonon has been (conditionally) excited, we can apply a laser pulse to
the $j$th ion that is tuned to the transition $\bigl | g\bigr \rangle_j \bigl |
1 \bigr \rangle_{\rm CM}\longmapsto \bigl | e'\bigr \rangle_j \bigl | 0 \bigr
\rangle_{\rm CM}$, where $\bigl | e'\bigr \rangle$ is another excited state
(different than $\bigl | e\bigr \rangle$) of the ion. The effect of a $2\pi$
pulse is to induce  the transformation
\begin{equation}
V^{(j)}:\>\> \left\{\begin{array}{lll}&\bigl | g\bigr\rangle_j|
0\bigr\rangle_{\rm CM}\longmapsto &| g\bigr\rangle_j| 0\bigr\rangle_{\rm
CM}\nonumber\\
&\bigl | e\bigr\rangle_j| 0\bigr\rangle_{\rm CM}\longmapsto &| e\bigr\rangle_j|
0\bigr\rangle_{\rm CM}\nonumber\\
& \bigl | g\bigr\rangle_j| 1\bigr\rangle_{\rm CM}\longmapsto -&|
g\bigr\rangle_j| 1\bigr\rangle_{\rm CM}\nonumber\\
&\bigl | e\bigr\rangle_j| 1\bigr\rangle_{\rm CM}\longmapsto &| e\bigr\rangle_j|
1\bigr\rangle_{\rm CM}\end{array}\right\}\nonumber\\
\end{equation}
Only the phase of the state $| g\bigr\rangle_j| 1\bigr\rangle_{\rm CM}$ is
affected by the $2\pi$ pulse, because this is the only state that is on
resonance for a transition when the laser is switched on.  (It would not have
had the same effect if we had tuned the laser to the transition from  $\bigl
|g\bigr\rangle| 1\bigr\rangle_{\rm CM}$ to $ | e\bigr\rangle|
0\bigr\rangle_{\rm CM}$, because then the state  $ | e\bigr\rangle|
0\bigr\rangle_{\rm CM}$ would also have been modified by the pulse.)   Applying
$W^{(i)}_{\rm phon}$ again removes the phonon, and we find that
\begin{equation}
\label{phase_gate}
V^{(i,j)}\>\equiv\>  W^{(i)}_{\rm phon}\cdot V^{(j)} \cdot W^{(i)}_{\rm
phon}:\>\> \bigl|\epsilon\bigr\rangle_i \bigl|\eta\bigr\rangle_j \longmapsto
(-1)^{\epsilon\eta} \bigl|\epsilon\bigr\rangle_i \bigl|\eta\bigr\rangle_j \ .
\end{equation}
is a {\it conditional phase} gate; it multiplies the quantum state by $(-1)$ if
the qubits $\bigl|\cdot\bigr\rangle_i$ and $\bigl|\cdot\bigr\rangle_j$ both
have the value 1, and acts trivially otherwise. A remarkable and convenient
feature of this construction is that the two qubits that interact need not be
in neighboring positions in the linear trap.  In principle, the ions on which
the gate acts could be arbitrarily far apart.

This gate can be generalized so that the conditional phase $(-1)$ is replaced
by an arbitrary phase $e^{i\theta}$---we replace the $2\pi$ pulse directed at
ion $j$ by two $\pi$ pulses with differing values of the laser phase, and
modify the laser phase for one of $\pi$ pulses directed at ion $i$.  Thus, with
4 pulses, we construct the conditional phase transformation $V^{(i,j)}(\theta)$
defined in Eq.\ (\ref{cond_phase}) that is needed to implement the Fourier
transform $\hat {FT}$.  The $L$-qubit Fourier transform, then, requiring
$L(L-1)/2$ conditional phase gates and $L$ single-qubit rotations, can be
implemented with altogether $L(2L-1)$ laser pulses.

Actually, we confront one annoying little problem when we attempt to implement
the Fourier transform.  The single-qubit rotations that can be simply induced
by shining the laser on an ion are unitary transformations with determinant one
(the exponential of an off-diagonal Hamiltonian), while the rotation $U^{(j)}$
defined in Eq.\ (\ref{qubit_rot}) actually has determinant $(-1)$.  We can
replace $U^{(j)}$ in the construction of the $\hat {FT}$ operator (Eq.\
(\ref{hat_ft})) by the transformation
\begin{equation}
\label{det_one_rot}
\tilde U^{(j)}: \>\> \pmatrix{\bigl|0\bigr\rangle_j\cr \bigl|1\bigr\rangle_j}
\longmapsto {1\over\sqrt{2}}\pmatrix{1&1\cr -1& 1}
\pmatrix{\bigl|0\bigr\rangle_j\cr \bigl|1\bigr\rangle_j}
\end{equation}
(which {\it can} be induced by a single laser pulse with properly chosen laser
phase).  However, the transformation $\tilde {FT} $ thus constructed differs
from $\hat{FT}$ according to
\begin{equation}
\bigl\langle y \bigr | \tilde{FT} \bigl | x\bigr \rangle \> =\>  \left(
-1\right) ^{{\rm Par} (y)}\bigl\langle y \bigr | \hat{FT} \bigl | x\bigr
\rangle
\end{equation}
where ${\rm Par}(y)$ is the {\it parity} of $y$, the number of 1's appearing in
its binary expansion.  Fortunately, the additional phase ${\rm Par}(y)$ has no
effect on the probability distribution Eq.\ (\ref{ft_prob}), so this
construction is adequate for the purpose of carrying out the factorization
algorithm.

\subsection{Controlled$^k$-NOT gate}
\label{sec:NOTpulses}

The conditional $(-1)$ phase gate Eq.\ (\ref{phase_gate}) differs from a {\it
controlled-NOT} gate by a mere change of basis \cite{cirac}.  The
controlled-NOT operation $C_{{\lbrack\!\lbrack i \rbrack\!\rbrack},j}$ acts as
\begin{equation}
C_{{\lbrack\!\lbrack i \rbrack\!\rbrack},j}:\>\>
\bigl|\epsilon\bigr\rangle_i \bigl|\eta\bigr\rangle_j \longmapsto
\bigl|\epsilon\bigr\rangle_i \bigl|\eta\oplus\epsilon\bigr\rangle_j \ ,
\end{equation}
where $\oplus$ denotes the logical XOR operation (binary addition mod 2).  Thus
$C_{{\lbrack\!\lbrack i \rbrack\!\rbrack},j}$ flips the value of the target
qubit $\bigl|\cdot\bigr\rangle_j$ if the control qubit
$\bigl|\cdot\bigr\rangle_i$ has the value 1, and acts trivially otherwise.  We
see that the controlled-NOT can be constructed as
\begin{equation}
\label{trap_CN}
C_{{\lbrack\!\lbrack i \rbrack\!\rbrack},j}\>\equiv \left[\tilde
U^{(j)}\right]^{-1}\cdot V^{(i,j)}\cdot \tilde U^{(j)}\> = \> \left[\tilde
U^{(j)}\right]^{-1}\cdot W^{(i)}_{\rm phon}\cdot V^{(j)} \cdot W^{(i)}_{\rm
phon}\cdot \tilde U^{(j)}
\end{equation}
where $\tilde U^{(j)}$ is the single-qubit rotation defined in Eq.\
(\ref{det_one_rot}).
Since $\tilde U^{(j)}$ (or its inverse) can be realized by directing a $\pi/2$
pulse at ion $j$, we see that the controlled-NOT operation can be implemented
in the ion trap with altogether 5 laser pulses.

The controlled-NOT gate can be generalized to an operation that has a string of
$k$ control qubits; we will refer to this operation as the controlled$^k$-NOT
operation.  (For $k=2$, it is often called the Toffoli gate.)  Its action is
\begin{equation}
C_{\lbrack\!\lbrack{i_1,\dots,i_k}\rbrack\!\rbrack,j} : \>\>
\bigl |{\epsilon_1}\bigr\rangle_{i_1} \cdots \bigl |{\epsilon_k}
\bigr\rangle_{i_k} \bigl | \epsilon \bigr\rangle_j
 \longmapsto
 \bigl | {\epsilon_1} \bigr\rangle_{i_1} \cdots \bigl | {\epsilon_k}
\bigr\rangle_{i_k}
 \bigl |{\epsilon \oplus (\epsilon_1 \wedge \cdots \wedge \epsilon_k)}
\bigr\rangle_j \ ,
\end{equation}
where $\wedge$ denotes the logical AND operation (binary multiplication).  If
all $k$ of the control qubits labeled $i_1,\dots,i_k$ take the value 1, then
$C_{\lbrack\!\lbrack{i_1,\dots,i_k}\rbrack\!\rbrack,j}$ flips the value of the
target qubit labeled $j$; otherwise,
$C_{\lbrack\!\lbrack{i_1,\dots,i_k}\rbrack\!\rbrack,j}$ acts trivially.  To
implement this gate in the ion trap, we will make use of an operation
$V^{(i)}_{\rm phon}$ that is induced by directing a $\pi$ pulse at ion $i$
tuned to the
transition $\bigl | g\bigr \rangle_i \bigl | 1 \bigr \rangle_{\rm
CM}\longmapsto \bigl | e'\bigr \rangle_j \bigl | 0 \bigr \rangle_{\rm CM}$; its
action is
\begin{equation}
V^{(i)}_{\rm phon}:\>\> \left\{\begin{array}{lll}&\bigl | g\bigr\rangle_i|
0\bigr\rangle_{\rm CM}\longmapsto &| g\bigr\rangle_i| 0\bigr\rangle_{\rm
CM}\nonumber\\
&\bigl | e\bigr\rangle_i| 0\bigr\rangle_{\rm CM}\longmapsto &| e\bigr\rangle_i|
0\bigr\rangle_{\rm CM}\nonumber\\
& \bigl | g\bigr\rangle_i| 1\bigr\rangle_{\rm CM}\longmapsto -i&|
e'\bigr\rangle_i| 0\bigr\rangle_{\rm CM}\nonumber\\
&\bigl | e\bigr\rangle_i| 1\bigr\rangle_{\rm CM}\longmapsto &| e\bigr\rangle_i|
1\bigr\rangle_{\rm CM}\end{array}\right\}\nonumber\\
\end{equation}
The pulse has no effect unless the initial state is $\bigl | g\bigr\rangle_i|
1\bigr\rangle_{\rm CM}$, in which case the phonon is absorbed and ion $i$
undergoes a transition to the state $| e'\bigr\rangle_i$.  We thus see that the
controlled$^k$-NOT gate can be constructed as\cite{cirac}
\begin{equation}
C_{{\lbrack\!\lbrack i_1,\dots,i_k \rbrack\!\rbrack},j}\>\equiv \> \left[\tilde
U^{(j)}\right]^{-1}\cdot  W^{(i_1)}_{\rm phon}\cdot V^{(i_2)}_{\rm phon} \cdots
V^{(i_{k})}_{\rm phon}\cdot V^{(j)} \cdot V^{(i_{k})}_{\rm phon} \cdots
V^{(i_{2})}_{\rm phon}\cdot W^{(i_1)}_{\rm phon}\cdot \tilde U^{(j)} \ .
\end{equation}
To understand how the construction works, note first of all that if
$\epsilon_{1}=0$, no phonon is ever excited and none of the pulses have any
effect.  If $\epsilon_1=\epsilon_2=\cdots =\epsilon_{m-1}=1$ and $\epsilon_m=0$
  ($m\le k$), then the first $W^{(i_1)}_{\rm phon}$ produces a phonon that is
absorbed during the
first $V^{(i_{m})}_{\rm phon}$ operation, reemited during the second
$V^{(i_{m})}_{\rm phon}$ operation, and finally absorbed again during the
second $W^{(i_1)}_{\rm phon}$; the other pulses have no effect.  Since each of
the four pulses that is on resonance advances the phase of the state by
$\pi/2$, there is no net change of phase.  If $\epsilon_1=\epsilon_2=\cdots
=\epsilon_{k}=1$, then a phonon is excited by the first $W^{(i_1)}_{\rm phon}$,
and all of the $V^{(i_{m})}_{\rm phon}$'s act trivially; hence in this case
$C_{{\lbrack\!\lbrack i_1,\dots,i_k \rbrack\!\rbrack},j}$ has the same action
as $C_{{\lbrack\!\lbrack i_1 \rbrack\!\rbrack},j}$.

We find then, that the controlled$^k$-NOT gate ($k=1,2,\dots$) can be
implemented in the ion trap with altogether $2k+3$ laser pulses.  These gates
are the fundamental operations that we will use to build the modular
exponential function.\footnote{In fact, the efficiency of our algorithms could
be improved somewhat if we adopted other fundamental gates that can also be
simply implemented with the ion trap.  Implementations of some alternative
gates are briefly discussed in Appendix A.}

\section{Modular exponentiation:  Some general features}

In the next section, we will describe in detail several algorithms for
performing modular exponentiation on a quantum computer.  These algorithms
evaluate the function
\begin{equation}
\label{expfunction}
f_{N,x}(a)=x^a ~({\rm mod} ~N)\ ,
\end{equation}
where $N$ and $x$ are $K$-bit classical numbers ({\it $c$-numbers}) and $a$ is
an $L$-qubit quantum number ({\it $q$-number}).  Our main motivation, of
course, is that the evaluation of $f_{N,x}$ is the bottleneck of Shor's
factorization algorithm.

Most of our algorithms require a ``time'' (number of elementary quantum gates)
of order $K^3$ for large $K$.  In fact, for asymptotically large $K$, faster
algorithms (time of order $K^2\log(K) \log\log (K)$) are possible---these take
advantage of tricks for performing efficient multiplication of very large
numbers \cite{fast_mult}.  We will not consider these asymptotically faster
algorithms in any detail here.  Fast multiplication requires additional storage
space.  Furthermore, because fast multiplication carries a high overhead cost,
the advantage in speed is realized only when the numbers being multiplied are
enormous.

We will concentrate instead on honing the efficiency of algorithms requiring
$K^3$ time, and will study the tradeoff of computation time versus storage
space for these algorithms.  We will also briefly discuss an algorithm that
takes considerably longer ($K^5$ time), but enables us to compress the storage
space further.

Finally, we will describe a ``customized'' algorithm that is designed to
evaluate $f_{N,x}$ in the case $N=15$, the smallest value of $N$ for which
Shor's algortihm can be applied.  Unsurprisingly, this customized algorithm is
far more efficient, both in terms of computation time and memory use, than our
general purpose algorithms that apply for any value of $N$ and $x$.

\subsection{The model of computation}
\label{model}

{\bf A classical computer and a quantum computer}: The machine that runs our
program can be envisioned as a quantum computer controlled by a classical
computer.  The input that enters the machine consists of both classical data (a
string of classical bits) and quantum data (a string of qubits prepared in a
particular quantum state).  The classical data takes a definite fixed value
throughout the computation, while for the quantum data coherent superpositions
of different basis states may be considered (and quantum entanglement of
different qubits may occur). The classical computer processes the classical
data, and produces an output that is a program for the quantum computer.

The quantum computer is a quantum gate network of the sort described by Deutsch
\cite{deutsch}.  The program prepared by the classical computer is a list of
elementary unitary transformations that are to be applied sequentially to the
input state in the quantum register.  (Typically, these elementary
transformations act on one, two, or three qubits at a time; their precise form
will vary depending on the design of the quantum computer.) Finally, the
classical computer calls a routine that measures the state of a particular
string of qubits, and the result is recorded. The result of this final
measurement is the output of our device.

This division between classical and quantum data is not strictly necessary.
Naturally, a $c$-number is just a special case of a $q$-number, so we could
certainly describe the whole device as a quantum gate network (though of
course, our classical computer, unlike the quantum network, can perform
irreversible operations).  However, if we are interested in how a practical
quantum computer might function, the distinction between the quantum computer
and the classical computer is vitally important.  In view of the difficulty of
building and operating a quantum computer, if there is any operation performed
by our device that is intrinsically classical, it will be highly advantageous
to assign this operation to the classical computer; the quantum computer should
be reserved for more important work. (This is especially so since it is likely
to be quite a while before a quantum computer's ``clock speed'' will approach
the speed of contemporary classical computers.)


{\bf Counting operations:} Accordingly, when we count the operations that our
algorithms require, we will be keeping track only of the elementary gates
employed by the quantum computer, and will not discuss in detail the time
required for the classical computer to process the classical data.  Of course,
for our device to be able to perform efficient factorization, the time required
for the classical computation must be bounded above by a polynomial in $K$. In
fact, the classical operations take a time of order $K^3$; thus, the operation
of the quantum computer is likely to dominate the total computation time even
for a very long computation.\footnote{Indeed, one important reason that we
insist that the quantum computer is controlled by a classical computer is that
we want to have an honest definition of computational complexity; if it
required an exponentially long classical computation to figure out how to
program the quantum computer, it would be misleading to say that the quantum
computer could solve a problem efficiently.}

In the case of the evaluation of the modular exponential function $f_{N,x}(a)$,
the classical input consists of $N$ and $x$, and the quantum input is $a$
stored in the quantum register; in addition, the quantum computer will require
some additional qubits (initially in the state $|0\rangle$) that will be used
for scratch space.   The particular sequence of elementary quantum gates that
are are applied to the quantum input will depend on the values of the classical
variables.  In particular, the number of operations is actually a complicated
function of $N$ and $x$.  For this reason, our statements about the number of
operations performed by the quantum computer require clarification.

We will report the number of operations in two forms, which we will call the
``worst case'' and the ``average case.''  Our classical computer will typically
compute and read a particular classical bit (or sequence of bits) and then
decide on the basis of its value what operation to instruct the quantum
computer to perform next.  For example, the quantum computer might be
instructed to apply a particular elementary gate if the classical bit reads 1,
but to do nothing if it reads 0.  To count the number of operations in the
worst case, we will assume that the classical control bits always assume the
value that maximizes the number of operations performed.  This worst case
counting will usually be a serious overestimate.  A much more realistic
estimate is obtained if we assume that the classical control bits are random (0
50\% of the time and 1 50\% of the time).  This is how the ``average case''
estimate is arrived at.

{\bf The basic machine and the enhanced machine:}  Our quantum computer can be
characterized by the elementary quantum gates that are ``hard-wired'' in the
device.  We will consider two different possibilities.  In our ``basic
machine'' the elementary operations will be the single-qubit NOT operation, the
two-qubit controlled-NOT operation, and the three-qubit
controlled-controlled-NOT operation (or Toffoli gate).  These elementary gates
are not computationally universal (we cannot construct arbitrary unitary
operations by composing them), but they will suffice for our purposes; our
machine won't need to be able to do anything else.\footnote{That is, these
operations suffice for evaluation of the modular exponential function.  Other
gates will be needed to perform the discrete Fourier transform, as described in
Sec. \ref{sec:ft}.}  Our ``enhanced machine'' is equipped with these gates plus
two more---a 4-qubit controlled$^3$-NOT gate and a 5-qubit controlled$^4$-NOT
gate.

In fact, the extra gates that are standard equipment for the enhanced machine
can be simulated by the basic machine.  However, this simulation is relatively
inefficient, so that it might be misleading to quote the number of operations
required by the basic machine when the enhanced machine could actually operate
much faster.  In particular, Cirac and Zoller described how to execute a
controlled$^k$-NOT ($k\ge 1$) operation using $2k+3$ laser pulses in the linear
ion trap; thus, {\it e.g.}, the controlled$^4$-NOT operation can be performed
much more quickly in the ion trap than if it had to be constructed from
controlled$^k$-NOT gates with $k=0,1,2$.

To compare the speed of the basic machine and the enhanced machine, we must
assign a relative cost to the basic operations.  We will do so by expressing
the number of operations in the currency of laser pulses under the Cirac-Zoller
scheme:  1 pulse for a NOT, 5 for a controlled-NOT, 7 for a controlled$^2$-NOT,
9 for a controlled$^3$-NOT, and 11 for a controlled$^4$-NOT.  We realize that
this measure of speed is very crude.  In particular, not all laser pulses are
really equivalent.  Different pulses may actually have differing frequencies
and differing durations.  Nevertheless, for the purpose of comparing the speed
of different algorithms, we will make the simplifying assumption that the
quantum computer has a fixed clock speed, and administers a laser pulse to an
ion in the trap once in each cycle.

The case of the (uncontrolled) NOT operation requires special comment.  In the
Cirac-Zoller scheme, the single qubit operations always are $2\times 2$ unitary
operations of determinant one (the exponential of an off-diagonal $2\times 2$
Hamiltonian).  But the NOT operation has determinant $(-1)$.  A simple solution
is to use the operation $i\cdot$(NOT) instead (which does have determinant 1
and can be executed with a single laser pulse).  The overall phase $(i)$ has no
effect on the outcome of the computation.  Hence, we take the cost of a NOT
operation to be one pulse.

In counting operations, we assume that the controlled$^k$-NOT operation can be
performed on any set of $k+1$ qubits in the device.  Indeed, a beautiful
feature of the Cirac-Zoller proposal is that the efficiency of the gate
implementation is unaffected by the proximity of the ions.  Accordingly, we do
not assign any cost to ``swapping'' the qubits before they enter a quantum
gate.\footnote{For a different type of hardware, such as the device envisioned
by Lloyd,\cite{lloyd}, swapping of qubits would be required, and the number of
elementary operations would be correspondingly larger.}

\subsection{Saving space}
\label{sec:space}
A major challenge in programming a quantum computer is to minimize the
``scratchpad space'' that the device requires.  We will repeatedly appeal to
two basic tricks (both originally suggested by C. Bennett
\cite{bennett,bennett_trade}) to make efficient use of the available space.

{\bf Erasing garbage:}  Suppose that a unitary transformation $F$ is
constructed that computes a (not necessarily invertible) function $f$ of a
$q$-number input $b$.  Typically, besides writing the result $f(b)$ in the
output register, the transformation $F$ will also fill a portion of the
scratchpad with some expendable garbage $g(b)$; the action of $F$ can be
expressed as
\begin{equation}
F_{\alpha,\beta,\gamma}: |b\rangle_\alpha|0\rangle_\beta|0\rangle_\gamma
\longmapsto |b\rangle_\alpha|f(b)\rangle_\beta
|g(b)\rangle_\gamma \ ,
\end{equation}
where $|\cdot\rangle_\alpha$, $|\cdot\rangle_\beta$, $|\cdot\rangle_\gamma$
denote the input, output, and scratch registers, respectively.  Before
proceeding to the next step of the computation, we would like to clear $g(b)$
out of the scratch register, so that the space $|\cdot\rangle_\gamma$ can be
reused.  To erase the garbage, we invoke a unitary operation
$COPY_{\alpha,\delta}$ that copies the contents of $|\cdot\rangle_\alpha$ to an
additional register $|\cdot\rangle_{\delta}$, and then we apply the {\it
inverse} $F^{-1}$ of the unitary operation $F$.  Thus, we have
\begin{equation}
XF_{\alpha,\beta,\gamma,\delta}\equiv F^{-1}_{\alpha,\beta,\gamma}\cdot
COPY_{\alpha,\delta} \cdot F_{\alpha,\beta,\gamma}: |b\rangle_\alpha
|0\rangle_\beta |0\rangle_\gamma |0\rangle_\delta \longmapsto |b\rangle_\alpha
|0\rangle_\beta |0\rangle_\gamma |f(b)\rangle_\delta
\ .
\end{equation}
The composite operation $XF$ uses both of the registers $|\cdot\rangle_\beta$
and $|\cdot\rangle_\gamma$ as scratch space, but it cleans up after itself.
Note that $XF$ preserves the value of $b$ in the input register.  This is
necessary, for a general function $f$, if the operation $XF$ is to be
invertible.

{\bf Overwriting invertible functions:}  We can clear even more scratch space
in the special case where $f$ is an invertible function.  In that case, we can
also construct another unitary operation $XFI$ that computes the inverse
function $f^{-1}$; that is,
\begin{equation}
\label{xfi}
XFI_{\alpha,\beta}: |b\rangle_\alpha|0\rangle_\beta \longmapsto
|b\rangle_\alpha |f^{-1}(b)\rangle_\beta \ .
\end{equation}
or, equivalently,
\begin{equation}
\label{xfi2}
XFI_{\beta,\alpha}: |0\rangle_\alpha|f(b)\rangle_\beta \longmapsto
|b\rangle_\alpha
|f(b)\rangle_\beta \ .
\end{equation}
($XFI$, like $XF$, requires scratchpad space.  But since $XFI$, like $XF$,
leaves the state of the scratchpad unchanged, we have suppressed the scratch
registers in Eq.\ (\ref{xfi}) and Eq.\ (\ref{xfi2}).)  By composing $XF$ and
$XFI^{-1}$, we obtain an operation $OF$ that evaluates the function $f(b)$ and
``overwrites'' the input $b$ with the result $f(b)$:
\begin{equation}
OF_{\alpha,\beta}\equiv XFI^{-1}_{\beta,\alpha}\cdot XF_{\alpha,\beta}:
|b\rangle_\alpha|0\rangle_\beta\longmapsto |0\rangle_\alpha |f(b)\rangle_\beta\
{}.
\end{equation}
(Strictly speaking, this operation does not ``overwrite'' the input; rather, it
erases the input register $|\cdot\rangle_\alpha$ and writes $f(b)$ in a
different register $|\cdot\rangle_\beta$.  A genuinely overwriting version of
the evaluation of of $f$ can easily be constructed, if desired, by following
$OF$ with a unitary $SWAP$ operation that interchanges the contents of the
$|\cdot\rangle_\alpha$ and $|\cdot\rangle_\beta$ registers.  Even more simply,
we can merely swap the {\it labels} on the registers, a purely classical
operation.)

In our algorithms for evaluating the modular exponentiation function, the
binary arithmetic operations that we perform have one classical operand and one
quantum operand.  For example, we evaluate the product $y\cdot b ~({\rm
mod}~N)$, where $y$ is a $c$-number and $b$ is a $q$-number.  Evaluation of the
product can be viewed as the evaluation of a {\it function} $f_y(b)$ that is
determined by the value of the $c$-number $y$.  Furthermore, since the positive
integers less than $N$ that are relatively prime to $N$ form a group under
multiplication, the function $f_y$ is an {\it invertible} function if
$gcd(y,N)=1$.  Thus, for $gcd(y,N)=1$, we can (and will) use the above trick to
overwrite the $q$-number $b$ with a new $q$-number $y\cdot b ~({\rm mod} N)$.

\subsection{Multiplexed Adder}
\label{multiplexed}
The basic arithmetic operation that we will need to perform is addition (mod
$N$)---we will evaluate $y+b ~ ({\rm mod}~ N)$ where $y$ is a $c$-number and
$b$ is a $q$-number.  The most efficient way that we have found to perform this
operation is to build a {\it multiplexed} mod $N$ adder.

Suppose that $N$ is a $K$-bit $c$-number, that  $y$ is a $K$-bit $c$-number
less than $N$, and that $b$ is a $K$-qubit $q$-number, also less than $N$.
Evaluation of $y+b ~ ({\rm mod}~ N)$ can be regarded as a function, determined
by the $c$-number $y$, that acts on the $q$-number $b$.  This function can be
described by the ``pseudo-code''
\begin{eqnarray}
{\tt if} & \ (N-y> b) & \quad ADD \quad y \ ,\nonumber\\
{\tt if} & \ (N-y \le b) & \quad ADD \quad y -N \ .
\end{eqnarray}
Our multiplexed adder is designed to evaluate this function.  First a
comparison is made to determine if the $c$-number $N-y$ is greater than the
$q$-number $b$, and the result of the comparison is stored as a ``select
qubit.''  The adder then reads the select qubit, and performs an ``overwriting
addition'' operation on the the $q$-number $b$, replacing it by either $y+b$
(for $N-y>b$), or $y+b-N$ (for $N-y \le b$).  Finally, the comparison operation
is run backwards to erase the select qubit.

Actually, a slightly modified version of the above pseudo-code is implemented.
Since it is a bit easier to add a positive $c$-number than a negative one, we
choose to add $2^K+y-N$ to $b$ for $N-y \le b$.  The $(K+1)$st bit of the sum
(which is guaranteed to be 1 in this case), need not be (and is not) explicitly
evaluated by the adder.

\subsection{Enable bits}
\label{sec:enable}
Another essential feature of our algorithms is the use of ``enable'' qubits
that control the arithmetic operations.  Our multiplexed adder, for example,
incorporates such an enable qubit.  The adder reads the enable qubit, and if it
has the value 1, the adder replaces the input $q$-number $b$ by the sum
$y+b~({\rm mod}~N)$ (where $y$ is a $c$-number).  If the enable qubit has the
value 0, the adder leaves the input $q$-number $b$ unchanged.

Enable qubits provide an efficient way to multiply a $q$-number by a
$c$-number.  A $K$-qubit $q$-number $b$ can be expanded in binary notation as
\begin{equation}
b=\sum_{i=0}^{K-1} b_i 2^i \ ,
\end{equation}
and the product of $b$ and a $c$-number $y$ can be expressed as
\begin{equation}
b\cdot y ~({\rm mod} ~N)= \sum_{i=0}^{K-1} b_i\cdot [2^i y ~({\rm mod}~N)] \ .
\end{equation}
This product can be built by running the pseudo-code:
\begin{equation}
{\tt For} \ i=0 \ {\tt to} \ K-1\ , \quad {\tt if} \ b_i=1\ , \quad ADD \quad
2^i y ~({\rm mod}~N)\ ;
\end{equation}
multiplication is thus obtained by performing $K$ {\it conditional}  mod $N$
additions.  Hence our multiplication routine calls the multiplexed adder $K$
times; in the $i$th call, $b_i$ is the enable bit that controls the addition.

In fact, to compute the modular exponential function as described below, we
will need {\it conditional} multiplication; the multiplication routine will
have an enable bit of its own.  Our multiplier will replace the $q$-number $b$
by the product $b\cdot y ~({\rm mod} ~N)$ (where $y$ is a $c$-number) if the
enable qubit reads $1$, and will leave $b$ unchanged if the enable qubit reads
0.  To construct a multiplier with an enable bit, we will need an adder with a
{\it pair} of enable bits---that is, an adder that is switched on only when
both enable qubits read 1.

The various detailed algorithms that we will describe differ according to how
enable qubits are incorporated into the arithmetic operations.  The most
straightforward procedure (and the most efficient, in the linear ion trap
device of Cirac and Zoller) is that underlying the design of our ``enhanced
machine.''  We will see that a multiplexed adder can be constructed from the
elementary gates NOT, controlled-NOT and controlled$^2$-NOT.  One way to
promote this adder to an adder with two enable bits is to replace each
controlled$^k$-NOT by a controlled$^{(k+2)}$-NOT, where the two enable bits are
added to the list of control bits in each elementary gate.  We thus construct a
routine that performs (multiplexed) addition when both enable bits read 1, and
does nothing otherwise. The routine is built from elementary controlled$^k$-NOT
gates with $k=4$ or less.

In fact, it will turn out that we will not really need to add enable bits to
the control list of every gate.  But following the above strategy does require
controlled$^k$ gates for $k$=0,1,2,3,4.  This is how our enhanced machine
performs mod $N$ addition with two enable bits (and mod $N$ multiplication with
one enable bit).

Because controlled$^4$-NOT and controlled$^3$-NOT gates are easy to implement
on the linear ion trap, the above procedure is an efficient way to compute the
modular exponential function with an ion trap.  However, for a different type
of quantum computing hardware, these elementary gates might not be readily
constructed.  Therefore, we will also consider a few other algorithms, which
are built from elementary controlled$^k$-NOT gates for only $k=0,1,2$.  These
algorithms for our ``basic machine'' follow the same general design as the
algorithm for the ``enhanced machine,'' except that the controlled$^3$-NOT and
the controlled$^4$-NOT gates are expanded out in terms of the simpler
elementary operations.  (The various algorithms for the basic machine differ in
the amount of scratch space that they require.)

\subsection{Repeated squaring}
\label{sec:repeated}
One way to evaluate the modular exponential $x^a ~ ({\rm mod} ~N)$ is to
multiply by $x$ a total of $a-1$ times, but this would be terribly inefficient.
 Fortunately, there is a well-known trick, {\it repeated squaring}, that speeds
up the computation enormously.

If $a$ is an $L$-bit number with the binary expansion $\sum_{i=0}^{L-1} a_i
2^i$, we note that
\begin{equation}
x^a=x^{\left(\sum_{i=0}^{L-1} a_i
2^i\right)}=\prod_{i=0}^{L-1}\left(x^{2^i}\right)^{a_i} \ .
\end{equation}
Furthermore, since
\begin{equation}
x^{2^i}=\left(x^{2^{i-1}}\right)^2 \ ,
\end{equation}
we see that $x^{2^i} ~ ({\rm mod} ~N)$, can be computed by
squaring $x^{2^{i-1}}$. We conclude that $x^a ~({\rm mod} ~N)$ can be obtained
from at most $2(L-1)$ mod $N$ multiplications (fewer if some of the $a_i$
vanish).  If ordinary ``grade school'' multiplication is used (rather than a
fast multiplication algorithm), this evaluation of $x^a ~ ({\rm mod} ~N)$
requires of order $L\cdot K^2$ elementary bit operations (where $N$ and $x<N$
are $K$-bit numbers).  Our algorithms for evaluating $x^a$, where $a$ is an
$L$-bit $q$-number and $x$ is a $K$-bit $c$-number, are based on ``grade
school'' multiplication, and will require of order $L\cdot K^2$ elementary
quantum gates.

Since $x$ is a $c$-number, the ``repeated squaring'' to evaluate $x^{2^i} ~
({\rm mod} ~N)$ can be performed by our classical computer.  Once these
$c$-numbers are calculated and stored, then $x^a ~ ({\rm mod} ~N)$ can be found
by running the pseudo-code
\begin{equation}
{\tt For} \ i=0 \ {\tt to} \ L-1\ , \quad {\tt if} \ a_i=1\ , \quad MULTIPLY
\quad x^{2^i} ~({\rm mod}~N)\ .
\end{equation}
Thus, the modular exponential function is obtained from $L$ conditional
multiplications.  It is for this reason that our mod $N$ multiplier comes
equipped with an enable bit.  Our modular exponentiation algorithm calls the
mod $N$ multiplier $L$ times; in the $i$th call, $a_{i-1}$ is the enable bit
that controls the multiplication.

\section{Modular Exponentiation in Detail}
\label{sec:detail}

\subsection{Notation}

Having described above the central ideas underlying the algorithms, we now
proceed to discuss their detailed implementation.  We will be evaluating $x^a
{}~({\rm mod}~ N)$, where $N$ is a $K$-bit $c$-number, $x$ is a $K$-bit
$c$-number less than $N$, and $a$ is an $L$-bit $q$-number.  For the
factorization algorithm, we will typically choose $L\approx 2K$.

We will use the ket notation $\bigl |\cdot\bigr\rangle$ to denote the quantum
state of a single {\it qubit}, a two-level quantum system.  The two basis
states of a qubit are denoted $\bigl |0\bigr\rangle$ and $\bigl
|1\bigr\rangle$.  Since most of the $q$-numbers that will be manipulated by our
computer will be $K$ qubits long, we will use a shorthand notation for
$K$-qubit registers; such registers will be denoted by a ket that carries a
lowercase Greek letter subscript, {\it e.g.}, $\bigl |b\bigr\rangle_\alpha$,
where $b$ is a $K$-bit string that represents the number $\sum_{i=0}^{K-1} b_i
2^i$ in binary notation.  Single qubits are denoted by kets that carry a
numeral subscript, {\it e.g} $\bigl |c\bigr\rangle_1$, where $c$ is 0 or 1.
Some registers will be $L$ bits long; these will be decorated by asterisk
superscripts, {\it e.g.} $\bigl |a\bigr\rangle_\alpha^*$.

\begin{figure}[htb]
\centerline{\epsfxsize=8truecm \epsfbox{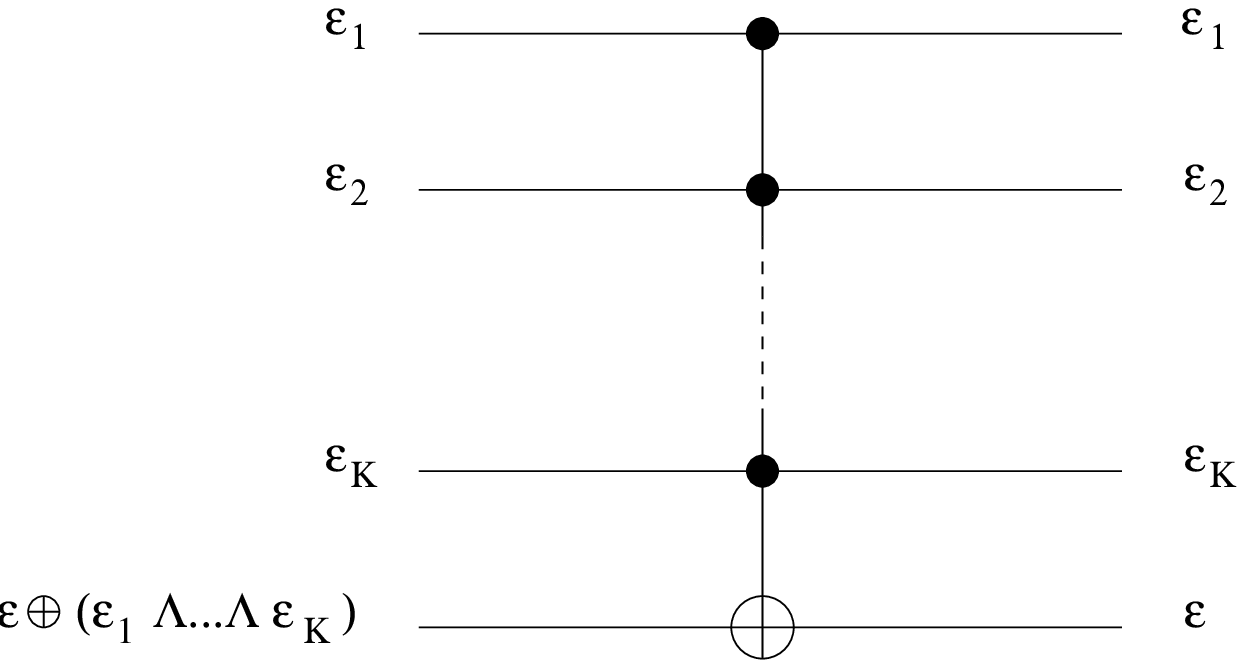}}
\vskip .5truecm
\caption{The controlled$^k$-NOT gate.  Input values of the qubits are shown on
the right and output values on the left.  This gate flips the value of the
target qubit if all $k$ control qubits take the value 1; otherwise, the gate
acts trivially.}
\label{figA}
\end{figure}

The fundamental operation that our quantum computer performs is the
controlled$^k$-NOT operation.  This is the $(k+1)$-qubit quantum gate that acts
on a basis according to
\begin{equation}
C_{\lbrack\!\lbrack{i_1,\dots,i_k}\rbrack\!\rbrack,j} : \>\>
\bigl |{\epsilon_1}\bigr\rangle_{i_1} \cdots \bigl |{\epsilon_k}
\bigr\rangle_{i_k} \bigl | \epsilon \bigr\rangle_j
 \longmapsto
 \bigl | {\epsilon_1} \bigr\rangle_{i_1} \cdots \bigl | {\epsilon_k}
\bigr\rangle_{i_k}
 \bigl |{\epsilon \oplus (\epsilon_1 \wedge \cdots \wedge \epsilon_k)}
\bigr\rangle_j \, .
\end{equation}
Here, each of $\epsilon_1,\dots,\epsilon_k,\epsilon$ takes the value 0 or 1,
$\wedge$ denotes the logical AND operation (binary multiplication) and $\oplus$
denotes the logical XOR operation (binary addition mod 2).  Thus, the gate
$C_{\lbrack\!\lbrack{i_1,\dots,i_k}\rbrack\!\rbrack,j}$ acts on $k$ ``control''
qubits labeled $i_1,\dots,i_k$ and on one ``target qubit'' labeled $j$.  If all
$k$ of the control qubits take the value 1, then
$C_{\lbrack\!\lbrack{i_1,\dots,i_k}\rbrack\!\rbrack,j}$ flips the value of the
target qubit; otherwise,
$C_{\lbrack\!\lbrack{i_1,\dots,i_k}\rbrack\!\rbrack,j}$ acts trivially.  In
order to represent our quantum circuits graphically, we will use Feynman's
notation for the controlled$^k$-NOT, shown in Fig.\ \ref{figA}.  Note that
$C^{-1}_{\lbrack\!\lbrack{i_1,\dots,i_k}\rbrack\!\rbrack,j} =
C_{\lbrack\!\lbrack{i_1,\dots,i_k}\rbrack\!\rbrack,j}$,
so a computation composed of controlled$^k$-NOT's can be inverted by simply
executing the controlled$^k$-NOT's in the reverse order.

As we explained above, our ``basic machine'' comes with the NOT, controlled-NOT
and controlled$^2$-NOT gates as standard equipment.  Our enhanced machine is
equipped with these fundamental gates and, in addition, the controlled$^3$-NOT
and controlled$^4$-NOT gates.

\subsection{Addition}
\label{sec:addition}

{}From the controlled$^k$-NOT gates, we can build (reversible) arithmetic
operations.  The basic operation in (classical) computer arithmetic is the full
adder.  Given two addend bits $a$ and $b$, and an input carry bit $c$, the full
adder computes the the sum bit
\begin{equation} s=a\oplus b \oplus c \label{simplesum}\end{equation}
and the output carry bit
\begin{equation} c' = (a \wedge b) \vee (c \wedge (a \vee b))\, .
\label{carry}\end{equation}
The addition that our quantum computer performs always involves adding a
$c$-number to a $q$-number.  Thus, we will use two different types of quantum
full adders, distinguished by the value of the classical addend bit.  To add
the classical bit $a=0$, we construct
\begin{equation}
\label{FA0}
FA(a=0)_{1,2,3}\> \equiv \> C_{\lbrack\!\lbrack 1 \rbrack\!\rbrack, 2}
C_{\lbrack\!\lbrack 1,2 \rbrack\!\rbrack, 3} \, ,
\end{equation}
which acts on a basis according to
\begin{equation}
FA(a=0)_{1,2,3}: \>\> \bigl| b\bigr\rangle_1 \bigl| c \bigr\rangle_2 \bigl| 0
\bigr\rangle_3 \longmapsto \bigl| b\bigr\rangle_1 \bigl| b\oplus c
\bigr\rangle_2 \bigl| b\wedge c \bigr\rangle_3 \, .
\end{equation}
Here, the string of controlled$^k$-NOT's defining $FA$ is to be read from right
to left; that is, the gate furthest to the right acts on the kets first.  The
operation $FA(a=0)$ is shown diagramatically in Fig.\ \ref{figB}a, where, in
keeping with our convention for operator ordering, the gate on the right acts
first; hence, in the diagram, ``time'' runs from right to left.
To add the classical bit $a=1$, we construct
\begin{equation}
\label{FA1}
FA(a=1)_{1,2,3}\> \equiv \> C_{\lbrack\!\lbrack 1 \rbrack\!\rbrack, 2}
C_{\lbrack\!\lbrack 1,2 \rbrack\!\rbrack, 3}C_2 C_{\lbrack\!\lbrack 2
\rbrack\!\rbrack, 3} \, ,
\end{equation}
(see Fig.\ \ref{figB}b) which acts as
\begin{equation}
FA(a=1)_{1,2,3}: \>\> \bigl| b\bigr\rangle_1 \bigl| c \bigr\rangle_2 \bigl| 0
\bigr\rangle_3 \longmapsto \bigl| b\bigr\rangle_1 \bigl| b\oplus c\oplus
1\bigr\rangle_2 \bigl| c'=b\vee c \bigr\rangle_3 \, .
\end{equation}
Eqs.\ (\ref{FA0}) and (\ref{FA1}) provide an elementary example that
illustrates the concept of a quantum computer controlled by a classical
computer, as discussed in Sec.\ \ref{model}.  The classical computer reads the
value of the classical bit $a$, and then directs the quantum computer to
execute either $FA(0)$ or $FA(1)$.

As we have already remarked in Sec.\ \ref{multiplexed}, to perform modular
arithmetic efficiently, we will construct a ``multiplexed'' full adder.  The
multiplexed full adder will choose as its classical addend {\it either one} of
two classical bits $a_0$ and $a_1$, with the choice dictated by the value of a
``select qubit'' $\ell$. That is, if $\ell=0$, the classical addend will be
$a_0$, and if $\ell=1$ the classical addend will be $a_1$.  Thus the
multiplexed full adder operation, which we denote $MUXFA'$, will actually be 4
distinct unitary transformations acting on the qubits of the quantum computer,
depending on the four possible values of the classical bits $(a_0,a_1)$.  The
action of $MUXFA'$ is
\begin{equation}
MUXFA'(a_0,a_1)_{1,2,3,4} : \>\>
 \bigl |{\ell}\bigr\rangle_1 \bigl |b \bigr\rangle_2 \bigl |c \bigr\rangle_3
\bigl | 0 \bigr\rangle_4
 \longmapsto
 \bigl |{\ell} \bigr\rangle_1 \bigl |b \bigr\rangle_2 \bigl |s \bigr\rangle_3
\bigl |{c'} \bigr\rangle_4 \, ;
\end{equation}
here $s$ and $c'$ are the sum and carry bits defined in Eqs.\ (\ref{simplesum})
and (\ref{carry}), but where now $a\equiv a_1\wedge \ell \vee a_0\wedge
\mathchar"0218\ell =a_{\ell}$.

\begin{figure}[htb]
\centerline{\epsfysize=2.5truecm \epsfbox{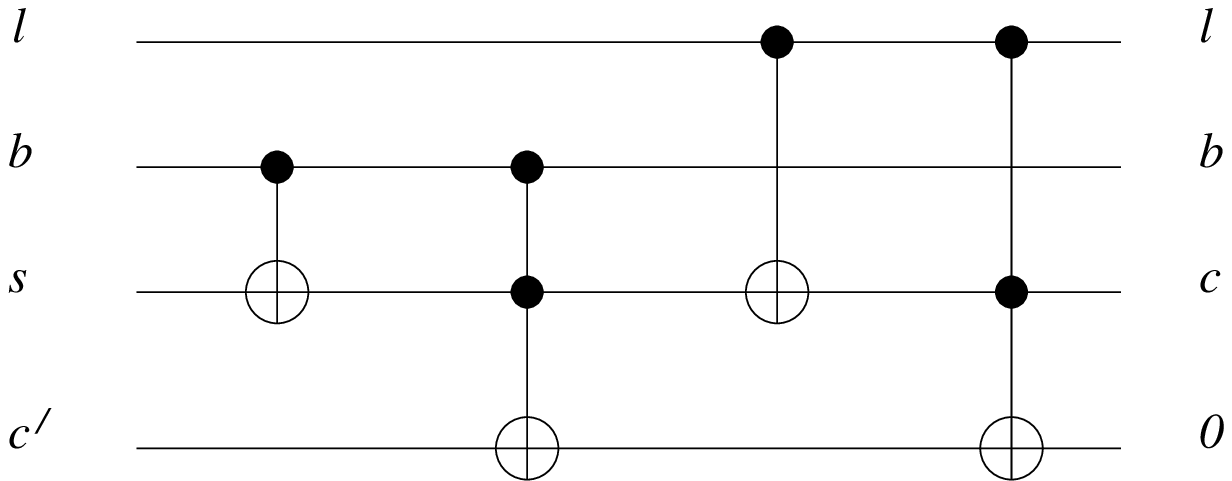} \hskip 1truecm
	\epsfysize=2.5truecm \epsfbox{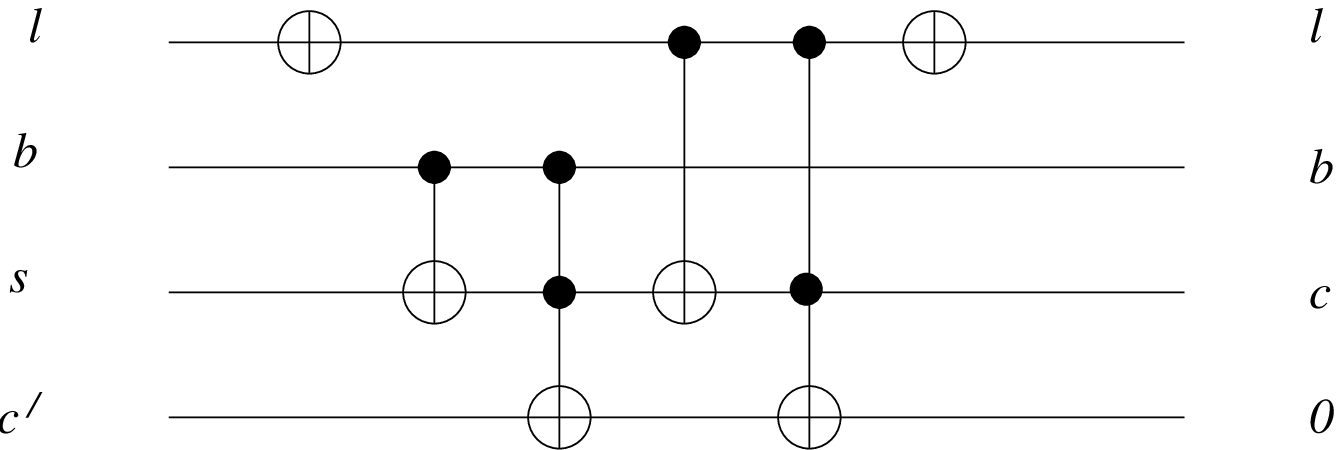} }
\leftline{\hskip 4truecm (a) \hskip 7.5truecm (b)}
\vskip .5truecm
\caption{The multiplexed full adder $MUXFA'(a_0,a_1)$.  Here $\ell$ is the {\it
select bit} that determines whether $a_0$ or $a_1$ is chosen as the classical
addend.  In (a), the case $a_0=0, a_1=1$ is shown; the gate array adds the
qubit $\ell$--which is the same as $a_0$ for $\ell=0$ and $a_1$ for $\ell=1$.
In (b), the case $a_0=1, a_1=0$ is shown; the array adds $\mathchar"0218\ell$.}
\label{figD}
\end{figure}

In fact, for  $a_0=a_1$, the value of the select qubit $\ell$ is irrelevant,
and $MUXFA'$ reduces to the $FA$ operation that we have already constructed:
\begin{eqnarray}
\label{muxfa_zero}
MUXFA'(a_0=0,a_1=0)_{1,2,3,4}&\equiv& FA(0)_{2,3,4}\nonumber \\
MUXFA'(a_0=1,a_1=1)_{1,2,3,4}&\equiv& FA(1)_{2,3,4} \, .
\end{eqnarray}
For $a_0=0$ and $a_1=1$, $MUXFA'$ adds $\ell$, while for $a_0=1$ and $a_1=0$,
it adds $\mathchar"0218 \ell$.  This is achieved by the construction (Fig.\
\ref{figD})
\begin{eqnarray}
\label{muxfa_one}
MUXFA'(a_0=0,a_1=1)_{1,2,3,4}\> &\equiv& \> C_{\lbrack\!\lbrack 2
\rbrack\!\rbrack, 3}C_{\lbrack\!\lbrack 2,3 \rbrack\!\rbrack, 4}
C_{\lbrack\!\lbrack 1 \rbrack\!\rbrack, 3} C_{\lbrack\!\lbrack 1,3
\rbrack\!\rbrack, 4} \, ,\nonumber \\
MUXFA'(a_0=1,a_1=0)_{1,2,3,4}\> &\equiv&  \> C_1C_{\lbrack\!\lbrack 2
\rbrack\!\rbrack, 3}C_{\lbrack\!\lbrack 2,3 \rbrack\!\rbrack, 4}
C_{\lbrack\!\lbrack 1 \rbrack\!\rbrack, 3} C_{\lbrack\!\lbrack 1,3
\rbrack\!\rbrack, 4}C_1 \, .
\end{eqnarray}
(The second operation is almost the same as the first; the difference is that
the qubit $\ell$ is flipped at the beginning and the end of the operation.)

The full adder that we will actually use in our algorithms will be denoted
$MUXFA$ (without the ${}'$). As noted in Sec.\ \ref{sec:enable}, to perform
multiplication and modular exponentiation, we will need a (multiplexed) full
adder that is controlled by an {\it enable bit}, or a string of enable bits.
Thus $MUXFA$ will be an extension of the $MUXFA'$ operation defined above that
incorporates enable bits.  If all the enable bits have the value 1, $MUXFA$
acts just like $MUXFA'$.
But if one or more enable bit is 0, $MUXFA$ will choose the classical addend to
be 0, irrespective of the values of $a_0$ and $a_1$.  We will use the symbol
$\mathchar"024C$ to denote the full list of enable bits for the operation.
Thus the action of $MUXFA$ can be expressed as
\begin{equation}
MUXFA(a_0,a_1)_{\lbrack\!\lbrack \mathchar"024C \rbrack\!\rbrack, 1,2,3,4} :
\>\>
 \bigl |{\ell}\bigr\rangle_1 \bigl |b \bigr\rangle_2 \bigl |c \bigr\rangle_3
\bigl | 0 \bigr\rangle_4
 \longmapsto
 \bigl |{\ell} \bigr\rangle_1 \bigl |b \bigr\rangle_2 \bigl |s \bigr\rangle_3
\bigl |{c'} \bigr\rangle_4 \, ;
\end{equation}
here $s$ and $c'$ are again the sum and carry bits defined in Eqs.\ (\ref{sum})
and (\ref{carry}), but this time $a\equiv \mathchar"024C \wedge \left(a_1\wedge
\ell \vee a_0\wedge \mathchar"0218\ell\right)$; that is, it is 0 unless all
bits of $\mathchar"024C$ take the value 1. The list $\mathchar"024C$ may not
include the bits $1$, $2$, $3$, or $4$.

\begin{figure}[htb]
\centerline{\epsfxsize=10truecm \epsfbox{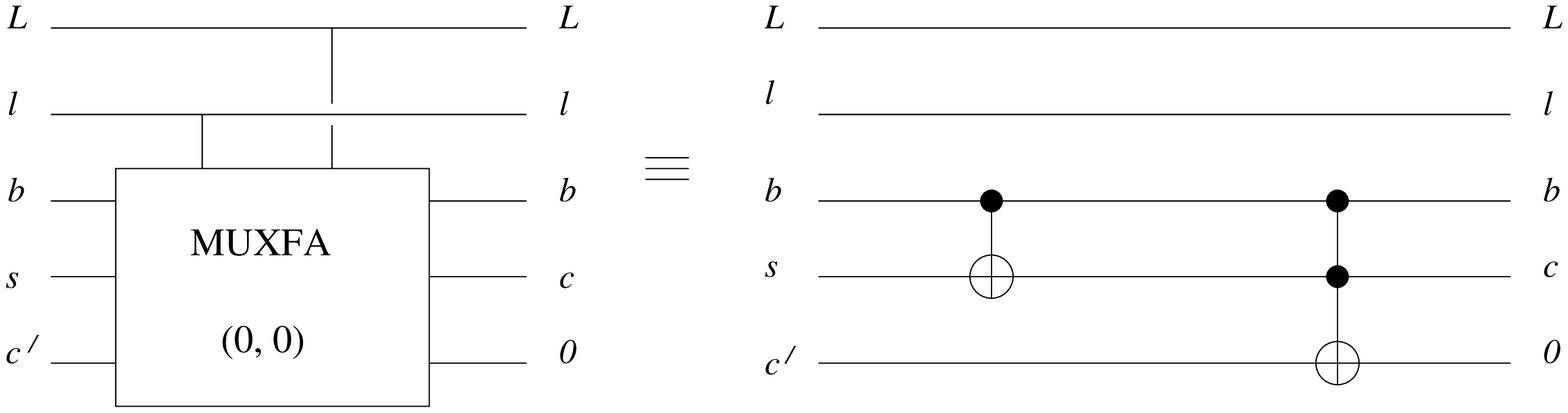}} \vskip .5truecm
\centerline{\epsfxsize=10truecm \epsfbox{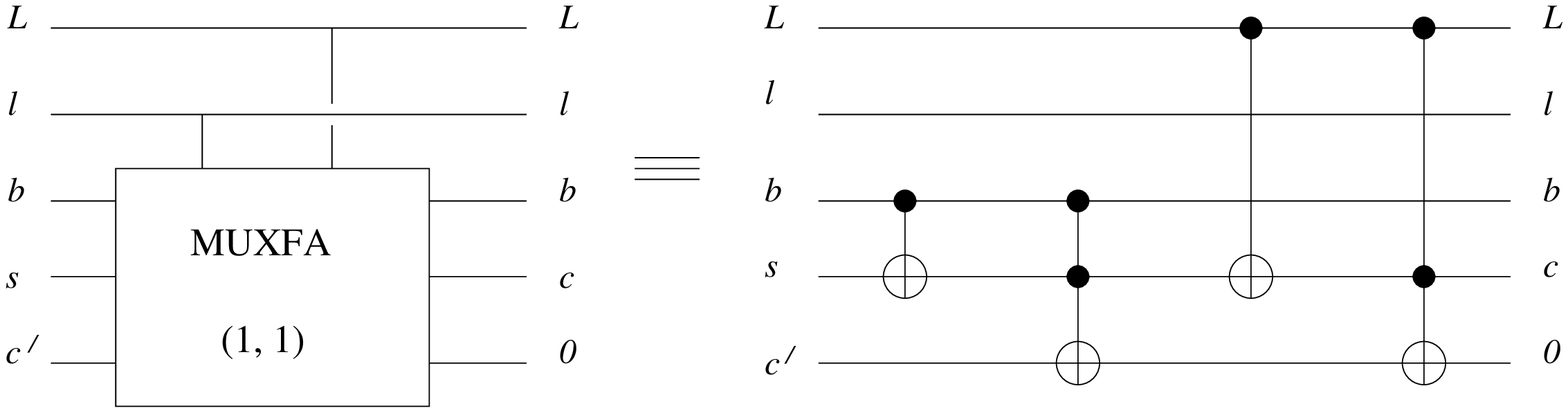}} \vskip .5truecm
\centerline{\epsfxsize=10truecm \epsfbox{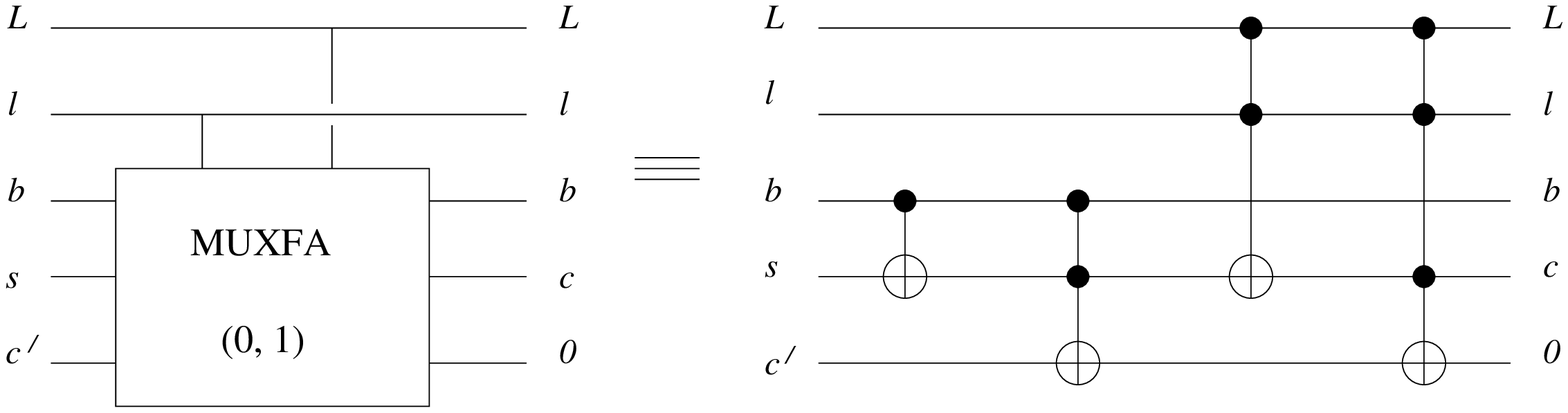}} \vskip .5truecm
\centerline{\epsfxsize=10truecm \epsfbox{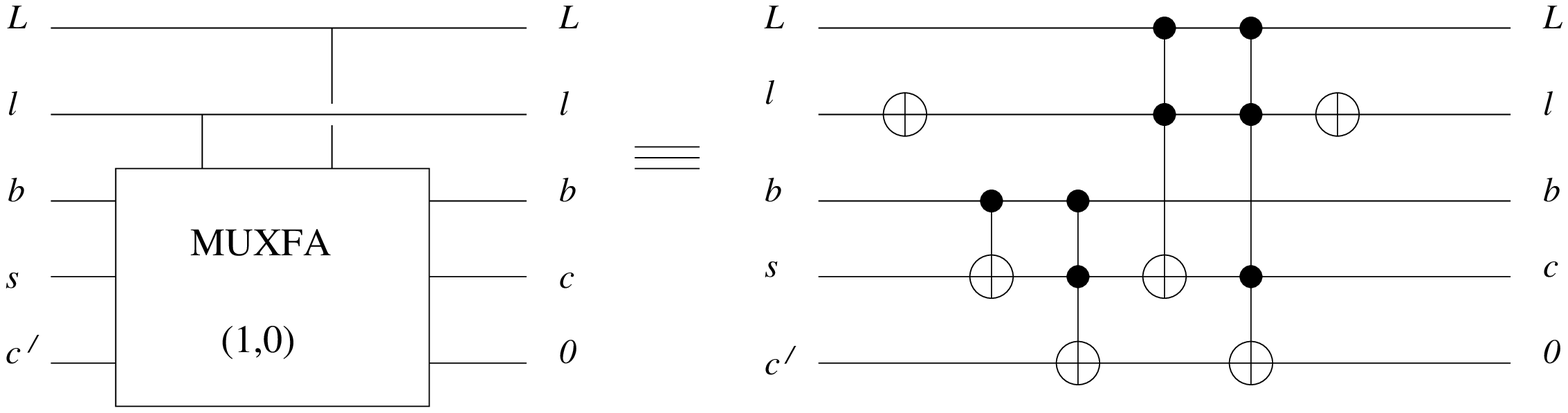}} \vskip .5truecm
\caption{The multiplexed full adder $MUXFA(a_0,a_1)$ has a select bit $\ell$
and an {\it enable string}  $\mathchar"024C$.  If all the bits of
$\mathchar"024C$ take the value 1, then $MUXFA$ acts in the same way as
$MUXFA'$ defined in Fig.\ 3.  Otherwise, the classical addend is chosen to be
0.}
\label{figE}
\end{figure}

In our algorithms, the number of enable bits will be either 1 or 2.  Hence,
there is a simple way to construct the $MUXFA$ operation on our ``enhanced
machine'' that comes equipped with controlled$^3$-NOT and controlled$^4$-NOT
gates.  To carry out the construction, we note by inspecting Eq.\
(\ref{muxfa_zero},\ref{muxfa_one}) (or Fig.\ \ref{figD}) that $MUXFA'(a_0,a_1)$
has the form $MUXFA'(0,0) \cdot F(a_0,a_1)$; thus, by adding $\mathchar"024C$
to the list of control bits for each of the gates in $F(a_0,a_1)$, we obtain an
operation that acts as $MUXFA'$ when $\mathchar"024C$ is all 1's, and adds 0
otherwise.  Explicitly, we have
\begin{eqnarray}
\label{muxfa}
MUXFA(a_0=0, a_1=0)_{\lbrack\!\lbrack \mathchar"024C
\rbrack\!\rbrack,1,2,3,4}\> &\equiv& \> C_{\lbrack\!\lbrack 2 \rbrack\!\rbrack,
3} C_{\lbrack\!\lbrack 2,3 \rbrack\!\rbrack, 4} \, ,\nonumber \\
MUXFA(a_0=1, a_1=1)_{\lbrack\!\lbrack \mathchar"024C
\rbrack\!\rbrack,1,2,3,4}\> &\equiv& \> C_{\lbrack\!\lbrack 2 \rbrack\!\rbrack,
3} C_{\lbrack\!\lbrack 2,3 \rbrack\!\rbrack, 4} C_{\lbrack\!\lbrack
\mathchar"024C \rbrack\!\rbrack, 3} C_{\lbrack\!\lbrack \mathchar"024C,3
\rbrack\!\rbrack, 4} \, ,\nonumber \\
MUXFA(a_0=0,a_1=1)_{\lbrack\!\lbrack \mathchar"024C \rbrack\!\rbrack,1,2,3,4}\>
&\equiv& \> C_{\lbrack\!\lbrack 2 \rbrack\!\rbrack, 3}C_{\lbrack\!\lbrack 2,3
\rbrack\!\rbrack, 4} C_{\lbrack\!\lbrack \mathchar"024C,1 \rbrack\!\rbrack, 3}
C_{\lbrack\!\lbrack \mathchar"024C,1,3 \rbrack\!\rbrack, 4} \, ,\nonumber \\
MUXFA(a_0=1,a_1=0)_{\lbrack\!\lbrack \mathchar"024C \rbrack\!\rbrack,1,2,3,4}\>
&\equiv&  \> C_1C_{\lbrack\!\lbrack 2 \rbrack\!\rbrack, 3}C_{\lbrack\!\lbrack
2,3 \rbrack\!\rbrack, 4} C_{\lbrack\!\lbrack \mathchar"024C,1 \rbrack\!\rbrack,
3} C_{\lbrack\!\lbrack \mathchar"024C,1,3 \rbrack\!\rbrack, 4}C_1 \, .
\end{eqnarray}
(as indicated in Fig.\ \ref{figE}). Here, if $\mathchar"024C$ is a list of $j$
bits, then $C_{\lbrack\!\lbrack \mathchar"024C,1,3 \rbrack\!\rbrack, 4}$, for
example, denotes the controlled$^{(j+2)}$-NOT with $\mathchar"024C,1,3$ as its
control bits.  Evidently, Eq.\ (\ref{muxfa}) is a construction of a multiplexed
adder with $j$ enable bits in terms of controlled$^k$-NOT gates with $k\le
j+2$. In particular, we have constructed the adder with two enable bits that we
will need, using the gates that are available on our enhanced machine.

\begin{figure}[htb]
\centerline{\epsfxsize=10truecm \epsfbox{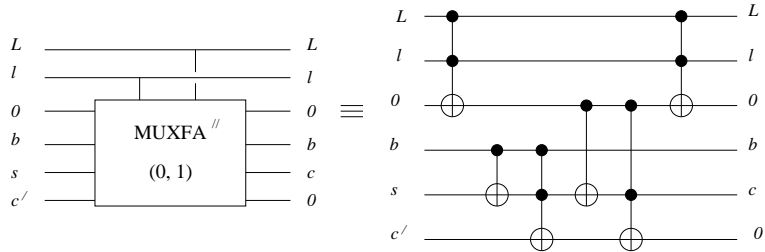}} \vskip .5truecm
\caption{The multiplexed full adder $MUXFA''(a_0,a_1)$ (shown here for $a_0=0,
a_1=1$) is a modification of $MUXFA$ that uses an extra bit of scratch space.
The first gate stores $\mathchar"024C \wedge \ell$ in the extra scratch qubit,
and subsequent gates use this scratch bit as a control bit.  The last gate
clears the scratch bit.  The advantage of $MUXFA''$ is that the longest control
string required by any gate is shorter by one bit than the longest control
string required in $MUXFA$.}
\label{figEa}
\end{figure}

The reader who is impatient to see how our algorithms work in detail is
encouraged to proceed now to the next subsection of the paper.  But first, we
would like to dispel any notion that the algorithms make essential use of the
elementary controlled$^3$-NOT and controlled$^4$-NOT gates.  So let us now
consider how the construction of the $MUXFA$ operation can be modified so that
it can be carried out on the basic machine (which is limited to
controlled$^k$-NOT gates with $k\le 2$).  The simplest such modification
requires an extra bit (or two) of scratch space.  Suppose we want to build a
$MUXFA''$ operation with a single enable bit, without using the
controlled$^3$-NOT gate.  For $a_0=a_1$, the construction in Eq.\ (\ref{muxfa})
need not be modified; in those cases, the action of the operation is
independent of the select bit $\ell$, and therefore no controlled$^3$-NOT gates
were needed.   For $a_0\ne a_1$, controlled$^3$-NOT gates are used, but we note
that the control string of these controlled$^3$-NOT gates includes both the
enable bit and the select bit.  Hence, we can easily eliminate the
controlled$^3$-NOT gate $C_{\lbrack\!\lbrack \mathchar"024C,1,3
\rbrack\!\rbrack, 4}$ by using a controlled$^2$-NOT to compute (and store) the
logical AND ($\mathchar"024C \wedge \ell$) of the enable and select bits, and
then replacing the controlled$^3$-NOT by a controlled$^2$-NOT that has the
scratch bit as one of its control bits.  Another controlled$^2$-NOT at the end
of the operation clears the scratch bit.  In an equation:
\begin{eqnarray}
\label{muxfaprimeprime}
MUXFA''(a_0=0,a_1=1)_{\lbrack\!\lbrack \mathchar"024C
\rbrack\!\rbrack,1,2,3,4,5}\> &\equiv& \> C_{\lbrack\!\lbrack \mathchar"024C,1
\rbrack\!\rbrack, 5}C_{\lbrack\!\lbrack 2 \rbrack\!\rbrack,
3}C_{\lbrack\!\lbrack 2,3 \rbrack\!\rbrack, 4} C_{\lbrack\!\lbrack 5
\rbrack\!\rbrack, 3} C_{\lbrack\!\lbrack 5,3 \rbrack\!\rbrack,
4}C_{\lbrack\!\lbrack \mathchar"024C,1 \rbrack\!\rbrack, 5} \, ,\nonumber \\
MUXFA''(a_0=1,a_1=0)_{\lbrack\!\lbrack \mathchar"024C
\rbrack\!\rbrack,1,2,3,4,5}\> &\equiv& \>
C_1 C_{\lbrack\!\lbrack \mathchar"024C,1 \rbrack\!\rbrack,
5}C_{\lbrack\!\lbrack 2 \rbrack\!\rbrack, 3}C_{\lbrack\!\lbrack 2,3
\rbrack\!\rbrack, 4} C_{\lbrack\!\lbrack 5 \rbrack\!\rbrack, 3}
C_{\lbrack\!\lbrack 5,3 \rbrack\!\rbrack, 4}C_{\lbrack\!\lbrack
\mathchar"024C,1 \rbrack\!\rbrack, 5} C_1\, ,
\end{eqnarray}
as illustrated in Fig.\ \ref{figEa}.  If the scratch bit $\bigl
|\cdot\bigr\rangle_5$ starts out in the state $\bigl |0\bigr\rangle_5$,
$MUXFA''$ has the same action as $MUXFA$, and it returns the scratch bit to the
state $\bigl |0\bigr\rangle_5$ at the end.  By adding yet another bit of
scratch space, and another controlled$^2$-NOT at the beginning and the end, we
easily construct a $MUXFA$ operation with two enable bits.

\begin{figure}[htb]
\centerline{\epsfxsize=10truecm \epsfbox{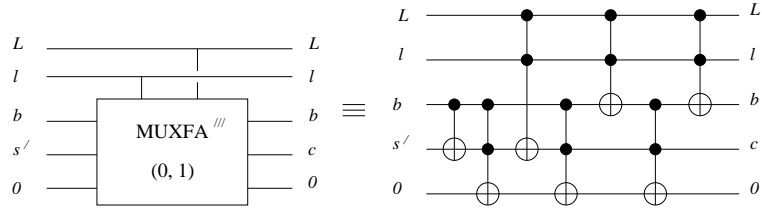}} \vskip .5truecm
\caption{The multiplexed full adder $MUXFA'''(a_0,a_1)$ (shown here for
$a_0=0,a_1=1$) uses simpler gates than those required by $MUXFA$, but unlike
$MUXFA''$, it does not need an extra scratch bit.}
\label{figF}
\end{figure}

At the alternative cost of slightly increasing the number of elementary gates,
the extra scratch bit in $MUXFA''$ can be eliminated.  That is, an operation
with precisely the same action as $MUXFA$ can be constructed from
controlled$^k$-NOT gates with $k\le 2$, and without the extra scratch bit.
This construction uses an idea of Barenco {\it et al.} \cite{barenco}, that a
controlled$^k$-NOT can be constructed from two controlled$^{(k-1)}$-NOT's and
two controlled$^2$-NOT's (for any $k\ge 3$) by employing an extra bit.  This
idea differs from the construction described above, because the extra bit,
unlike our scratch bit, is not required to be preset to 0 at the beginning of
the operation.  Hence, to construct the $C_{\lbrack\!\lbrack \mathchar"024C,1,3
\rbrack\!\rbrack, 4}$ gate needed in MUXFA, we can use $\bigl | b\bigr
\rangle_2$ as the extra bit.  That is, we may use the Barenco {\it et al.}
identity
\begin{equation}
\label{barenco}
C_{\lbrack\!\lbrack \mathchar"024C,1,3 \rbrack\!\rbrack, 4}=
C_{\lbrack\!\lbrack 2,3\rbrack\!\rbrack, 4} C_{\lbrack\!\lbrack
\mathchar"024C,1\rbrack\!\rbrack, 2} C_{\lbrack\!\lbrack 2,3\rbrack\!\rbrack,
4} C_{\lbrack\!\lbrack \mathchar"024C,1\rbrack\!\rbrack, 2}
\end{equation}
to obtain, say,
\begin{equation}
\label{muxfappp}
MUXFA'''(a_0=0,a_1=1)_{\lbrack\!\lbrack \mathchar"024C
\rbrack\!\rbrack,1,2,3,4}\> \equiv \> C_{\lbrack\!\lbrack 2 \rbrack\!\rbrack,
3}C_{\lbrack\!\lbrack 2,3 \rbrack\!\rbrack, 4} C_{\lbrack\!\lbrack
\mathchar"024C,1 \rbrack\!\rbrack, 3} C_{\lbrack\!\lbrack 2,3\rbrack\!\rbrack,
4} C_{\lbrack\!\lbrack \mathchar"024C,1\rbrack\!\rbrack, 2} C_{\lbrack\!\lbrack
2,3\rbrack\!\rbrack, 4} C_{\lbrack\!\lbrack \mathchar"024C,1\rbrack\!\rbrack,
2}
\,
\end{equation}
(as in Fig.\ \ref{figF}).
This identity actually works irrespective of the number of bits in the enable
string $\mathchar"024C$, but we have succeeded in reducing the elementary gates
to those that can implemented on the basic machine only in the case of $MUXFA$
with a single enable bit.  To reduce the $MUXFA$ operation with two enable bits
to the basic gates, we can apply the same trick again, replacing each
controlled$^3$-NOT by four controlled$^2$-NOT's (using, say, the 4th bit as the
extra bit required by the Barenco {\it et al.} construction).  We will refer to
the resulting operation as $MUXFA''''$.

\begin{figure}
\centerline{\epsfxsize=10truecm \epsfbox{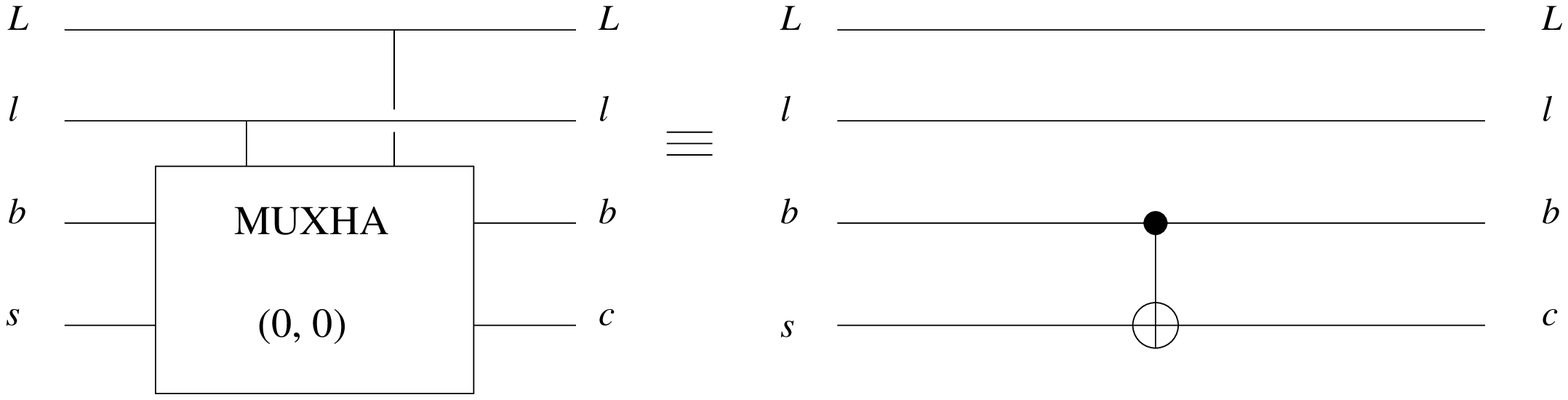}} \vskip .5truecm
\centerline{\epsfxsize=10truecm \epsfbox{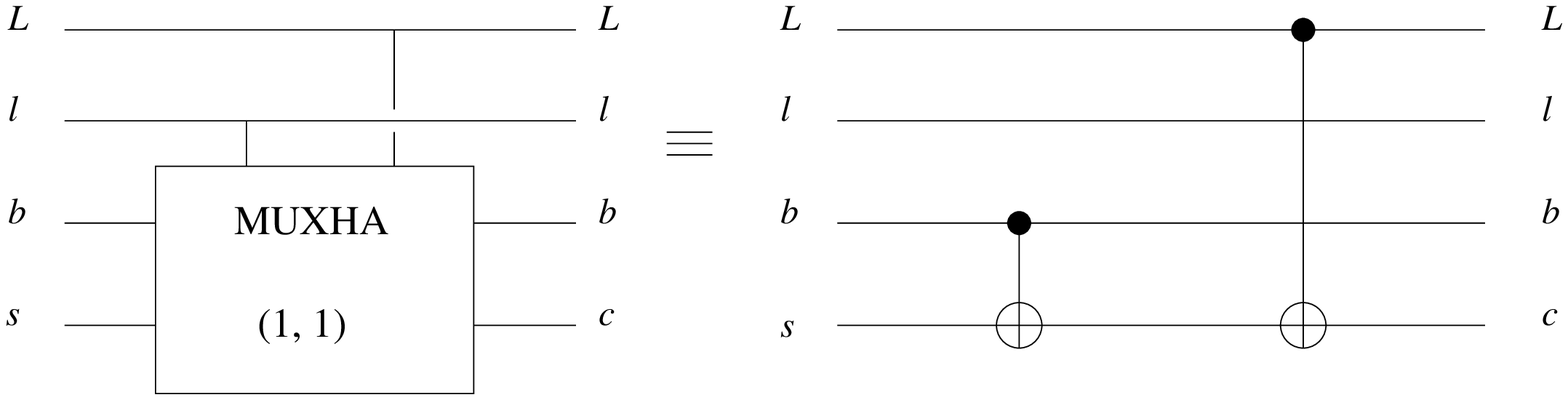}} \vskip .5truecm
\centerline{\epsfxsize=10truecm \epsfbox{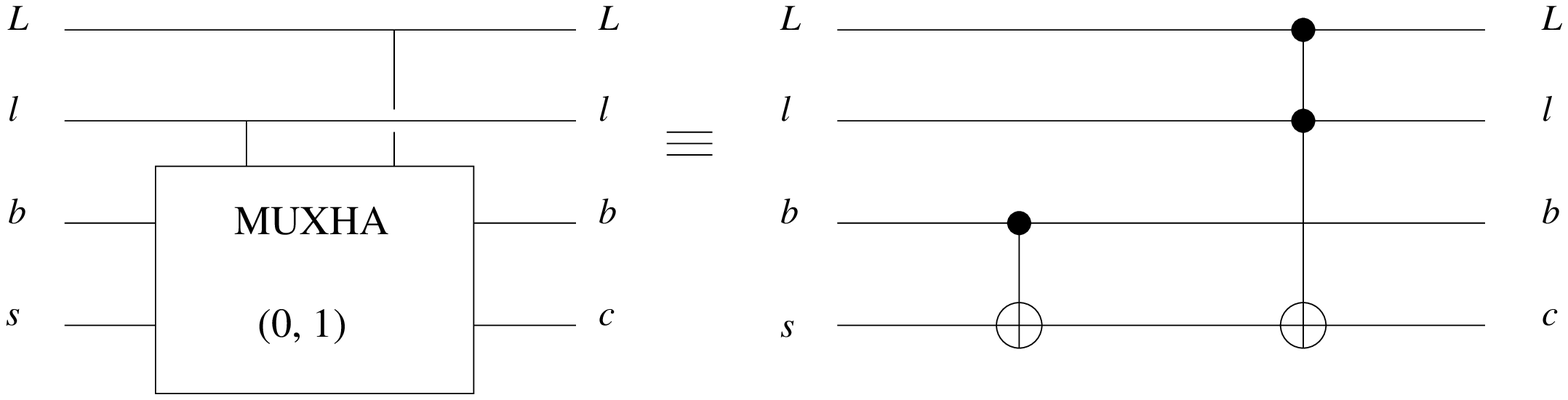}} \vskip .5truecm
\centerline{\epsfxsize=10truecm \epsfbox{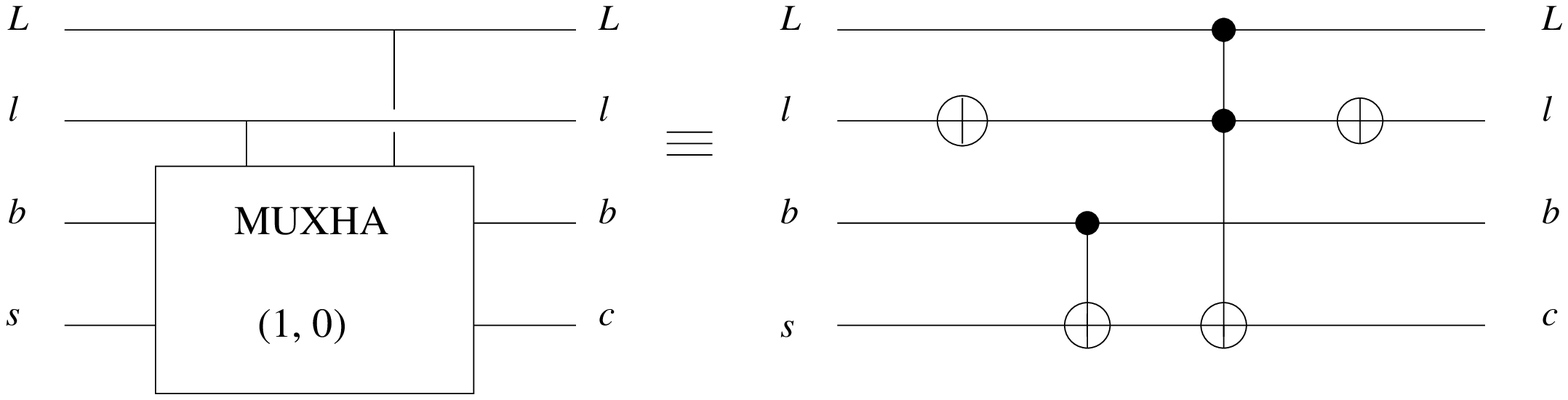}} \vskip .5truecm
\caption{The multiplexed half adder $MUXHA$ is simpler than $MUXFA$ because it
does not compute the output carry bit.}
\label{figG}
\end{figure}

Aside from the multiplexed full adder $MUXFA$, we will also use a multiplexed
{\it half adder} which we will call $MUXHA$.  The half adder does not compute
the final carry bit; it acts according to
\begin{equation}
MUXHA(a_0,a_1)_{\lbrack\!\lbrack \mathchar"024C \rbrack\!\rbrack,1,2,3} : \>\>
 \bigl |{\ell}\bigr\rangle_1 \bigl |b \bigr\rangle_2 \bigl |c \bigr\rangle_3
\longmapsto
 \bigl |{\ell} \bigr\rangle_1 \bigl |b \bigr\rangle_2 \bigl |s \bigr\rangle_3
\, ,
\end{equation}
where $s=a \oplus b \oplus c$, and $a=\mathchar"024C\wedge (a_1\wedge \ell \vee
a_0\wedge \mathchar"0218\ell)$.  (Note that, since the input qubit $b$ is
preserved, the final carry bit is not needed to ensure the reversibility of the
operation.) $MUXHA$ is constructed from elementary gates according to
\begin{eqnarray}
\label{muxha}
MUXHA(a_0=0, a_1=0)_{\lbrack\!\lbrack \mathchar"024C \rbrack\!\rbrack,1,2,3}\>
&\equiv& \> C_{\lbrack\!\lbrack 2 \rbrack\!\rbrack, 3}  \, ,\nonumber \\
MUXHA(a_0=1, a_1=1)_{\lbrack\!\lbrack \mathchar"024C \rbrack\!\rbrack,1,2,3}\>
&\equiv& \> C_{\lbrack\!\lbrack 2 \rbrack\!\rbrack, 3} C_{\lbrack\!\lbrack
\mathchar"024C \rbrack\!\rbrack, 3}\, ,\nonumber \\
MUXHA(a_0=0,a_1=1)_{\lbrack\!\lbrack \mathchar"024C \rbrack\!\rbrack,1,2,3}\>
&\equiv& \> C_{\lbrack\!\lbrack 2 \rbrack\!\rbrack, 3}C_{\lbrack\!\lbrack
\mathchar"024C,1 \rbrack\!\rbrack, 3} \, , \nonumber \\
MUXHA(a_0=1,a_1=0)_{\lbrack\!\lbrack \mathchar"024C \rbrack\!\rbrack,1,2,3}\>
&\equiv&  \> C_1C_{\lbrack\!\lbrack 2 \rbrack\!\rbrack, 3}C_{\lbrack\!\lbrack
\mathchar"024C,1 \rbrack\!\rbrack, 3}  C_1 \,
\end{eqnarray}
(see Fig.\ \ref{figG}).
For a single enable bit, this construction can be carried out on the basic
machine.  If there are two enable bits, the controlled$^3$-NOT's can be
expanded in terms of controlled$^2$-NOT's as described above.

\begin{figure}
\centerline{\epsfxsize=16truecm \epsfbox{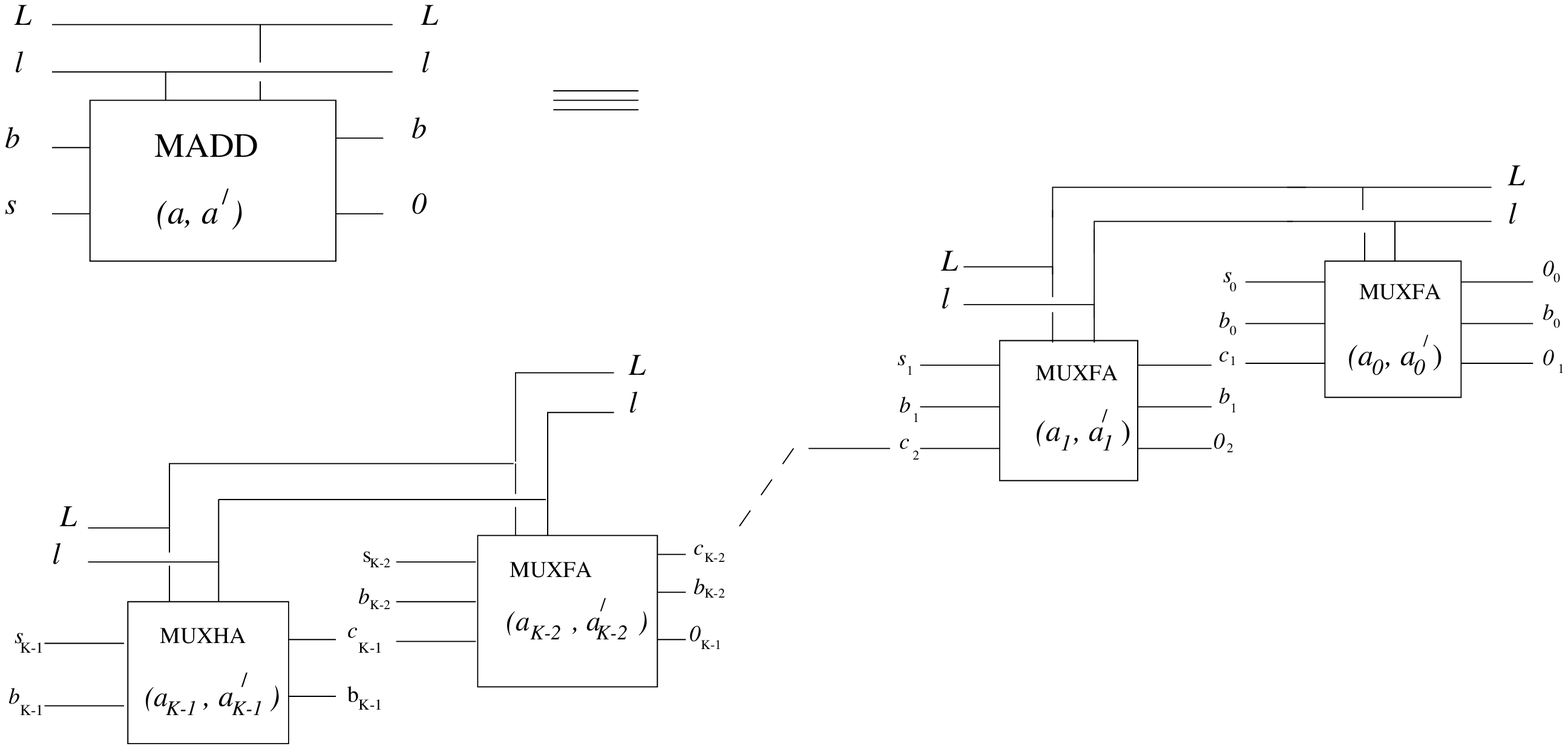}}\vskip .5truecm
\caption{The multiplexed $K$-bit adder $MADD(a,a')$ is constructed by chaining
together $K-1$ $MUXFA$ operations and one $MUXHA$ operation.  $MADD$ adds a
$K$-bit $c$-number to an input $K$-bit $q$-number and obtains an output $K$-bit
$q$-number (the final carry bit is not computed). If $MADD$ is enabled, the
classical addend is $a$ when the select bit has the value $\ell=0$ or is $a'$
when $\ell=1$. (When $MADD$ is not enabled, the classical addend is 0.)}
\label{figH}
\end{figure}

A multiplexed $K$-bit adder is easily constructed  by chaining together $(K-1)$
$MUXFA$ gates and one $MUXHA$ gate, as shown in Fig.\ \ref{figH}.  This
operation, which we denote $MADD$, depends on a pair of $K$-bit $c$-numbers $a$
and $a'$.  $MADD$ (if all enable bits read 1) adds either $a$ or $a'$ to the
$K$-bit $q$-number $b$, with the choice determined by the value of the select
bit $\ell$.  (That is, it adds $a$ for $\ell=0$ and adds $a'$ for $\ell=1$.)
Thus, $MADD$ acts according to:
\begin{equation}
MADD(a,a')_{\lbrack\!\lbrack \mathchar"024C \rbrack\!\rbrack,\beta,\gamma,1} :
\>\>
 \bigl |{b}\bigr\rangle_\beta \bigl |0\bigr\rangle_\gamma \bigl |\ell
\bigr\rangle_1
 \longmapsto
\bigl |{b}\bigr\rangle_\beta \bigl |s \bigr\rangle_\gamma \bigl |\ell
\bigr\rangle_1 \, ;
\end{equation}
where
\begin{equation}
\label{sum}
s=\left[b+ \mathchar"024C\wedge (a'\wedge \ell \vee a\wedge
\mathchar"0218\ell)\right]_{{\rm mod}~ 2^K} \, .
\end{equation}
The $[{\cdot}]_{{\rm mod}~ 2^K}$ notation in Eq.\ (\ref{sum}) indicates that
the sum $s$ residing in $\bigl |\cdot \bigr\rangle_\gamma$ at the end of the
operation is only $K$ bits long---$MADD$ does not compute the final carry bit.
Since we will not need the final bit to perform addition mod $N$, we save a few
elementary operations by not bothering to compute it.  (The $MADD$ operation is
invertible nonetheless.)

Transcribed as an equation, Fig.\ \ref{figH} says that $MADD$ is constructed as
\begin{eqnarray}
\label{madd}
MADD(a,a')_{\lbrack\!\lbrack \mathchar"024C \rbrack\!\rbrack,\beta,\gamma,1}
 \> \equiv \>&&
MUXHA(a_{K-1},a_{K-1}')_{\lbrack\!\lbrack \mathchar"024C
\rbrack\!\rbrack,1,\beta_{K-1},\gamma_{K-1}}\nonumber\\
&&\cdot~ \Bigl(\prod_{\mkern36mu i=0}^{K-2\mkern36mu}
 MUXFA(a_i,a_i')_{\lbrack\!\lbrack \mathchar"024C
\rbrack\!\rbrack,1,\beta_i,\gamma_i,\gamma_{i+1}}\Bigr)
\end{eqnarray}
We have skewed the subscript and superscript of $\prod$ in Eq.\ (\ref{madd}) to
remind the reader that the order of the operations is to be read from right to
left---hence the product has the operator with $i=0$ furthest to the {\it
right} (acting first).  Each $MUXFA$ operation reads the enable string
$\mathchar"024C$, and, if enabled, performs an elementary (multiplexed)
addition, passing its final carry bit on to the next operation in the chain.
The two classical bits used by the $j$th $MUXFA$ are $a_j$ and $a_j'$, the
$j$th bits of the $c$-numbers $a$ and $a'$.  The final elementary addition is
performed by $MUXHA$ rather than $MUXFA$, because the final carry bit will not
be needed.


\subsection{Comparison}
\label{sec:comparison}

In our algorithms, we need to perform addition mod $N$ of a $c$-number $a$ and
a $q$-number $b$.  An important step in modular addition is {\it
comparison}---we must find out whether $a+b \ge N$.  Thus, our next task is to
devise a unitary operation that compares a $c$-number and a $q$-number.  This
operation should, say, flip a target bit if the $c$-number is greater than the
$q$-number, and leave the target bit alone otherwise.

A conceptually simple way to compare a $K$-bit $c$-number $a$ and a $K$-bit
$q$-number $b$ is to devise an adder that computes the sum of the $c$-number
$2^K-1-a$ and the $q$-number $b$.  Since the sum is less than $2^K$ only for
$a>b$, the final carry bit of the sum records the outcome of the comparison.
This method works fine, but we will use a different method that turns out to be
slightly more efficient.

The idea of our method is that we can scan $a$ and $b$ from left to right, and
compare them one bit at a time.  If $a_{K-1}$ and $b_{K-1}$ are different, then
the outcome of the comparison is determined and we are done.  If $a_{K-1}$ and
$b_{K-1}$ are the same, we proceed to examine $a_{K-2}$ and $b_{K-2}$ and
repeat the procedure, {\it etc}.  We can represent this routine in pseudo-code
as
\begin{eqnarray}
\label{pseudolt}
{\tt if}&\ a_{K-1}=0:\quad& \left\{\begin{array}{l}
b_{K-1}=0 \Longrightarrow {\tt PROCEED} \nonumber\\
			     b_{K-1}=1 \Longrightarrow b\ge a\ {\tt END}\end{array}\right\}
\nonumber\\
{\tt if}&\ a_{K-1}=1:\quad& \left\{\begin{array}{l}b_{K-1}=0 \Longrightarrow
b<a \ {\tt END}\nonumber\\
			     b_{K-1}=1 \Longrightarrow {\tt PROCEED}\end{array}\right\}\nonumber\\
{\tt if}&\ a_{K-2}=0:\quad& \left\{\begin{array}{l}b_{K-2}=0 \Longrightarrow
{\tt PROCEED}\nonumber\\
			     b_{K-2}=1 \Longrightarrow b\ge a\ {\tt
END}\end{array}\right\}\nonumber\\
{\tt if}&\ a_{K-2}=1:\quad& \left\{\begin{array}{l}b_{K-2}=0 \Longrightarrow
b<a\ {\tt END}\nonumber\\
			     b_{K-2}=1 \Longrightarrow {\tt PROCEED}\end{array}\right\}\nonumber\\
&\cdot\nonumber\\
&\cdot \\
{\tt if}&\ a_{0}=0:\quad& 	\hskip .9in ~		  b\ge a\ {\tt END}\nonumber\\
{\tt if}&\ a_{0}=1:\quad& \left\{\begin{array}{l}b_{0}=0 \Longrightarrow  b<a\
{\tt END}\nonumber\\
			     b_{0}=1 \Longrightarrow b\ge a\ {\tt END}\end{array}\right\}
\end{eqnarray}
To implement this pseudo-code as a unitary transformation, we will use enable
qubits in each step of the comparison.  Once the comparison has ``ended,'' all
subsequent enable bits will be switched off, so that the subsequent operations
will have no effect on the outcome.  Unfortunately, to implement this strategy
reversibly, we seem to need a new enable bit for (almost) every step of the
comparison, so the comparison operation will fill $K-1$ bits of scratch space
with junk.  This need for scratch space is not really a big deal, though.  We
can immediately clear the scratch space, which will be required for subsequent
use anyway.

As in our construction of the adder, our comparison operation is a sequence of
elementary quantum gates that depends on the value of the $K$-bit $c$-number
$a$.  We will call the operation $LT$ (for ``less than'').  Its action is
\begin{equation}
LT(a)_{\beta,1,{\hat\gamma}} : \>\>
 \bigl |{b}\bigr\rangle_\beta \bigl  |0 \bigr\rangle_1
|0\bigr\rangle_{\hat\gamma}\longmapsto
\bigl |{ b'}\bigr\rangle_\beta \bigl |\ell \bigr\rangle_1\bigl |{\rm junk}
\bigr\rangle_{\hat\gamma}  \, ,
\end{equation}
where $\ell$ takes the value 1 for $b<a$ and the value 0 for $b\ge a$.  Here
the register labeled $|\cdot\bigr\rangle_{\hat\gamma}$ is actually $K-1$ rather
than $K$ qubits long.  The junk that fills this register has a complicated
dependence on $a$ and $b$, the details of which are not of interest.  In
passing, the $LT$ operation also modifies the $q$-number $b$, replacing it by
$b'$.  ($b'$ is almost the {\it negation} of $b$, $b$ with all of its qubits
flipped, except that $b_0$ is not flipped unless $a_0=1$).  We need not be
concerned about this either, as we will soon run the $LT$ operation backwards
to repair the damage.

The $LT$ operation is constructed from elementary gates as:
{\samepage
\begin{eqnarray}
\label{lt_long}
LT(a)_{\beta,1,{\hat\gamma}} \> \equiv \>
&&\left\{
 {\tt if} \ (a_{0} = 1) \
 C_{\lbrack\!\lbrack {\hat\gamma}_0,\beta_0\rbrack\!\rbrack,1} \> C_{\beta_0}
\right\} \\
 \cdot  \prod_{i=1\mkern36mu}^{\mkern36mu K-2} &&\left\{
\begin{array}{ll}
 {\tt if} \ (a_{i} = 0) \
 &
C_{\lbrack\!\lbrack{{\hat\gamma}_i,\beta_i}\rbrack\!\rbrack,{\hat\gamma}_{i-1}}
\> C_{\beta_i} \nonumber\\
 {\tt if }\ (a_{i} = 1) \
 & C_{\lbrack\!\lbrack{{\hat\gamma}_i,\beta_i}\rbrack\!\rbrack,1} \>
C_{\beta_i}
\>
C_{\lbrack\!\lbrack{{\hat\gamma}_i,\beta_i}\rbrack\!\rbrack,{\hat\gamma}_{i-1}}
\end{array} \right\}\nonumber\\
\cdot \> &&\left\{\begin{array}{ll}
{\tt if} \ (a_{K-1} = 0) \
  & C_{\lbrack\!\lbrack \beta_{K-1}\rbrack\!\rbrack,{\hat\gamma}_{K-2}}  \>
C_{\beta_{K-1}} \nonumber\\
 {\tt if} \ (a_{K-1}= 1) \
& C_{\lbrack\!\lbrack \beta_{K-1}\rbrack\!\rbrack,1} \> C_{\beta_{K-1}} \>
 C_{\lbrack\!\lbrack \beta_{K-1}\rbrack\!\rbrack,{\hat\gamma}_{K-2}}
\end{array}\right\}
\end{eqnarray}
}
As usual, the gates furthest to the right act first.  We have skewed the
subscript and superscript of $\prod$ here to indicate that the  operator with
$i=1$ is furthest to the {\it left} (and hence acts last).  The first step of
the $LT$ algorithm is different from the rest because it is not conditioned on
the value of any ``switch.''  For each of the $K-2$ intermediate steps ($i=K-2,
K-1, \dots, 1$), the switch ${\hat\gamma}_i$ is read, and if the switch is on,
the comparison of $a_i$ and $b_i$ is carried out.  If $a_i\ne b_i$, then the
outcome of the comparison of $a$ and $b$ is settled; the value of $\ell$ is
adjusted accordingly, and the switch ${\hat\gamma}_{i-1}$ is {\it not} turned
on.  If $a_i=b_i$, then ${\hat\gamma}_{i-1}$ {\it is} switched on, so that the
comparison can continue.  Finally, the last step can be simplified, as in Eq.\
(\ref{pseudolt}).

We can now easily construct a comparison operator that cleans up the scratch
space, and restores the original value of $b$, by using the trick mentioned in
Sec.\ \ref{sec:space}---we run $LT$, copy the outcome $\ell$ of the comparison,
and then run $LT$ in reverse.  We will actually want our comparison operator to
be enabled by a string $\mathchar"024C$, which we can achieve by controlling
the copy operation with $\mathchar"024C$. The resulting operator, which we call
$XLT$, flips the target qubit if $b<a$:
\begin{eqnarray}
XLT(a)_{\lbrack\!\lbrack \mathchar"024C
\rbrack\!\rbrack,\beta,1,2,{\hat\gamma}} \>
 \equiv \>
&LT(a)^{-1}_{\beta,2,{\hat\gamma}} \> C_{\lbrack\!\lbrack \mathchar"024C, 2
\rbrack\!\rbrack, 1}
 \> LT(a)_{\beta,2,{\hat\gamma}} :\nonumber\\
&\bigl |b\bigr\rangle_\beta \bigl |x\bigr\rangle_1 \bigl |0\bigr\rangle_2 \bigl
|0\bigr\rangle_{\hat\gamma}
 \longmapsto
\bigl |b\bigr\rangle_\beta \bigl |x\oplus y\bigr\rangle_1 \bigl
|0\bigr\rangle_2 \bigl |0\bigr\rangle_{\hat\gamma}
\end{eqnarray}
where $y$ is 1 if $b<a$ and 0 otherwise.  We recall that the register $\bigl
|\cdot\bigr\rangle_{\hat\gamma}$ is actually $K-1$ qubits long, so the $XLT$
routine requires $K$ qubits of scratch space.

\subsection{Addition mod $N$}

Now that we have constructed a multiplexed adder and a comparison operator, we
can easily perform addition mod $N$.  First $XLT$ compares the $c$-number $N-a$
with the $q$-number $b$, and switches on the select bit $\ell$ if $a+b<N$. Then
the multiplexed adder adds either $a$ (for $a+b<N$) or $2^K +a - N$ (for
$a+b\ge N$) to $b$.  Note that $2^K +a -N$ is guaranteed to be positive ($N$
and $a$ are $K$-bit numbers with $a<N$).  In the case where $2^K+a-N$ is added,
the desired result $a+b ~({\rm mod}~ N)$ is obtained by subtracting $2^K$ from
the sum; that is, by dropping the final carry bit.  That is why our $MADD$
routine does not bother to compute this final bit.

\begin{figure}
\centerline{\epsfxsize=12truecm \epsfbox{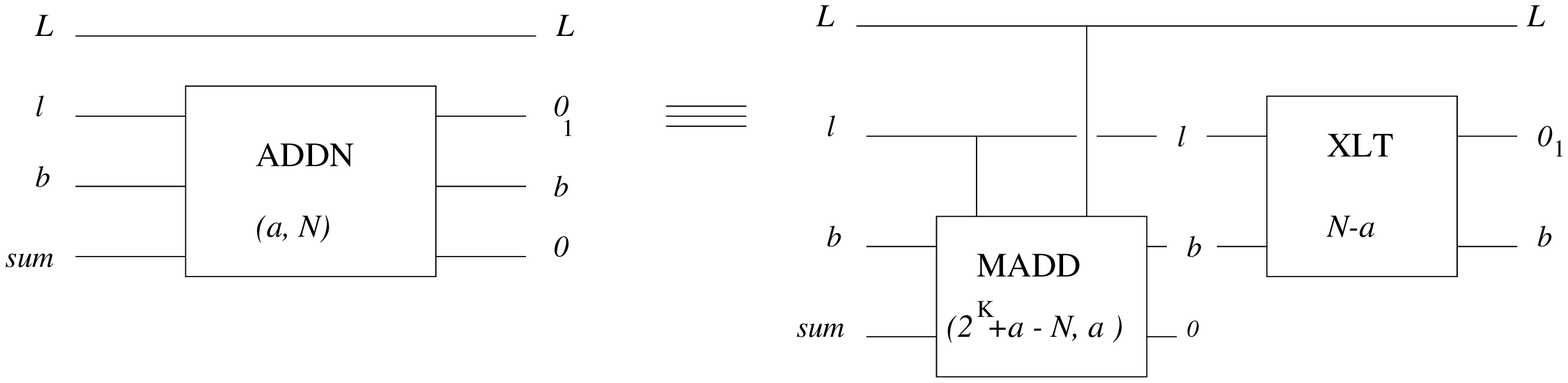}} \vskip .5truecm
\caption{The mod $N$ addition operator $ADDN(a,N)$ computes $a+b~({\rm
mod}~N)$, where $a$ is a $K$-bit $c$-number and $b$ is a $K$-bit $q$-number.
When $ADDN$ is enabled, the comparison operator $XLT(N-a)$ flips the value of
the select bit to $\ell=1$ if $a+b<N$; then the multiplexed adder
$MADD(2^K+a-N,a)$ chooses the $c$-number addend to be $a$ for $\ell=1$ and
$2^K+a-N$ for $\ell=0$.  $XLT$ uses and then clears $K$-bits of scratch space
before $MADD$ writes the mod $N$ sum there.}
\label{figI}
\end{figure}

We call our mod $N$ addition routine $ADDN$; it acts as
\begin{eqnarray}
\label{addn_act}
ADDN(a,N)_{\lbrack\!\lbrack \mathchar"024C \rbrack\!\rbrack,\beta,1,\gamma} :
\>\>
&& \bigl|{b}\bigr\rangle_\beta \bigl |{0}\bigr\rangle_1 \bigl
|{0}\bigr\rangle_\gamma   \nonumber\\
 &&\longmapsto
\bigl|{b}\bigr\rangle_\beta \bigl |{\ell \equiv \mathchar"024C \wedge(a+b <
N)}\bigr\rangle_1 \bigl |b+ \mathchar"024C \wedge a~({\rm mod}~
N)\bigr\rangle_\gamma    \, .
\end{eqnarray}
(Here the notation $\ell \equiv \mathchar"024C \wedge(a+b < N)$ means that the
qubit $\ell$ reads 1 if the statement $\mathchar"024C \wedge(a+b < N)$ is true
and reads 0 otherwise.)  If enabled, this operator computes $a+b ~({\rm mod}~
N)$; if not, it merely copies $b$.\footnote{Thus, if $ADDN$ is {\it not}
enabled, Eq.\ (\ref{addn_act}) is valid only for $b<N$. We assume here and in
the following that $b<N$ is satisfied; in the evaluation of the modular
exponential function, our operators will always be applied to $q$-numbers that
satisfy this condition.} $ADDN$ is constructed from $MADD$ and $XLT$ according
to
\begin{equation}
ADDN(a,N)_{\lbrack\!\lbrack \mathchar"024C \rbrack\!\rbrack,\beta,1,\gamma} \>
 \equiv \>
MADD(2^K +a -N, a)_{\lbrack\!\lbrack \mathchar"024C
\rbrack\!\rbrack,\beta,\gamma,1}\cdot
XLT(N-a)_{\lbrack\!\lbrack \mathchar"024C \rbrack\!\rbrack,\beta,1,\gamma}
\end{equation}
(see Fig.\ \ref{figI}).
Note that $XLT$ uses and then clears the $K$ bits of scratch space in the
register $\bigl | \cdot\bigr \rangle_\gamma$, before $MADD$ writes the mod $N$
sum there.

The $ADDN$ routine can be viewed as the computation of an invertible function
(specified by the $c$-numbers $a$ and $N$) of the $q$-number $b$.  (Note that
the output of this function is the sum $a+b ~({\rm mod}~ N)$ {\it and} the
comparison bit $\ell$---the comparison bit is needed to ensure invertibility,
since it is possible that $b\ge N$).  Thus, we can use the trick mentioned in
Sec.\ \ref{sec:space} to devise an ``overwriting'' version of this function.
Actually, since we will not need to know the value of $\ell$ (or worry about
the case $b\ge N$), we can save a qubit by modifying the trick slightly.

\begin{figure}
\centerline{\epsfxsize=12truecm \epsfbox{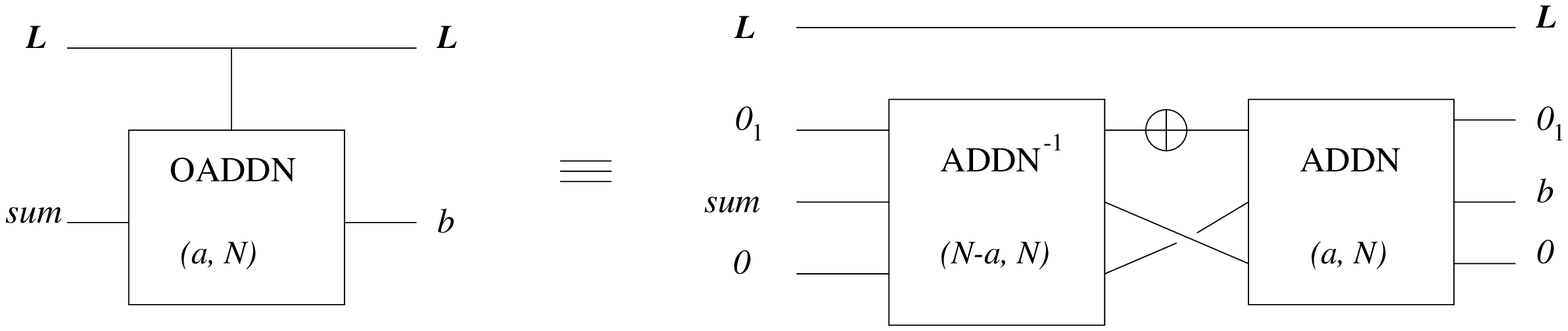}} \vskip .5truecm
\caption{The {\it overwriting} mod $N$ addition operator $OADDN(a,N)$ (when
enabled) adds the $c$-number $a$ to the $q$-number $b$, and then erases $b$.
The ``swapping of the leads'' is a classical operation, not a quantum gate.
$OADDN$ uses and then clears $K+1$ bits of scratch space; this scratch space is
suppressed on the left side of the figure.}
\label{figJ}
\end{figure}

The overwriting addition routine $OADDN$ is constructed as
\begin{eqnarray}
\label{oaddn}
OADDN(a,N)_{\lbrack\!\lbrack \mathchar"024C \rbrack\!\rbrack,\beta,1,\gamma} \>
\equiv \> && SWAP_{\beta,\gamma} \> ADDN^{-1}(N-a,N)_{\lbrack\!\lbrack
\mathchar"024C \rbrack\!\rbrack,\gamma,1,\beta}\nonumber\\
&&\cdot \> C_{\lbrack\!\lbrack \mathchar"024C\rbrack\!\rbrack,1} \>
ADDN(a,N)_{\lbrack\!\lbrack \mathchar"024C \rbrack\!\rbrack,\beta,1,\gamma}
\end{eqnarray}
(see Fig\ \ref{figJ}),
and acts (for $b<N$) according to
\begin{eqnarray}
\label{oaddnact}
OADDN(a,N)_{\lbrack\!\lbrack \mathchar"024C \rbrack\!\rbrack,\beta,1,\gamma} :
\>\>
&& \bigl|{b}\bigr\rangle_\beta \bigl |{0}\bigr\rangle_1 \bigl
|{0}\bigr\rangle_\gamma   \nonumber\\
 && \longmapsto
\bigl|{b}\bigr\rangle_\beta \bigl |{\ell \equiv \mathchar"024C \wedge(a+b <
N)}\bigr\rangle_1 \bigl |b+ \mathchar"024C \wedge a~({\rm mod}~
N)\bigr\rangle_\gamma     \nonumber\\
 && \longmapsto
\bigl|{b}\bigr\rangle_\beta \bigl |{\ell \equiv \mathchar"024C \wedge(a+b \ge
N)}\bigr\rangle_1 \bigl |b+ \mathchar"024C \wedge a~({\rm mod}~
N)\bigr\rangle_\gamma    \nonumber\\
 && \longmapsto
\bigl|{0}\bigr\rangle_\beta \bigl |0\bigr\rangle_1 \bigl |b+ \mathchar"024C
\wedge a~({\rm mod}~ N)\bigr\rangle_\gamma \nonumber\\
&& \longmapsto
\bigl |b+ \mathchar"024C \wedge a~({\rm mod}~ N)\bigr\rangle_\beta
|0\bigr\rangle_1\bigl|{0}\bigr\rangle_\gamma \, .
\end{eqnarray}
Here, in Eq.\ (\ref{oaddnact}), we have indicated the effect of each of the
successive operations in Eq.\ (\ref{oaddn}).  We can easily verify that
applying $ADDN(N-a,N)_{\lbrack\!\lbrack \mathchar"024C
\rbrack\!\rbrack,\gamma,1,\beta}$ to the second-to-last line of Eq.\
(\ref{oaddnact}) yields the preceding line.  If the enable string
$\mathchar"024C$ is false, the verification is trivial, for $b<N$.  (It was in
order to ensure that this would work that we needed the $XLT$ operation to be
enabled by $\mathchar"024C$.) When $\mathchar"024C$ is true, we need only
observe that $N-a +[b+a ~({\rm mod}~ N)]<N$ if and only if $a+b\ge N$ (assuming
that $b<N$).

The $SWAP$ operation in Eq.\ (\ref{oaddn}) is not a genuine quantum operation
at all; it is a mere {\it relabeling} of the $\bigl|{\cdot}\bigr\rangle_\beta$
and $\bigl|{\cdot}\bigr\rangle_\gamma$ registers that is performed by the {\it
classical} computer.  We have included the $SWAP$ because it will be convenient
for the sum to be stored in the $\bigl|{\cdot}\bigr\rangle_\beta$
register when we chain together $OADDN$'s to construct a multiplication
operator.  We see that $OADDN$ uses and then clears $K+1$ qubits of scratch
space.

\subsection{Multiplication mod $N$}

We have already explained in Sec.\ \ref{sec:enable} how mod $N$ multiplication
can be constructed from {\it conditional} mod $N$ addition.  Implementing the
strategy described there, we can construct a conditional multiplication
operator $MULN$ that acts according to
\begin{eqnarray}
\label{muln_act}
MULN(a,N)_{\lbrack\!\lbrack \mathchar"024C
\rbrack\!\rbrack,\beta,\gamma,1,\delta}: \>\> &&\bigl|{b}\bigr\rangle_\beta
\bigl |{0}\bigr\rangle_\gamma \bigl |{0}\bigr\rangle_1 \bigl
|{0}\bigr\rangle_\delta   \nonumber\\
&&\longmapsto
\bigl|{b}\bigr\rangle_\beta \bigl |\mathchar"024C\wedge a\cdot b ~({\rm mod}~N)
\bigr\rangle_\gamma\bigl |0\bigr\rangle_1 \bigl |0\bigr\rangle_\delta \, .
\end{eqnarray}
If enabled, $MULN$ computes the product mod $N$ of the $c$-number $a$ and the
$q$-number $b$; otherwise, it acts trivially.

\begin{figure}
\centerline{\epsfxsize=12truecm \epsfbox{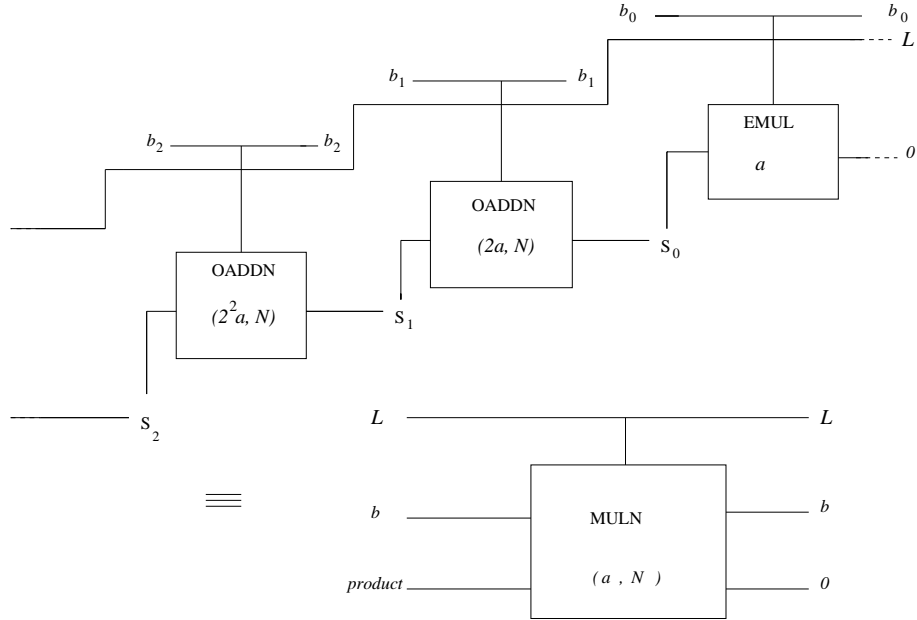}} \vskip .5truecm
\caption{The mod $N$ multiplication operator $MULN(a,N)$ (when enabled)
computes $a\cdot b ~({\rm mod}~N)$, where $a$ is a $c$-number and $b$ is a
$q$-number; it is constructed by chaining together $K-1$ $OADDN$ operators and
one $EMUL$ operator.}
\label{figK}
\end{figure}

We could construct $MULN$ by chaining together $K$ $OADDN$ operators. The first
$ADDN$ loads $a\cdot b_0$, the second adds $a\cdot 2b_1$, the third adds
$a\cdot 2^2 b_2$, and so on.  But we can actually save a few elementary
operations by simplifying the first operation in the chain.  For this purpose
we introduce an elementary multiplication operator $EMUL$ that multiplies a
$c$-number $a$ by a single qubit $b_0$:
\begin{equation}
EMUL(a)_{\lbrack\!\lbrack \mathchar"024C \rbrack\!\rbrack,1,\gamma}: \>\>
\bigl|{b_0}\bigr\rangle_1 \bigl |{0}\bigr\rangle_\gamma \longmapsto
\bigl|{b_0}\bigr\rangle_1 \bigl |\mathchar"024C\wedge{a\cdot
b_0}\bigr\rangle_\gamma \, ,
\end{equation}
which is constructed according to
\begin{equation}
EMUL(a)_{\lbrack\!\lbrack \mathchar"024C \rbrack\!\rbrack,1,\gamma}\>\equiv\>
\prod_{\mkern36mu i=0}^{ K-1\mkern36mu}
{\tt if \ (a_{\tt{i}} = 1) \ }
C_{\lbrack\!\lbrack{\mathchar"024C,1}\rbrack\!\rbrack, \gamma_i} \, .
\end{equation}
Now we can construct $MULN$ as
\begin{eqnarray}
\label{muln}
MULN(a,N)_{\lbrack\!\lbrack \mathchar"024C
\rbrack\!\rbrack,\beta,\gamma,1,\delta}\>\equiv\>
\prod_{\mkern36mu i=1}^{ K-1\mkern36mu} && OADDN(2^i\cdot a~({\rm
mod}~N),N)_{\lbrack\!\lbrack \mathchar"024C,
\beta_i\rbrack\!\rbrack,\gamma,1,\delta} \nonumber\\
&&\cdot EMUL(a)_{\lbrack\!\lbrack \mathchar"024C
\rbrack\!\rbrack,\beta_0,\gamma}
\end{eqnarray}
(see Fig.\ \ref{figK}).  Note that the computation of $2^i\cdot a ~({\rm mod} ~
N)$ is carried out by the classical computer. (It can be done efficiently by
``repeated doubling.'')

\begin{figure}
\centerline{\epsfxsize=12truecm \epsfbox{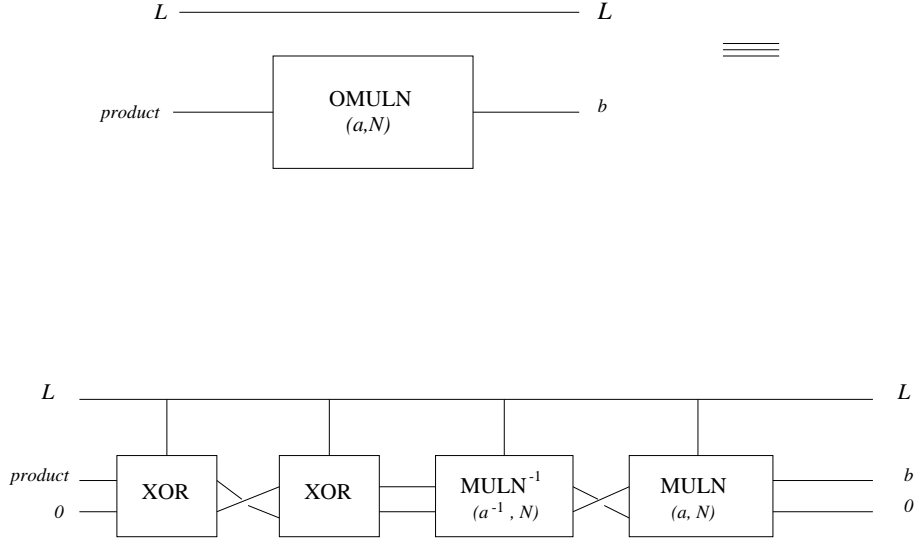}} \vskip .5truecm
\caption{The overwriting mod $N$ multiplication operator $OMULN(a,N)$ (when
enabled) computes $a\cdot b ~({\rm mod}~N)$ and then erases the $q$-number $b$.
 The $XOR$ gates at the end (when enabled) swap the contents of the two
registers.  $OMULN$ uses and then clears $2K+1$ qubits of scratch space, of
which only $K$ bits are indicated in the figure.}
\label{figL}
\end{figure}

As long as $a$ and $N$ have no common divisor ($gcd(a,N)=1$), the operation of
multiplying by $a$ (mod $N$) is invertible.  In fact, the multiplicative
inverse $a^{-1}$ (mod $N$) exists, and $MULN(a)$ is inverted by $MULN(a^{-1})$.
 Thus, we can use the trick discussed in Sec.\ \ref{sec:space} to construct an
overwriting version of the multiplication operator.  This operator, denoted
$OMULN$, acts according to
\begin{eqnarray}
\label{omuln_act}
OMULN(a,N)_{\lbrack\!\lbrack \mathchar"024C
\rbrack\!\rbrack,\beta,\gamma,1,\delta}: \>\> &&\bigl|{b}\bigr\rangle_\beta
\bigl |{0}\bigr\rangle_\gamma \bigl |{0}\bigr\rangle_1 \bigl
|{0}\bigr\rangle_\delta   \nonumber\\
&&\longmapsto
\bigl |\mathchar"024C\wedge a\cdot b ~({\rm mod}~N)
\vee\mathchar"0218\mathchar"024C\wedge
b\bigr\rangle_\beta\bigl|{0}\bigr\rangle_\gamma\bigl |0\bigr\rangle_1 \bigl
|0\bigr\rangle_\delta \, ,
\end{eqnarray}
Note that $OMULN$ acts trivially when not enabled. It can be constructed as
\begin{eqnarray}
\label{omuln}
OMULN(a,N)_{\lbrack\!\lbrack \mathchar"024C
\rbrack\!\rbrack,\beta,\gamma,1,\delta} \> \equiv\> &&
XOR_{\lbrack\!\lbrack \mathchar"024C \rbrack\!\rbrack,\beta,\gamma}\> \cdot \>
XOR_{\lbrack\!\lbrack \mathchar"024C \rbrack\!\rbrack,\gamma,\beta}\nonumber\\
&& \cdot \> MULN^{-1}(a^{-1},N)_{\lbrack\!\lbrack \mathchar"024C
\rbrack\!\rbrack,\gamma,\beta,1,\delta}\>
\cdot \>
MULN(a,N)_{\lbrack\!\lbrack \mathchar"024C
\rbrack\!\rbrack,\beta,\gamma,1,\delta}
\end{eqnarray}
(see Fig.\ \ref{figL}). Here, the (conditional) $XOR$ operation is
\begin{equation}
XOR_{\lbrack\!\lbrack \mathchar"024C \rbrack\!\rbrack,\alpha,\beta} \> \equiv
\> \prod_{i=0}^{L-1}C_{\lbrack\!\lbrack
\mathchar"024C,\alpha_i\rbrack\!\rbrack, \beta_i} :\>\> \bigl| a
\bigr\rangle_\alpha \bigl| b \bigr\rangle_\beta \longmapsto \bigl| a
\bigr\rangle_\alpha \bigl| b\oplus (a\wedge \mathchar"024C)  \bigr\rangle_\beta
\
\end{equation}
where $\oplus$ denotes bitwise addition modulo 2.  It is easy to verify that,
when enabled, $OMULN$ acts as specified in Eq.\ (\ref{omuln_act}); the two
$XOR's$ at the end are needed to swap $\bigl|{0}\bigr\rangle_\beta$ and $\bigl
| a\cdot b ~({\rm mod}~N) \bigr\rangle_\gamma$.  To verify Eq.\
(\ref{omuln_act}) when $OMULN$ is {\it not} enabled, we need to know that
$MULN$, when not enabled, acts according to
\begin{eqnarray}
\label{funny_omuln}
MULN(a,N)_{\lbrack\!\lbrack \mathchar"024C\ne 1
\rbrack\!\rbrack,\beta,\gamma,1,\delta}: \>\> &&\bigl|{0}\bigr\rangle_\beta
\bigl |{b}\bigr\rangle_\gamma \bigl |{0}\bigr\rangle_1 \bigl
|{0}\bigr\rangle_\delta   \nonumber\\
&&\longmapsto
\bigl|{0}\bigr\rangle_\beta \bigl | b \bigr\rangle_\gamma\bigl |0\bigr\rangle_1
\bigl |0\bigr\rangle_\delta \, .
\end{eqnarray}
Though Eq.\ (\ref{funny_omuln}) does not follow directly from the defining
action of $MULN$ specified in Eq.\ (\ref{muln_act}),
it can be seen to be a consequence of Eq.\ (\ref{muln},\ref{oaddnact}).
Note that the computation of $a^{-1}$ is performed by the classical computer.
(This is, in fact, the most computationally intensive task that our classical
computer will need to perform.)

We will require the $OMULN$ operator with an enable string $\mathchar"024C$
that is only a single qubit.  Thus the construction that we have described can
be implemented on our enhanced machine. So constructed, the $OMULN$ operator
uses (and then clears) $2K+1$ qubits of scratch space.  This amount is all of
the scratch space that will be required to compute the modular exponential
function.

\begin{figure}
\centerline{\epsfxsize=12truecm \epsfbox{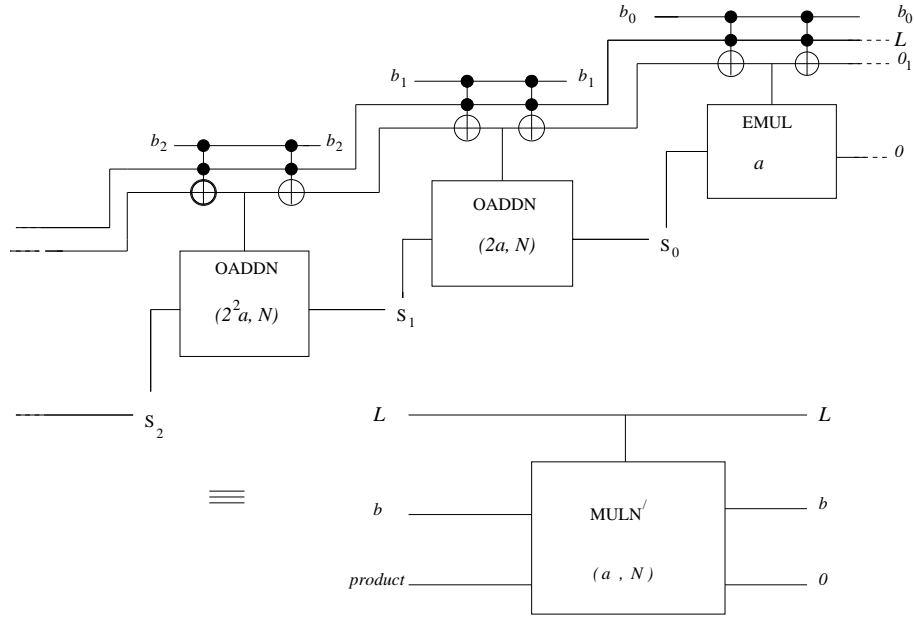}} \vskip .5truecm
\caption{The modified mod $N$ multiplication routine $MULN'(a,N)$ uses simpler
elementary gates than those used by $MULN$, but $MULN'$ requires an extra bit
of scratch space. Instead of calling the $OADDN$ routine with two enable bits,
$MULN'$ first stores the AND of the two enable bits in the extra scratch bit.
Then $OADDN$ with one enable bit can be called instead, where the scratch bit
is the enable bit.}
\label{figLa}
\end{figure}

If we wish to construct $OMULN$ on the basic machine (using
controlled$^k$-NOT's with $k=0,1,2$), there are several alternatives.  One
alternative (that requiring the fewest elementary gates) is to use two
additional qubits of scratch space ($2K+3$ scratch qubits altogether).  Then,
when $MULN$ calls for $OADDN$ with two enable bits, we use one of the scratch
qubits to store the logical AND of the two enable bits.  Now $OADDN$ with one
enable bit can be called instead, where the scratch bit is the enable bit.
(See Fig.\ \ref{figLa}.)  When $OADDN$ eventually calls for $MUXFA$ with a
single enable bit, we can use the second extra scratch qubit to construct
$MUXFA''$ as in Eq.\ (\ref{muxfaprimeprime}) and Fig.\  \ref{figEa}.  Of
course, another alternative is to use the Barenco {\it et al.} identity Eq.\
(\ref{barenco}) repeatedly to expand all the controlled$^3$-NOT and
controlled$^4$-NOT gates in terms of controlled$^k$-NOT gates with $k=0,1,2$.
Then we can get by with $2K+1$ bits of scratch space, but at the cost of
sharply increasing the number of elementary gates.

\subsection{Modular exponentiation}

\begin{figure}
\centerline{\epsfxsize=12truecm \epsfbox{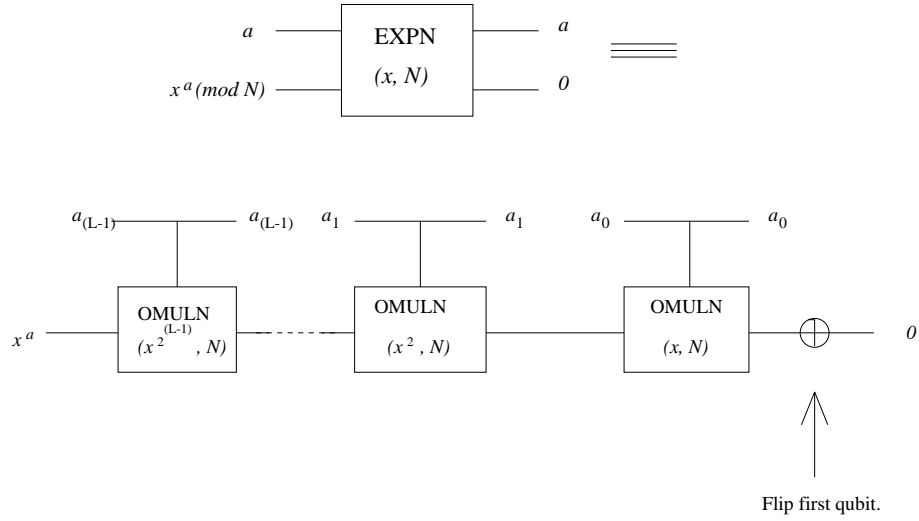}} \vskip .5truecm
\caption{The mod $N$ exponentiation operator $EXPN(x,N)$ computes $x^a ~({\rm
mod}~N)$, where $x$ is a $K$-bit $c$-number and $a$ is an $L$-bit $q$-number.
It is constructed by chaining together $L$ $OMULN$ operators and a NOT.  The
$2K+1$ qubits of scratch space used by $EXPN$ are suppressed in the figure.
The first $OMULN$ in the chain can be replaced by a simpler operation, as is
discussed in the text.}
\label{figM}
\end{figure}

The operator $EXPN$ that computes the modular exponentiation operator can now
be constructed from the conditional overwriting multiplication operator, as
outlined in Sec.\ \ref{sec:repeated}.  Its action is:
\begin{eqnarray}
EXPN(x,N)_{\alpha,\beta,\gamma,1,\delta}: \>\> &&\bigl|{a}\bigr\rangle_\alpha^*
\bigl |{0}\bigr\rangle_\beta \bigl |{0}\bigr\rangle_\gamma \bigl
|{0}\bigr\rangle_1 \bigl |{0}\bigr\rangle_\delta   \nonumber\\
&&\longmapsto
\bigl|{a}\bigr\rangle_\alpha^* \bigl |{x^a ~({\rm mod} ~ N)}\bigr\rangle_\beta
\bigl |{0}\bigr\rangle_\gamma \bigl |{0}\bigr\rangle_1 \bigl
|{0}\bigr\rangle_\delta
 \, .
\end{eqnarray}
(Recall that $\bigl|{\cdot}\bigr\rangle_\alpha^*$ denotes a register that is
$L$ qubits long; $N$ and $x$ are $K$-bit $c$-numbers.)  It is constructed as
\begin{equation}
\label{expn_from_muln}
EXPN(x,N)_{\alpha,\beta,\gamma,1,\delta}\> \equiv\>\left(
\prod_{\mkern36mu i=0}^{ L-1\mkern36mu} OMULN(x^{2^i}~({\rm
mod}~N),N)_{\lbrack\!\lbrack \alpha_i\rbrack\!\rbrack,\beta,\gamma,1,\delta} \>
\right) C_{\beta_0}\,
\end{equation}
(Fig.\ \ref{figM}).  Note that the $C_{\beta_0}$ is necessary at the beginning
to set the register $\bigl |{\cdot}\bigr\rangle_\beta$ to 1 (not 0).  The
classical computer must calculate each $x^{2^i}$ and each inverse $x^{-2^i}$.
The computation of $x^{-1}$ (mod $N$) can be performed using Euclid's algorithm
in O($K^3$) elementary bit operations using ``grade school'' multiplication, or
more efficiently using fast multiplication tricks.  Fortunately, only one
inverse need be computed---the $x^{-2^i}$'s, like the $x^{2^i}$'s, are
calculated by repeated squaring.

Actually, it is possible to reduce the number of quantum gates somewhat if the
NOT and the first $OMULN$ in Eq.\ (\ref{expn_from_muln}) are replaced by the
simpler operation
\begin{equation}
\left( C_{\alpha_0}C_{\lbrack\!\lbrack \alpha_0\rbrack\!\rbrack,\beta_0}
C_{\alpha_0}\right) \cdot EMUL(x)_{\alpha_0,\beta} \, .
\end{equation}
It is easy to verify that this operator has the same action on the state
$\bigl|{a_0}\bigr\rangle_{\alpha_0} \bigl |{0}\bigr\rangle_\beta$ as
$OMULN(x,N)_{\lbrack\!\lbrack \alpha_0\rbrack\!\rbrack,\beta,\gamma,1,\delta}
\cdot C_{\beta_0}$.  With this substitution, we have defined the $EXPN$
operation whose complexity will be analyzed in the following section.

\section{Space versus time}
Now that we have spelled out the algorithms in detail, we can count the number
of elementary quantum gates that they use.

\subsection{Enhanced machine}
\label{sec:enhanced_count}
We will use the notation
\begin{equation}
\label{enhanced_count}
[OPERATOR]=[c_0,c_1,c_2,c_3,c_4]
\end{equation}
to indicate that $OPERATOR$ is implemented using $c_0$ NOT gates, $c_1$
controlled-NOT gates, $c_2$ controlled$^2$-NOT gates, $c_3$ controlled$^3$-NOT
gates, and $c_4$ controlled$^4$-NOT gates on the enhanced machine, or
\begin{equation}
[OPERATOR]=[c_0,c_1,c_2]
\label{basic_count}
\end{equation}
to indicate that $OPERATOR$ is implemented using $c_0$ NOT gates, $c_1$
controlled-NOT gates, and $c_2$ controlled$^2$-NOT gates on the basic machine.
By inspecting the network constructed in Sec.\ \ref{sec:detail}, we see that
the following identities hold:
\begin{eqnarray}
\label{expn_count}
&& [EXPN] \> = \>(L-1) \cdot \left [OMULN_{[1]}\right] \> + \> [EMUL]\> + \>
\left[{\rm controlled}{\rm -NOT}\right] \> + \>2\cdot [{\rm NOT}] \, ;
\nonumber\\
&& \left [OMULN_{[1]}\right] \> = \>  2\cdot \left [MULN_{[1]}\right]\> + \> 2
\cdot \left [XOR_{[1]}\right] \, ; \nonumber\\
&& \left [MULN_{[1]}\right]\> = \> (K-1)\cdot \left [OADDN_{[2]}\right] \> + \>
\left [EMUL_{[1]}\right] \, ; \nonumber\\
&& \left [OADDN_{[2]}\right] \> = \> 2\cdot \left [ADDN_{[2]}\right]\> + \>
\left[{\rm controlled}^2{\rm -NOT}\right] \, ; \nonumber\\
&& \left [ADDN_{[2]}\right] \> = \> \left [MADD_{[2]}\right] \> + \> \left
[XLT_{[2]}\right] \, ; \nonumber\\
&& \left [MADD_{[2]}\right] \> = \> (K-1)\cdot \left[MUXFA_{[2]}\right] \> + \>
\left[MUXHA_{[2]}\right] \, ; \nonumber\\
&& \left [XLT_{[2]}\right] = 2\cdot [LT] \> + \>\left[{\rm controlled}^3{\rm
-NOT} \right]\, .
\end{eqnarray}
These equations just say that $OMULN_{[1]}$, say, is constructed from $2~
MULN_{[1]}$'s and $2~XOR_{[1]}$'s, and so forth.
The subscript ${}_{[\cdot]}$ indicates the length of the string of enable bits
for each operator.  By combining these equations, we find the following
expression for the total number of elementary gates called by our $EXPN$
routine:
\begin{eqnarray}
[EXPN] \> =  \> (L-1)\cdot &&
\Bigl \{ \>
4(K-1)^2 \cdot \left[MUXFA_{[2]}\right] \> + \> 4(K-1)
\cdot\left[MUXHA_{[2]}\right] \> \nonumber\\
&& + \> 8(K-1)\cdot  [LT] \>
 + \> 4(K-1)\cdot \left[{\rm controlled}^3{\rm -NOT}\right]\nonumber\\
&& + \> 2(K-1)\cdot \left[{\rm controlled}^2{\rm -NOT}\right] \>  + \> 2\cdot
\left [EMUL_{[1]}\right] \> + \>  2\cdot \left [XOR_{[1]}\right] \>
\Bigr \}\nonumber\\
&& \> + \>  \> [EMUL]\> + \> \left[{\rm controlled}{\rm -NOT}\right] \> +
\>2\cdot [{\rm NOT}] \, .
\end{eqnarray}
By plugging in the number of elementary gates used by $MUXFA$, $MUXHA$, $LT$,
$EMUL$, and $XOR$, we can find the number of controlled$^k$-NOT gates used in
the $EXPN$ network.

For large $K$, the leading term in our expression for the number of gates is of
order $LK^2$.  Only the $MUXFA$ and $LT$ operators contribute to this leading
term; the other operators make a subleading contribution.  Thus
\begin{equation}
\label{expn_ops}
[EXPN] \> =  \> \Bigl (\> 4LK^2\cdot \left[MUXFA_{[2]}\right] + 8LK\cdot [LT]
\>\Bigr ) \Bigl( \> 1 + O(1/K)\>\Bigr ) \, .
\end{equation}
We will now discuss how this leading term varies as we change the amount of
available scratch space, or replace the enhanced machine by the basic machine.

The numbers of elementary gates used by $MUXFA$ and by $LT$ actually depend on
the particular values of the classical bits in the binary expansions of $2^j
x^{\pm 2^i}$ (mod $N$) and $2^K -N + 2^jx^{\pm 2^i}$ (mod $N$), where
$j=1,\dots, K-1$ and $i=0,1,\dots,L-1$.  We will estimate the number of gates
in two different ways.  To count the gates in the ``worst case,'' we always
assume that the classical bits take values that maximize the number of gates.
To count in the ``average case,'' we make the much more reasonable assumption
that the classical bits take the value 0 with probability $1\over 2$ and take
the value 1 with probability $1\over 2$.

For example, in the case of the implementation of $MUXFA_{[2]}$ on the enhanced
machine described in Eq.\ (\ref{muxfa}), counting the operations yields
\begin{eqnarray}
&&\left[MUXFA(0,0)_{[2]}\right] \> =\>[0,1,1,0,0] \, ,\nonumber\\
&&\left[MUXFA(1,1)_{[2]}\right] \> =\>[0,1,2,1,0] \, ,\nonumber\\
&&\left[MUXFA(0,1)_{[2]}\right] \> =\>[0,1,1,1,1] \, ,\nonumber\\
&&\left[MUXFA(1,0)_{[2]}\right] \> =\>[2,1,1,1,1] \, ,
\end{eqnarray}
and thus
\begin{eqnarray}
& \left[MUXFA_{[2]}\right]^{\rm worst}\>&=\> [2,1,2,1,1] \, ,\nonumber\\
& \left[MUXFA_{[2]}\right]^{\rm ave}\>&=\> \left[{1\over 2},1,{5\over
4},{3\over 4},{1\over 2}\right] \, .
\end{eqnarray}
That is, the worst case is the {\it maximum} in each column, and the average
case is the {\it mean} of each column.  When we quote the number of gates
without any qualification, the average case is meant.  Similarly, for the $LT$
operation described in Eq.\ (\ref{lt_long}), we have
\begin{eqnarray}
& \left[LT\right]^{\rm worst}\>&=\> [K,2,2K-3,0,0] \, ,\nonumber\\
& \left[LT\right]^{\rm ave}\>&=\> \left[K-{1\over 2},{3\over 2},{3\over
2}K-{5\over 2},0,0\right] \, .
\end{eqnarray}
Note that $LT$ uses no controlled$^3$-NOT or controlled$^4$-NOT gates, and so
can be implemented as above on the basic machine.

Now, from Eq.\ (\ref{expn_ops}), we find the leading behavior of the number of
gates used by the $EXPN$ routine:
\begin{eqnarray}
\label{worst_enhanced}
&&[EXPN]_{{\rm enhanced},2K+1}^{\rm worst} \> =  \> \> LK^2\cdot [16,4,24,4,4]
\cdot \Bigl( \> 1 + O(1/K)\>\Bigr )\, ,\nonumber\\
&&[EXPN]_{{\rm enhanced},2K+1}^{\rm ave} \> =  \> \> LK^2\cdot [10,4,17,3,2]
\cdot \Bigl( \> 1 + O(1/K)\>\Bigr )\, ,
\end{eqnarray}
where the subscript ${}_{{\rm enhanced},2K+1}$ serves to remind us that this
count applies to the enhanced machine with $2K+1$ qubits of scratch space.  A
convenient (though quite crude) ``one-dimensional'' measure of the complexity
of the algorithm is the total number of laser pulses required to implement the
algorithm on a linear ion trap, following the scheme of Cirac and Zoller.
Assuming 1 pulse for a NOT and $2k+3$ pulses for a controlled$^k$-NOT,
$k=1,2,3,4$, we obtain
\begin{eqnarray}
&&[EXPN]_{{\rm enhanced},2K+1}^{\rm worst~ pulses} \> =  \> \> 256LK^2\cdot
\Bigl( \> 1 + O(1/K)\>\Bigr )\, ,\nonumber\\
&&[EXPN]_{{\rm enhanced},2K+1}^{\rm ave~ pulses} \> =  \> \> 198LK^2\cdot
\Bigl( \> 1 + O(1/K)\>\Bigr )\, .
\end{eqnarray}
(The estimate for the worst case is not obtained directly from Eq.\
(\ref{worst_enhanced}); instead we assume that $MUXFA$ is always called with
the argument $(a_0=1,a_1=0)$---this maximizes the number of pulses required,
though it does not maximize the number of controlled$^2$-NOT gates.)
Including the subleading contributions, the count of gates and pulses
used by our network in the average case is
\begin{eqnarray}
\label{pulses_enh_twoKone}
&[EXPN]_{{\rm enhanced},2K+1}^{\rm ave} \> =  \> \> (L-1)\cdot
&[10K^2-14K+4,4K^2+8K -12,17K^2-36K+22,\nonumber\\
&&3K^2-3,2K^2-4K+2]
+[2,{1\over 2}K+1,0,0,0]\, ,\nonumber\\
&[EXPN]_{{\rm enhanced},2K+1}^{\rm ave~ pulses} \> =  \> \>
(L-1)\cdot&\bigl(198K^2-270K+93\bigr)+{5\over 2}K +7 \, .
\end{eqnarray}

By allowing one extra qubit of scratch space, we can reduce the complexity
(measured in laser pulses) somewhat.  When $MULN_{[1]}$ calls for
$OADDN_{[2]}$, we may use a controlled$^2$-NOT to store the AND of the two
enable bits in the extra scratch qubit, and then call $OADDN_{[1]}$ instead,
with the scratch bit as the enable bit.  The extra controlled$^2$-NOT's that
compute and clear the AND bit do not affect the leading behavior of the count
of elementary gates. The only effect on the leading behavior is that
$MUXFA_{[2]}$ can be replaced by $MUXFA_{[1]}$, for which
\begin{eqnarray}
& \left[MUXFA_{[1]}\right]^{\rm worst}\>&=\> [2,2,2,1,0] \, ,\nonumber\\
& \left[MUXFA_{[1]}\right]^{\rm ave}\>&=\> \left[{1\over 2},{5\over 4},{7\over
4},{1\over 2},0\right] \, .
\end{eqnarray}
Hence we find
\begin{eqnarray}
&&[EXPN]_{{\rm enhanced},2K+2}^{\rm worst} \> =  \> \> LK^2\cdot [16,8,24,4,0]
\cdot \Bigl( \> 1 + O(1/K)\>\Bigr )\, ,\nonumber\\
&&[EXPN]_{{\rm enhanced},2K+2}^{\rm ave} \> =  \> \> LK^2\cdot [10,5,19,2,0]
\cdot \Bigl( \> 1 + O(1/K)\>\Bigr )\, ,
\end{eqnarray}
and
\begin{eqnarray}
&&[EXPN]_{{\rm enhanced},2K+2}^{\rm worst~ pulses} \> =  \> \> 240 LK^2\cdot
\Bigl( \> 1 + O(1/K)\>\Bigr )\, ,\nonumber\\
&&[EXPN]_{{\rm enhanced},2K+2}^{\rm ave~ pulses} \> =  \> \> 186 LK^2\cdot
\Bigl( \> 1 + O(1/K)\>\Bigr )\, .
\end{eqnarray}
The precise count in the average case is
\begin{eqnarray}
&[EXPN]_{{\rm enhanced},2K+2}^{\rm ave} \> =  \> \>
(L-1)\cdot&[10K^2-14K+4,5K^2+10K -14,19K^2-34K+21,\nonumber\\
&&2K^2-4K+2,0]+[2,{1\over2}K+1,0,0,0]\, ,\nonumber\\
&[EXPN]_{{\rm enhanced},2K+2}^{\rm ave~ pulses} \> =  \> \>
(L-1)\cdot&(186K^2-238K+99) +{5\over2}K+7\, .
\end{eqnarray}
Note that, in this version of the algorithm, no controlled$^4$-NOT gates are
needed.

\subsection{Basic machine}
\label{sec:basic_count}

Now we consider the basic machine, first with $2K+3$ bits of scratch space.  We
use one of our extra scratch bits to combine the enable bits for $OADDN$ as
explained above.  The other extra bit is used to replace $MUXFA_{[1]}$ by the
version $MUXFA''_{[1]}$ given in Eq.\ (\ref{muxfaprimeprime})---$MUXFA''_{[1]}$
uses only the gates available on the basic machine. The new count is
\begin{eqnarray}
& \left[MUXFA''_{[1]}\right]^{\rm worst}\>&=\> [2,2,4] \, ,\nonumber\\
& \left[MUXFA''_{[1]}\right]^{\rm ave}\>&=\> \left[{1\over 2},{7\over
4},{11\over 4}\right] \, .
\end{eqnarray}
The $LT$ operation need not be modified, as it requires no controlled$^3$-NOT
or controlled$^4$-NOT gates.  We therefore find
 \begin{eqnarray}
&&[EXPN]_{{\rm basic},2K+3}^{\rm worst} \> =  \> \> LK^2\cdot [16,8,32] \cdot
\Bigl( \> 1 + O(1/K)\>\Bigr )\, ,\nonumber\\
&&[EXPN]_{{\rm basic},2K+3}^{\rm ave} \> =  \> \> LK^2\cdot [10,7,23] \cdot
\Bigl( \> 1 + O(1/K)\>\Bigr )\, ,
\end{eqnarray}
and
\begin{eqnarray}
&&[EXPN]_{{\rm basic},2K+3}^{\rm worst~ pulses} \> =  \> \>  280 LK^2\cdot
\Bigl( \> 1 + O(1/K)\>\Bigr )\, ,\nonumber\\
&&[EXPN]_{{\rm basic},2K+3}^{\rm ave~ pulses} \> =  \> \> 206 LK^2\cdot \Bigl(
\> 1 + O(1/K)\>\Bigr )\, .
\end{eqnarray}
With the subleading corrections we have in the average case
\begin{eqnarray}
&[EXPN]_{{\rm basic},2K+3}^{\rm ave} \> =  \> \>
(L-1)\cdot&[10K^2-14K+4,7K^2+6K -12,23K^2-42K+25]\nonumber\\
&& +[2,{1\over 2}K+1,0]\, ,\nonumber\\
&[EXPN]_{{\rm basic},2K+3}^{\rm ave~ pulses} \> =  \> \> (L-1)\cdot
&(206K^2-278K+119)+{5\over 2}K+7 \, .
\end{eqnarray}

We can squeeze the scratch space down to $2K+2$ bits if we replace
$MUXFA''_{[1]}$ by $MUXFA'''_{[1]}$ given in Eq.\ (\ref{muxfappp}), which does
not require an extra scratch bit.  The gate count becomes
\begin{eqnarray}
& \left[MUXFA'''_{[1]}\right]^{\rm worst}\>&=\> [2,2,6] \, ,\nonumber\\
& \left[MUXFA'''_{[1]}\right]^{\rm ave}\>&=\> \left[{1\over 2},{5\over
4},{15\over 4}\right] \, ,
\end{eqnarray}
so that we now have
 \begin{eqnarray}
&&[EXPN]_{{\rm basic},2K+2}^{\rm worst} \> =  \> \> LK^2\cdot [16,8,40] \cdot
\Bigl( \> 1 + O(1/K)\>\Bigr )\, ,\nonumber\\
&&[EXPN]_{{\rm basic},2K+2}^{\rm ave} \> =  \> \> LK^2\cdot [10,5,27] \cdot
\Bigl( \> 1 + O(1/K)\>\Bigr )\, ,
\end{eqnarray}
and
\begin{eqnarray}
&&[EXPN]_{{\rm basic},2K+2}^{\rm worst~ pulses} \> =  \> \>  316 LK^2\cdot
\Bigl( \> 1 + O(1/K)\>\Bigr )\, ,\nonumber\\
&&[EXPN]_{{\rm basic},2K+2}^{\rm ave~ pulses} \> =  \> \> 224 LK^2\cdot \Bigl(
\> 1 + O(1/K)\>\Bigr )\, .
\end{eqnarray}
The precise count of gates and pulses in the average case is
\begin{eqnarray}
&[EXPN]_{{\rm basic},2K+2}^{\rm ave} \> =  \> \>
(L-1)\cdot&[10K^2-14K+4,5K^2+10K -14,27K^2-50K+29]+\nonumber\\
&&[2,{1\over 2} K+1,0]\, ,\nonumber\\
&[EXPN]_{{\rm basic},2K+2}^{\rm ave~ pulses} \> =  \> \> (L-1)\cdot
&(224K^2-314K+137) +{5\over 2}K+7\, .
\end{eqnarray}

To squeeze the scratch space by yet another bit, we must abandon the extra bit
used by $MULN$.  We then construct $MUXFA''''_{[2]}$ by expanding the
controlled$^3$-NOT and controlled$^4$-NOT gates in terms of controlled$^2$-NOT
gates, as discussed in Sec.\ \ref{sec:addition}.  We find that
\begin{eqnarray}
& \left[MUXFA''''_{[2]}\right]^{\rm worst}\>&=\> [2,1,15] \, ,\nonumber\\
& \left[MUXFA''''_{[2]}\right]^{\rm ave}\>&=\> \left[{1\over 2},1,{37\over
4}\right] \, ;
\end{eqnarray}
therefore,
\begin{eqnarray}
&&[EXPN]_{{\rm basic},2K+1}^{\rm worst} \> =  \> \> LK^2\cdot [16,4,76] \cdot
\Bigl( \> 1 + O(1/K)\>\Bigr )\, ,\nonumber\\
&&[EXPN]_{{\rm basic},2K+1}^{\rm ave} \> =  \> \> LK^2\cdot [10,4,49] \cdot
\Bigl( \> 1 + O(1/K)\>\Bigr )\, ,
\end{eqnarray}
and
\begin{eqnarray}
&&[EXPN]_{{\rm basic},2K+1}^{\rm worst~ pulses} \> =  \> \>  568 LK^2\cdot
\Bigl( \> 1 + O(1/K)\>\Bigr )\, ,\nonumber\\
&&[EXPN]_{{\rm basic},2K+1}^{\rm ave~ pulses} \> =  \> \> 373 LK^2\cdot \Bigl(
\> 1 + O(1/K)\>\Bigr )\, .
\end{eqnarray}
Including the subleading corrections the count in the average case is
\begin{eqnarray}
&[EXPN]_{{\rm basic},2K+1}^{\rm ave} \> =  \> \>(L-1)\cdot
&[10K^2-14K+4,4K^2+8K -12,49K^2-76K+30]\nonumber\\
&&+[2,{1\over 2} K+1,0]\, ,\nonumber\\
&[EXPN]_{{\rm basic},2K+1}^{\rm ave~ pulses} \> =  \> \>
(L-1)\cdot&(373K^2-506K+154)+{5\over 2}K+7 \, .
\end{eqnarray}

Our results for the average number of gates and pulses are summarized in the
following table:
\begin{center}
\begin{tabular}{|c|c|c||c|c|} \cline{2-5}
\multicolumn{1}{c|}{} & \multicolumn{2}{c||}{basic} &
\multicolumn{2}{c|}{enhanced} \\ \hline
scratch&gates & pulses& gates& pulses \\ \hline\hline
2K+1 &[10,4,49]&373&[10,4,17,3,2]& 198 \\
2K+2 & [10,5,27]& 224& [10,5,19,2,0]&186 \\ \cline{4-5}
2K+3 & [10,7,23]& 206 &\multicolumn{2}{c}{}\\ \cline{1-3}
\end{tabular}
\end{center}
\begin{equation}
\label{gate_table}
\end{equation}
Each entry in the table is the coefficient of $LK^2$ (the leading term) in the
number of gates or pulses, where the notation for the number of gates is that
defined in Eq.\ (\ref{enhanced_count},\ref{basic_count}).  Of course, the
numbers just represent our best effort to construct an efficient network.
Perhaps a more clever designer could do better.

\subsection{Unlimited Space}
\label{sec:unlimited}

The gate counts summarized in Eq.\ (\ref{gate_table}) provide a ``case study''
of the tradeoff between the amount of scratch space and the speed of
computation. But all of the algorithms described above are quite parsimonious
with scratch space.  We will now consider how increasing the amount of scratch
space considerably allows us to speed things up further.

First of all, recall that our $OADDN$ routine calls the comparison operator
$LT$ four times, twice running forwards and twice running in reverse.  The
point was that we wanted to clear the scratch space used by $LT$ before $MADD$
acted, so that space could be reused by $MADD$.  But if we were to increase the
scratch space by $K-1$ bits, it would not be necessary for $LT$ to run
backwards before $MADD$ acts.  Instead, a modified $OADDN$ routine could clear
the scratch space used by $LT$ and by $MADD$, running each subroutine only
twice (once forward and once backward).

Thus, with adequate space, we can replace Eq.\ (\ref{expn_ops}) with
\begin{equation}
[EXPN] \> =  \> \Bigl (\> 4LK^2\cdot \left[MUXFA_{[1]}\right] + 4LK\cdot [LT]
\>\Bigr ) \Bigl( \> 1 + O(1/K)\>\Bigr ) \, .
\end{equation}
Using this observation, we can modify our old network on the enhanced machine
(with $2K+2$ bits of scratch) to obtain
\begin{eqnarray}
&&[EXPN]_{{\rm enhanced},3K+1}^{\rm ave} \> =  \> \> LK^2\cdot [6,5,13,2,0]
\cdot \Bigl( \> 1 + O(1/K)\>\Bigr )\, ,\nonumber\\
&&[EXPN]_{{\rm enhanced},3K+1}^{\rm ave~ pulses} \> =  \> \> 140 LK^2\cdot
\Bigl( \> 1 + O(1/K)\>\Bigr )\, ,
\end{eqnarray}
about 25\% faster.

To do substantially better requires much more space.  Optimized for speed, our
algorithms will never clear the scratch space at intermediate stages of the
computation.  Instead, $EXPN$ will carry out of order $LK$ additions, filling
new space each time a comparison is performed or a sum is computed.  Once the
computation of $x^a ~ ({\rm mod} ~N)$ is complete, we copy the result and then
run the computation backwards to clear all the scratch space.  But with
altogether $\sim LK$ $ADDN$'s, each involving a comparison and a sum, we fill
about $2LK^2$ qubits of scratch space.  Combining the cost of running the gates
forward and backward, we have
\begin{equation}
[EXPN]\cdot [EXPN^{-1}] \> =  \> \Bigl (\> 2LK^2\cdot \left[MUXFA_{[1]}\right]
+ 2LK\cdot [LT] \>\Bigr ) \Bigl( \> 1 + O(1/K)\>\Bigr ) \, ,
\end{equation}
and therefore
\begin{eqnarray}
&&[EXPN]_{{\rm enhanced},\mathchar"0218 2LK^2}^{\rm ave} \> =  \> \> LK^2\cdot
\left[3,{5\over 2},{13\over 2},1,0\right] \cdot \Bigl( \> 1 + O(1/K)\>\Bigr )\,
,\nonumber\\
&&[EXPN]_{{\rm enhanced},\mathchar"0218 2LK^2}^{\rm ave~ pulses} \> =  \> \> 70
LK^2\cdot \Bigl( \> 1 + O(1/K)\>\Bigr )\, ,
\end{eqnarray}
another factor of 2 improvement in speed.

For asymptotically large $K$, further improvements are possible, for we can
invoke classical algorithms that multiply $K$-bit numbers in time less than
O($K^2$).   The fastest known, the Sch\"onhage-Strassen algorithm, requires
O($K\log K\log\log K$) elementary operations \cite{fast_mult}. It thus should
be possible to perform modular exponentiation on a quantum computer in a time
of order $LK\log K\log\log K$.  We have not worked out the corresponding
networks in detail, or determined the precise scratch space requirements for
such an algorithm.

\subsection{Minimal Space}
\label{sec:minimal}
Now consider the other extreme, where we disregard speed, and optimize our
algorithms to minimize space.

Since addition is an invertible operation, it is possible to construct a
unitary ``overwriting addition'' operator that adds a $c$-number to a
$q$-number and replaces the $q$ number addend with the sum.  But the
construction of our $OADDN$ operator involved two stages---first we performed
the addition {\it without} overwriting the input, and then ran the addition
routine backwards to erase the input.  Thus, our overwriting $OADDN$ routine
for adding a $K$-bit $c$-number to a $K$-bit $q$-number (mod $N$) required
$K+1$ bits of scratch space.

There is no reason in principle why this scratch space should be necessary
(though eliminating it may slow down the computation).  In fact, we will show
that it is possible to add without using any scratch space at all.  Of course,
we will still need a comparison bit to perform mod $N$ addition.  And there is
no obvious way to eliminate the need for a $K$-bit scratch register that stores
partial sums when we multiply.  Still, using overwriting addition, we can
construct an $EXPN$ operator that requires just $K+1$ bits of scratch space
(compared to $2K+1$ in our best previous effort).  The price we pay is that the
computation slows down considerably.

The key to adding without scratch space is to work from left to right instead
of right to left.  It is sufficient to see how to add a single-bit $c$-number
$a_0$ to a $K$-bit $q$-number $b$, obtaining a $(K+1)$-bit $q$-number.  Of
course, if the classical bit is 0, we do nothing.  If the classical bit is 1,
we perform addition by executing the pseudo-code:
\begin{eqnarray}
\label{pseudo_oadd}
{\tt if}&\ b_{K-1}=b_{K-2}=\cdots=b_1=b_0=1:\quad& {\tt flip}\  b_K\nonumber\\
{\tt if}&\ b_{K-2}=b_{K-3}=\cdots=b_1=b_0=1:\quad& {\tt flip}\
b_{K-1}\nonumber\\
&\cdot\nonumber\\
&\cdot\nonumber\\
{\tt if}&\ b_{1}=b_{0}=1:\quad& {\tt flip} \ b_2\nonumber\\
{\tt if}&\ b_{0}=1:\quad& {\tt flip} \ b_1\nonumber\\
& & {\tt flip}\  b_0
\end{eqnarray}
Thus, the operator
\begin{eqnarray}
ADD(a_0)_{\beta_{K},\beta}\>\> && \equiv\>\>
{\tt if} \ (a_0 = 1) \ \nonumber\\
 C_{\beta_0}\> && C_{\lbrack\!\lbrack \beta_0\rbrack\!\rbrack,\beta_1}
\>\dots\> C_{\lbrack\!\lbrack
\beta_0,\beta_1\dots\beta_{K-2}\rbrack\!\rbrack,\beta_{K-1}}\>
C_{\lbrack\!\lbrack \beta_0,\beta_1\dots\beta_{K-1}\rbrack\!\rbrack,\beta_K}
\end{eqnarray}
has the action
\begin{equation}
ADD(a_0)_{\beta_{K},\beta} : \>\> \bigl |0 \bigr\rangle_{\beta_K}\bigl
|{b}\bigr\rangle_\beta
\longmapsto \bigl |\left(b+a_0\right)_{K} \bigr\rangle_{\beta_K}\bigl
|{b+a_0}\bigr\rangle_\beta \, .
\end{equation}
It fills the $K+1$ qubits $|\cdot\bigr\rangle_{\beta_K}\bigl
|\cdot\bigr\rangle_\beta$ with the $(K+1)$-bit sum $b+a_0$.  To add a $K$-bit
$c$-number $a$ to the $K$-bit $q$-number $b$, we apply this procedure
iteratively.  After adding $a_0$ to $b$, we add $a_1$ to the $(K-1)$-qubit
number $b_{K-1}b_{K-2}\dots b_2 b_1$, then add $a_2$ to the $(K-2)$-qubit
number $b_{K-1}b_{K-2}\dots b_3 b_2$, and so on.  Thus, the computation of
$b+a$ requires in the worst case ($a=111\dots 11$) a total number of operations
\begin{equation}
[ADD(a)]\> = \>  [K,K,K-1,K-2,\dots,2,1] \> ;
\end{equation}
that is, $K$ NOT's, $K$ controlled-NOT's, $K-1$ controlled$^2$-NOT's, $\dots$,
2 controlled$^{K-1}$-NOT's, and 1 controlled$^K$-NOT. In the average case
(where half the bits of $a$ are zero), only half of these gates need to be
executed.  For the Cirac-Zoller device, figuring $2k+3$ laser pulses for a
controlled$^k$-NOT with $k\ge 1$, and one pulse for a NOT, this translates into
${1\over 6}K\left(2K^2 +15K+19\right)$ laser pulses for each $K$-bit addition,
in the worst case, or in the average case
\begin{equation}
[ADD]^{\rm ave~pulses}_{\rm no ~ scratch}={1\over 6} K^3 + {5\over 4} K^2 +
{19\over 12} K \, .
\end{equation}
We can easily promote this operation to a {\it conditional} $ADD$ with $\ell$
enable bits by simply adding the enable qubits to the control string of each
gate; the complexity then becomes
\begin{equation}
\left[ADD_{[\ell]}\right]^{\rm ave~pulses}_{\rm no ~ scratch}={1\over 6} K^3 +
\left({1\over 2}\ell+{5\over 4}\right) K^2 + \left({3\over 2}\ell + {31\over
12}\right) K \, ,\quad \ell\ge 1 \, .
\end{equation}

We will need to add mod $N$.  But if we can add, we can compare.  We can do the
comparison of $N-a$ and $b$ by adding $(2^K-N+a)$ to $b$; the final carry bit
will be 1 only for $a+b\ge N$.  Thus, we can use the overwriting addition
operation $ADD$ in place of $LT$ to fix the value of the select bit, and then
use a multiplexed version of $ADD$ to complete the mod $N$ addition. Following
this strategy, we construct an overwriting mod $N$ adder that uses just one
qubit of scratch space according to
\begin{eqnarray}
OADDN'&&(a,N)_{\lbrack\!\lbrack \mathchar"024C\rbrack\!\rbrack,  \beta,\beta_K}
\> \equiv \nonumber\\
\>&& ADD(a)_{\lbrack\!\lbrack \mathchar"024C\rbrack\!\rbrack,
\beta_K,\beta}\cdot MADD'(N-a, 2^K-a)_{\lbrack\!\lbrack
\mathchar"024C\rbrack\!\rbrack, \beta_K,\beta} \cdot
ADD(2^K-N+a)_{\lbrack\!\lbrack \mathchar"024C\rbrack\!\rbrack, \beta_K,\beta}:
\nonumber\\
&&  \qquad\bigl |0 \bigr\rangle_{\beta_K}\bigl |{b}\bigr\rangle_\beta
\longmapsto \bigl |0\bigr\rangle_{\beta_K}\bigl |{b+\mathchar"024C \wedge a ~
({\rm mod} ~ N)}\bigr\rangle_\beta \, .
\end{eqnarray}
Here each $ADD$ operation computes a $(K+1)$-bit sum as above, placing the
final carry bit in the qubit  $\bigl |\cdot\bigr\rangle_{\beta_K}$; however
$MADD'$ computes a $K$-bit sum -- it is a multiplexed adder that adds $N-a$ if
the select bit  $\bigl |\cdot\bigr\rangle_{\beta_K}$ reads 0, and adds $2^K -a$
if the select bit reads 1.  The construction of $MADD'$ follows the spirit of
the construction of $MADD$ described in Sec.\ \ref{sec:addition}. In the
average case, the number of laser pulses required to implement this $OADDN'$
operation is
\begin{equation}
\left[OADDN'_{[\ell]}\right]^{\rm ave ~ pulses}_{\rm 1~ scratch}= {7\over
12}K^3 + \left({7\over 4}\ell+{33\over 8}\right) K^2 + \left({15\over
4}\ell+{169\over 24}\right) K \, .
\end{equation}

The construction of the modular exponentiation operator $EXPN$ from this
$OADDN'$ operator follows the construction described in Sec.\ \ref{sec:detail}.
 Thus, using the expression for $[EXPN]$ in terms of $\left[OADDN_{[2]}\right]$
implicit in Eq.\ (\ref{expn_count}), we find that with $K+1$ qubits of scratch
space, the $EXPN$ function can be computed, in the average case, with a number
of laser pulses given by
\begin{equation}
[EXPN]^{\rm ave ~ pulses}_{K+1}=(L-1)\cdot \left( {7\over 6} K^4 + {169\over
12} K^3 + {83\over 6} K^2 - {97\over 12} K\right) + {5\over 2}K + 7 \, .
\end{equation}
For small values of $K$ ($K<7$), fewer pulses are required than for the
algorithms described in Sec.\ \ref{sec:enhanced_count} and
\ref{sec:basic_count}.

\section{$N=15$}
\label{sec:fifteen}

As we noted in Sec.\ \ref{sec:factoring}, Shor's factorization algorithm fails
if $N$ is even or a prime power ($N=p^\alpha$, $p$ prime).  Thus, the smallest
composite integer $N$ that can be successfully factored by Shor's method is
$N=15$.  Though factoring 15 is not very hard, it is amusing to consider the
computational resources that would be needed to solve this simplest of quantum
factoring problems on, say, a linear ion trap.

Appealing to Eq.\ (\ref{pulses_enh_twoKone}), with $K=4$ and $L=2K=8$, our
``average case'' estimate of the number of laser pulses required on a machine
with altogether $K+L+(2K+1)=21$ qubits of storage is 15,284.  With 22 qubits of
storage, our estimate improves to 14,878 pulses.  With another three qubits (25
total), we can use the technique described in Sec.\ \ref{sec:unlimited} to
achieve a further improvement in speed.

Several observations allow us to reduce these resources substantially further.
First of all, we notice that, for any positive integer $x$ with $x<15$ and
$gcd(x,15)=1$ ({\it i.e.}, for $x=1,2,4,7,8,11,13,14$), we have $x^4\equiv 1
{}~({\rm mod} ~15)$.  Therefore,
\begin{equation}
x^a=  x^{2a_1}\cdot x^{a_0}\ ;
\end{equation}
only the last two bits of $a$ are relevant in the computation of $x^a$.  Hence,
we might as well choose $L=2$ instead of $L=8$, which reduces the number of
elementary operations required by a factor of about 7.  (Even if the value of
$L$ used in the evaluation of the discrete Fourier transform is greater than 2,
there is still no point in using $L>2$ in the evaluation of the modular
exponential function.)

Second, we can save on storage space (and improve speed) by noting that the
overwriting addition routine described in Sec.\ \ref{sec:minimal} is reasonably
efficient for small values of $K$.  For $K=4$ and $L=2$, we need 11 qubits of
storage and an estimated 1406 laser pulses.

For $N=15$, the above is the most efficient algorithm we know that actually
computes $x^a$ on the quantum computer.  We can do still better if we are
willing to allow the classical computer to perform the calculation of $x^a$.
Obviously, this strategy will fail dismally for large values of $K$---the
classical calculation will require exponential time.  Still, if our goal is
merely to construct the entangled state
\begin{equation}
\label{mod_entangled}
{1\over 2^{L/2}}\sum_{a} |a \rangle_i |x^a (mod ~N)\rangle_o \ ,
\end{equation}
while using our quantum computational resources as sparingly as possible, then
classical computation of $x^a$ is the most efficient procedure for small $K$.

So we imagine that $x<15$ with $gcd(x,15)=1$ is randomly chosen, and that the
classical computer generates a ``lookup table'' by computing the four bit
number $x^a$ (mod 15) for $a=0,1,2,3$.  The classical computer then instructs
the quantum computer to execute a sequence of operations that prepares the
state Eq.\ (\ref{mod_entangled}).  These operations require no scratch space at
all, so only $L+K=6$ qubits of storage are needed to prepare the entangled
state.

The ``worst case'' (most complex lookup table) is $x=7$ or $x=13$.  The lookup
table for $x=7$ is:
\begin{center}
\begin{tabular}{|ll|llll|} \hline
\multicolumn{2}{|c|}{$a$}& \multicolumn{4}{c|}{$7^a ~({\rm mod}~ 15)$}  \\
\hline
0&0&0&0&0&1\\
0&1&0&1&1&1\\
1&0&0&1&0&0\\
1&1&1&1&0&1 \\ \hline
$a_1$&$a_0$&$b_3$&$b_2$&$b_1$&$b_0$\\ \hline
\end{tabular}
\end{center}
\begin{equation}
\end{equation}
An operator
\begin{equation}
EXPN(x=7,N=15)_{\alpha,\beta}: \>\> \bigl|{a}\bigr\rangle_\alpha^* \bigl
|{0}\bigr\rangle_\beta \longmapsto
\bigl|{a}\bigr\rangle_\alpha^* \bigl |{7^a ~({\rm mod} ~ 15)}\bigr\rangle_\beta
\end{equation}
that recreates this table can be constructed as
\begin{equation}
\label{expn_short}
EXPN(x,N)_{\alpha,\beta}\> \equiv\>  C_{\alpha_1}C_{\lbrack\!\lbrack
\alpha_1,\alpha_0\rbrack\!\rbrack,\beta_1}C_{\alpha_0}C_{\lbrack\!\lbrack
\alpha_1,\alpha_0\rbrack\!\rbrack,\beta_2}C_{\alpha_1}C_{\lbrack\!\lbrack
\alpha_1,\alpha_0\rbrack\!\rbrack,\beta_0}C_{\alpha_0}C_{\lbrack\!\lbrack
\alpha_1,\alpha_0\rbrack\!\rbrack,\beta_3}C_{\beta_2}C_{\beta_0} \ .
\end{equation}
The two NOT's at the beginning generate a ``table'' that is all 1's in the
$\beta_0$ and $\beta_2$ columns, and all 0's in the $\beta_1$ and $\beta_3$
columns.  The remaining operations fix the one incorrect entry in each row of
the table.  Thus, we have constructed an $EXPN$ operator with complexity
\begin{equation}
\left[EXPN(7,15)\right]= [6,0,4] \ ;
\end{equation}
it can be implemented with 34 laser pulses on the Cirac-Zoller device.  Since
two additional pulses suffice to prepare the input register in the
superposition state
\begin{equation}
\label{fifteen_input}
{1\over 2}\sum_{a=0}^{3} |a \rangle_i
\end{equation}
before $EXPN$ acts, we need 36 laser pulses to prepare the entangled state Eq.\
(\ref{mod_entangled}).

The $EXPN$ operator constructed in Eq.\ (\ref{expn_short}) acts trivially on
the input $q$-number $a$.  Of course, this feature is not necessary; as long as
the output state has the right correlations between the
$\bigl|{\cdot}\bigr\rangle_\alpha^*$ and $\bigl|{\cdot}\bigr\rangle_\beta$
registers, we will successfully prepare the entangled state  Eq.\
(\ref{mod_entangled}).  By exploiting this observation, we can achieve another
modest improvement in the complexity of $EXPN$; we see that
\begin{equation}
 C_{\lbrack\!\lbrack
\alpha_1,\alpha_0\rbrack\!\rbrack,\beta_3}C_{\alpha_0}C_{\lbrack\!\lbrack
\alpha_1,\alpha_0\rbrack\!\rbrack,\beta_0}C_{\alpha_1}C_{\lbrack\!\lbrack
\alpha_1,\alpha_0\rbrack\!\rbrack,\beta_2}C_{\alpha_0}C_{\lbrack\!\lbrack
\alpha_1,\alpha_0\rbrack\!\rbrack,\beta_1}C_{\beta_2}C_{\beta_0} \ .
\end{equation}
applied to the input Eq.\ (\ref{fifteen_input}) also produces the output Eq.\
(\ref{mod_entangled}), even though it flips the value of $a_1$.  Compared to
Eq.\ (\ref{expn_short}), we do without the final NOT gate, and hence save one
laser pulse.  We can do better still by invoking the ``custom gates'' described
in Appendix A; another implementation of the $EXPN$ operator is
\begin{equation}
\label{expn_custom}
EXPN'(x,N)_{\alpha,\beta}\> \equiv\>  C_{\lbrack\!\lbrack
\overline\alpha_1,\alpha_0\rbrack\!\rbrack,\beta_1}C_{\lbrack\!\lbrack
\overline\alpha_1,\overline\alpha_0\rbrack\!\rbrack,\beta_2}C_{\lbrack\!\lbrack
\alpha_1,\overline\alpha_0\rbrack\!\rbrack,\beta_0}C_{\lbrack\!\lbrack
\alpha_1,\alpha_0\rbrack\!\rbrack,\beta_3}C_{\beta_2}C_{\beta_0} \ .
\end{equation}
Here, $C_{\lbrack\!\lbrack
\overline\alpha_1,\overline\alpha_0\rbrack\!\rbrack,\beta_2}$, for example, is
a gate that flips the value of qubit $\beta_2$ if and only if both qubit
$\alpha_1$ and qubit $\alpha_0$ have the value {\it zero} rather than one (see
Appendix A).
Each custom gate in Eq.\ (\ref{expn_custom}) can be implemented with 7 laser
pulses.  Hence, compared to Eq.\ (\ref{expn_short}) we save 4 pulses, and the
state Eq.\ (\ref{mod_entangled}) can be prepared with just 32 pulses.

To complete the task of ``factoring 15,'' it only remains to perform the
Fourier transform on the input register and read it out.  The measured value,
the result of our quantum computation, will be a nonnegative integer $y<2^L$
satisfying
\begin{equation}
\label{y_fifteen}
{y\over 2^L} ={{\rm integer}\over r} \ ,
\end{equation}
where $r$ is the order of $x$ mod $N$ ($r$=4 in the case $N$=15 and $x$=7), and
the integer takes a random value ranging from 0 to $r-1$.  (Here the
probability distribution for $y$ is actually perfectly peaked at the values in
Eq.\ (\ref{y_fifteen}), because $r$ divides $2^L$.)  Thus, if we perform the
Fourier transform with $L=2$, the result for $y$ is a completely {\it random}
number ranging over $y=0,1,2,3$.  (Even so, by reducing $y$/4 to lowest terms,
we succeed in recovering the correct value of $r$ with probability 1/2.)

It is a bit disappointing to go to all the trouble to prepare the state Eq.\
(\ref{mod_entangled}) only to read out a random number in the end.  If we wish,
we can increase the number of qubits $L$ of the input register (though the
$EXPN$ operator will still act only on the last two qubits).  Then the outcome
of the calculation will be a random multiple of $2^{L-2}$.  But the probability
of recovering the correct value of $r$ is still $1/2$.

Once we have found $r=4$, a classical computer calculates $7^{(4/2)}\pm 1
\equiv 3,5 ~({\rm mod}~ N)$, which are, in fact, the factors of $N=15$.  Since
the $L=2$ Fourier transform can be performed using $L(2L-1)=6$ laser pulses on
the ion trap, we can ``factor 15'' with 38 pulses (not counting the final
reading out of the device). For values of $x$ other than 7 and 13, the number
of pulses required is even smaller.

\section{Testing the Fourier transform}

In Shor's factorization algorithm, a periodic function (the modular exponential
function) is computed, creating entanglement between the input register and the
output register of our quantum computer.  Then the Fourier transform is applied
to the input register, and the input register is read.  In Sec.\
\ref{sec:fifteen}, we noted that a simple demonstration of this procedure
(factorization of 15) could be carried out on a linear ion trap, requiring only
a modest number of laser pulses.

Here we point out an even simpler demonstration of the principle underlying
Shor's algorithm.  Consider the function
\begin{equation}
f_K(a)=a ~({\rm mod} ~2^K) \ .
\end{equation}
Evaluation of this function is very easy, since it merely copies the last $K$
bits of the argument $a$.  A unitary operator $MOD_{2^K}$ that acts according
to
\begin{equation}
MOD_{2^K}:\>\> \bigl|a\bigr\rangle_\alpha^* \bigl|0\bigr\rangle_\beta
\longmapsto \bigl|a\bigr\rangle_\alpha^* \bigl|a ~({\rm mod}
{}~2^K)\bigr\rangle_\beta
\end{equation}
can be constructed as
\begin{equation}
MOD_{2^K}\>\equiv\> C_{\lbrack\!\lbrack \alpha_{K-1}
\rbrack\!\rbrack,\beta_{K-1}}\cdots C_{\lbrack\!\lbrack \alpha_{1}
\rbrack\!\rbrack,\beta_{1}} C_{\lbrack\!\lbrack \alpha_{0}
\rbrack\!\rbrack,\beta_{0}}
\end{equation}
(where $\bigl|\cdot\bigr\rangle_\alpha^*$ is an $L$-qubit register and
$\bigl|0\bigr\rangle_\beta$ is a $K$-qubit register).  These $K$ controlled-NOT
operations can be accomplished with $5K$ laser pulses in the ion trap.
Including the $L$ single qubit rotations needed to prepare the input register,
then, the entangled state
\begin{equation}
{1\over 2^{L/2}}\sum_{a=0}^{2^L-1} \bigl|a\bigr\rangle_\alpha^* \bigl|a ~({\rm
mod} ~2^K)\bigr\rangle_\beta
\end{equation}
can be generated with $5K+L$ pulses.

Now we can Fourier transform the input register ($L(2L-1)$ pulses), and read it
out.  Since the period $2^K$ of $f_K$ divides $2^L$, the Fourier transform
should be perfectly peaked about values of $y$ that satisfy
\begin{equation}
y=2^{L-K}\cdot\left({\rm integer}\right)
\end{equation}
Thus, $y_{K-1},\dots,y_1,y_0$ should be identically zero, while
$y_{L-1},\dots,y_{K+1},y_K$ take random values.

The very simplest demonstration of this type ($L=2, K=1$) requires only three
ions.  Since $f_1$ has period 2, the two-qubit input register, after Fourier
transforming, should read $y_1={\rm random}, y_0=0$.  This demonstration can be
performed with 13 laser pulses (not counting the final reading out), and should
be feasible with current technology.

\acknowledgments
We thank Al Despain, Jeff Kimble, and Hideo Mabuchi for helpful discussions and
encouragement. This research was supported in part by DOE Grant No.
DE-FG03-92-ER40701, and by Caltech's Summer Undergraduate Research Fellowship
program.

\appendix
\section{Custom gates}

In the algorithms that we have described in this paper, we have used the
controlled$^k$-NOT operator as our fundamental quantum gate.  Of course, there
is much arbitrariness in this choice.  For example, instead of the operation
$C_{\lbrack\!\lbrack{i_1,\dots,i_k}\rbrack\!\rbrack,j}$, which flips qubit $j$
if and only if qubits $i_1,\dots,i_k$ all take the value 1, we could employ a
gate that flips qubit $j$ if and only if $i_1 i_2 \dots i_k$ is some other
specified string of $k$ bits.  This generalized gate, like
$C_{\lbrack\!\lbrack{i_1,\dots,i_k}\rbrack\!\rbrack,j}$ itself, can easily be
implemented on, say, a linear ion trap.  We remark here that using such
``custom gates'' can reduce the complexity of some algorithms (as measured by
the total number of laser pulses required).

To see how these generalized gates can be constructed using the ion trap, we
note first of all that if we apply an appropriately tuned $3\pi$ pulse (instead
of a $\pi$ pulse) to the $i$th ion,\footnote{Alternatively, we can implement
$\tilde W^{(i)}_{\rm phon}$ with a $\pi$ pulse if the laser phase is
appropriately adjusted.} then the operation $W^{(i)}_{\rm phon}$ defined in
Eq.\ (\ref{Wphonon}) is replaced by
\begin{equation}
\tilde W^{(i)}_{\rm phon}:\>\> \left\{\begin{array}{lll}&\bigl |
g\bigr\rangle_i| 0\bigr\rangle_{\rm CM}\longmapsto &| g\bigr\rangle_i|
0\bigr\rangle_{\rm CM}\nonumber\\
&\bigl | e\bigr\rangle_i| 0\bigr\rangle_{\rm CM}\longmapsto i&|
g\bigr\rangle_i| 1\bigr\rangle_{\rm CM}\end{array}\right\}\nonumber\\
\end{equation}
(whose nontrivial action differs by a sign from that of $W^{(i)}_{\rm phon}$).
With $W^{(i)}_{\rm phon}$ and $\tilde W^{(i)}_{\rm phon}$, we can construct an
alternative conditional phase gate
\begin{equation}
\tilde V^{(i,\overline j)}\>\equiv\>  W^{(i)}_{\rm phon}\cdot V^{(j)} \cdot
\tilde W^{(i)}_{\rm phon}:\>\> \bigl|\epsilon\bigr\rangle_i
\bigl|\eta\bigr\rangle_j \longmapsto (-1)^{\epsilon\wedge\mathchar"0218\eta}
\bigl|\epsilon\bigr\rangle_i \bigl|\eta\bigr\rangle_j \
\end{equation}
that acts nontrivially only if $\epsilon=1$ and $\eta=0$.  With an appropriate
change of basis, this conditional phase gate becomes
\begin{eqnarray}
C_{{\lbrack\!\lbrack \overline i \rbrack\!\rbrack},j}\>\equiv \left[\tilde
U^{(j)}\right]^{-1}\cdot V^{(j,\overline i)}\cdot \tilde U^{(j)}\> &&= \>
\left[\tilde U^{(j)}\right]^{-1}\cdot W^{(j)}_{\rm phon}\cdot V^{(i)} \cdot
\tilde W^{(j)}_{\rm phon}\cdot \tilde U^{(j)}:\nonumber\\
&&\bigl|\epsilon\bigr\rangle_i \bigl|\eta\bigr\rangle_j \longmapsto
\bigl|\epsilon\bigr\rangle_i \bigl|\eta\oplus\epsilon\oplus 1\bigr\rangle_j \,
,
\end{eqnarray}
a modified controlled-NOT gate that flips the target qubit if and only if the
control qubit reads {\it zero} (compare Eq.\ (\ref{trap_CN})).  Like the
controlled-NOT gate, then, $C_{{\lbrack\!\lbrack \overline i
\rbrack\!\rbrack},j}$ can be implemented with 5 laser pulses.  Following the
discussion in Sec.\ \ref{sec:NOTpulses}, it is straightforward to construct a
modified controlled$^k$-NOT gate with a specified ``custom'' control string,
for any $k\ge 1$.

\begin{figure}[htb]
\centerline{\epsfysize=3truecm \epsfbox{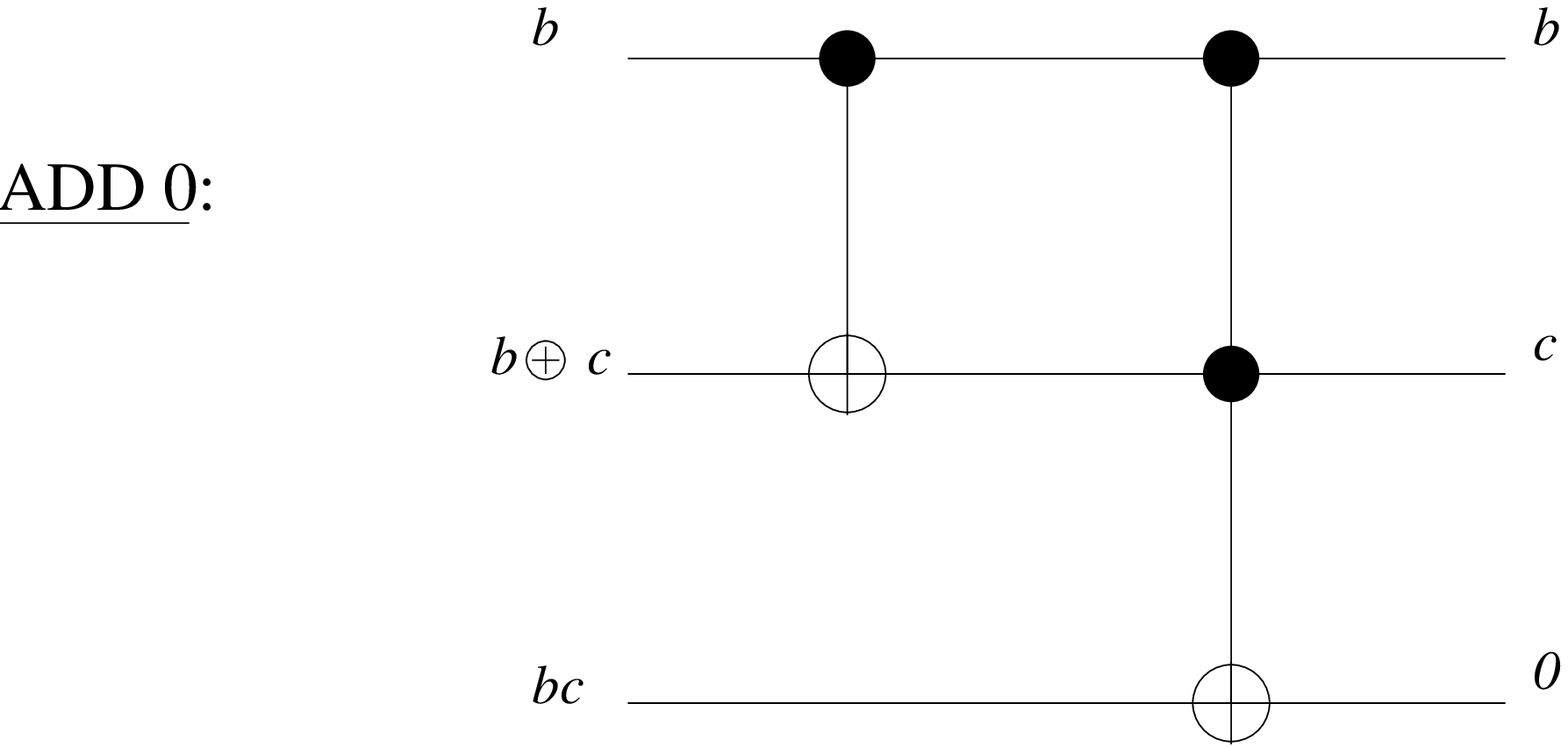}
\hskip 1truecm \epsfysize=3truecm \epsfbox{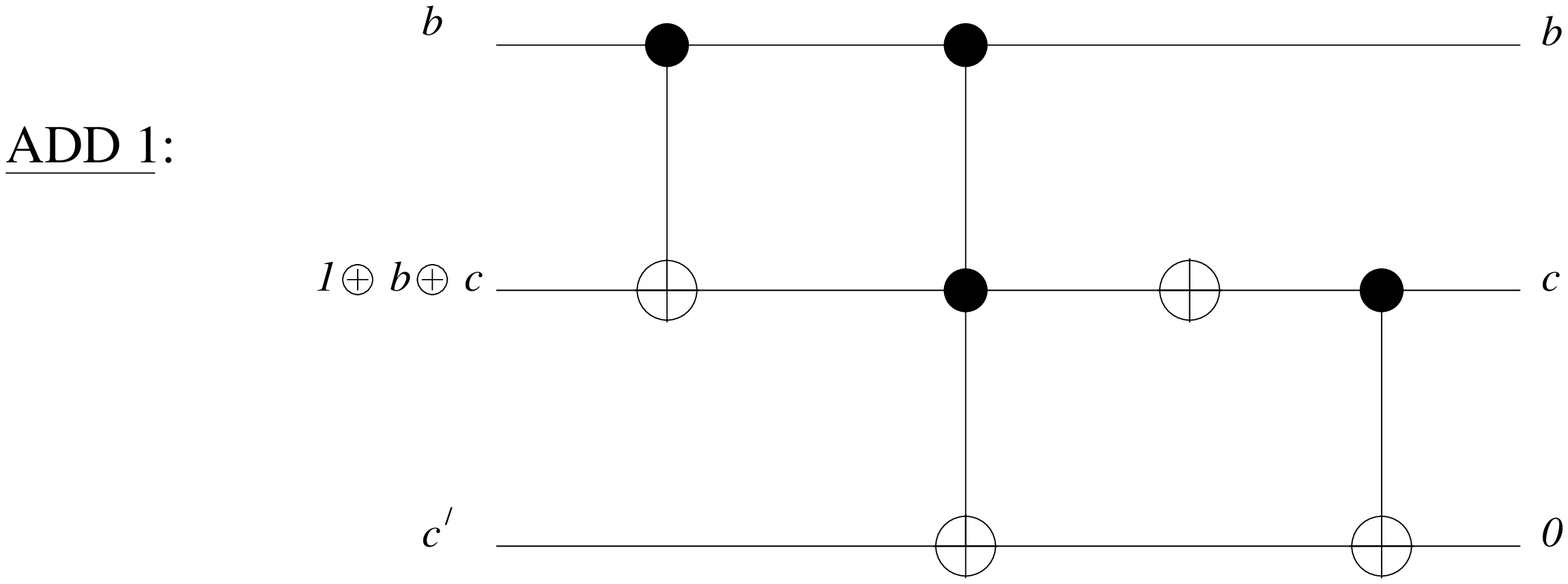} }
\leftline{\hskip 4.5truecm (a) \hskip 8truecm (b)}
\vskip .5cm
\caption{The full adder $FA(a)$.  The order of the gates (here and in all of
the following figures) is to be read from right to left.  The gate array shown
in (a)  adds the classical bit $a=0$; the second qubit carries the output sum
bit and the third qubit carries the output carry bit.  The gate array shown in
(b) adds the classical bit $a=1$.}
\label{figB}
\end{figure}

As a simple illustration of how a reduction in complexity can be achieved by
using custom gates, consider the full adder $FA(a)$ defined by Eq.\
(\ref{FA0},\ref{FA1}) and shown in Fig.\ \ref{figB}.  We can replace $FA(1)$ by
the alternative implementation
\begin{equation}
FA'(a=1)_{1,2,3}\> \equiv \> C_{\lbrack\!\lbrack \overline 1 \rbrack\!\rbrack,
2} C_{\lbrack\!\lbrack \overline 1,2\rbrack\!\rbrack, 3}
C_{\lbrack\!\lbrack 1\rbrack\!\rbrack, 3}
\end{equation}
(where the $\overline i$ indicates that qubit $i$ must have the value 0 (not 1)
for the gate to act nontrivially).
This saves one NOT gate, and hence one laser pulse, compared to the
implementation in Eq.\ (\ref{FA1}).  Another example of the use of custom gates
is described in Sec.\ \ref{sec:fifteen}.


\vfill\eject

\vfill\eject

\vfill\eject

\vfill\eject

\end{document}